\documentclass[manuscript,dvipsnames,nonacm]{acmart}

\AtBeginDocument{%
  \providecommand\BibTeX{{%
    \normalfont B\kern-0.5em{\scshape i\kern-0.25em b}\kern-0.8em\TeX}}}

\usepackage{tikz}
\usepackage{makecell}
\usetikzlibrary{positioning}
\usepackage{amsmath}
\usepackage{varwidth}
\usepackage{xcolor}
\usepackage{soul}
\usepackage{harveyballs}
\usepackage{array}
\usepackage{xspace}
\usepackage{soul}
\usepackage{subcaption}
\usepackage[export]{adjustbox}
\usepackage{fancyvrb}
\usepackage{vcell}
\usepackage{pgfplots}
\pgfplotsset{compat=1.15}
\usepgfplotslibrary{statistics}
\usepackage{graphicx}
\usepackage{lscape}
\usepackage{float}
\usepackage{enumitem}

\graphicspath{ {./images/} }

\usepackage{hyperref}
\usepackage{cleveref}

\crefname{lstlisting}{listing}{listings}
\Crefname{lstlisting}{Listing}{Listings}
\crefalias{stepenumi}{step}
\crefname{figure}{Figure}{Figures}
\crefname{equation}{Equation}{Equations}
\crefname{step}{Step}{Steps}
\crefname{appsec}{Appendix}{Appendices}
\crefname{line}{Line}{Lines}

\usetikzlibrary{shapes,arrows}
\tikzset{%
  base/.style = {inner sep=5pt,
                 text centered,
                 thin,
                 font=\rmfamily}
}

\tikzset{%
	square/.style = {shape=rectangle,
		text width=2.75cm,
		minimum height=1.75cm,
		text centered,
		font=\sffamily\small,
		very thick,
		draw=black}
}

\tikzset{%
	tab/.style = {shape=rectangle,
		text width=.25cm,
		minimum height=.25cm,
		text centered,
		font=\sffamily\small,
		very thick,
		draw=black}
}

\newcommand{\authoricon}{\protect\includegraphics[width=.15in, trim=0 .09in 0 0]{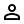}} 
\newcommand{\codeicon}{\protect\includegraphics[width=.15in, trim=0 .09in 0 0]{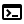}}
\newcommand{\rallyicon}{\protect\includegraphics[width=.13in, trim=0 .07in 0 0]{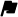}}
\newcommand{\webscienceicon}{\protect\includegraphics[width=.15in, trim=0 .071in 0 0]{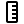}}

\acmConference[CHI '23]{ACM CHI Conference on Human Factors in Computing Systems}{April 23-28, '23}{Hamburg, Germany}
\acmBooktitle{CHI '23: ACM CHI Conference on Human Factors in Computing Systems}
\acmPrice{15.00}
\acmISBN{978-1-4503-XXXX-X/18/06}

\newcommand\WebScience{WebScience}
\newcommand\sdkName{WebScience}

\newcommand\Rally{Rally}
\newcommand\pioneer{Rally}

\newcommand\mozilla{Mozilla}
\newcommand\Mozilla{Mozilla}
\newcommand\firefox{Firefox}

\newcommand\redact[1]{#1\xspace}
\newcommand{\bnote}[1]{{}}
\newcommand{\knote}[1]{{}}
\newcommand{\anote}[1]{{}}

\definecolor{cb1}{RGB}{51, 34, 136}
\definecolor{cb2}{RGB}{17, 199, 51}
\definecolor{cb3}{RGB}{136, 204, 238}
\definecolor{cb4}{RGB}{221, 204, 119}
\definecolor{cb5}{RGB}{204, 102, 119}
\definecolor{cb6}{RGB}{136, 34, 85}

\begin{document}

\title{Rally and WebScience: A Platform and Toolkit for Browser-Based Research on Technology and Society Problems}

\author{Anne Kohlbrenner}
\email{akohlbrenner@princeton.edu}
\author{Ben Kaiser}
\email{bkaiser@princeton.edu}
\author{Kartikeya Kandula}
\email{kartkand@princeton.edu}
\author{Rebecca Weiss}
\email{rebecca.weiss@princeton.edu}
\author{Jonathan Mayer}
\email{jonathan.mayer@princeton.edu}
\affiliation{%
  \institution{Princeton University}
  \city{Princeton}
  \state{New Jersey}
  \country{USA}
}
\author{Ted Han}
\email{than@mozilla.com}
\author{Robert Helmer}
\email{rhelmer@mozilla.com}
\affiliation{%
  \institution{Mozilla}
  \city{San Francisco}
  \state{California}
  \country{USA}
}


\begin{abstract}
Empirical technology and society research is in a methodological crisis. Problems increasingly involve closed platforms, targeted content, and context-specific behavior. Prevailing research methods, such as surveys, tasks, and web crawls, pose design and ecological validity limitations.

Deploying studies in participant browsers and devices is a promising direction. These vantage points can observe individualized experiences and implement UI interventions in real settings.

We survey scholarship that uses these methods, annotating 284 sampled papers. Our analysis demonstrates their potential, but also recurring implementation barriers and shortcomings.

We then present \Rally{} and \sdkName{}, a platform and toolkit for browser-based research. These systems lower implementation barriers and advance the science of measuring online behavior. 

Finally, we evaluate \Rally{} and \sdkName{} against our design goals. We report results from a one-month pilot study on news engagement, analyzing 4,466,200 webpage visits from 1,817 participants. We also present observations from interviews with researchers using these systems.

\end{abstract}


\maketitle

\section{Introduction}

How can society understand the problems that arise from our interactions with online content and services? The challenges have never been more urgent: Harmful, toxic, and inauthentic content is rapidly and widely proliferating on the internet~\cite{cinelli_covid-19_2020, phadke_many_2020, edelson_2021_understanding-engagement}, while content moderation responses can be inconsistent and  inequitable~\cite{york_how_2019}. Personalized content and targeted advertisements distort individual beliefs and behaviors~\cite{chaney_how_2018} and permit potentially illegal discrimination~\cite{10.1145/3359301, tobin_hud_2019, faife_credit_2021}. Dark patterns manipulate individual decision making about privacy and purchasing~\cite{mathur21}. Platform companies self-preference and unfairly impede competition~\cite{jeffries_how_2020}.
And internet connectivity itself is unequally distributed, reflecting and reinforcing social injustices~\cite{vogels_digital_2021}.

Research on technology and society problems like these is in a methodological crisis~\cite{shapiro21, bak-coleman_2021_stewardship-collective,lazer2021meaningful}. The problems increasingly involve individual experiences, perceptions, and decisions both within the context of proprietary online services and across online services, targeted and personalized content, and black-box recommender and decision-making systems. Prevailing research methods
impose significant research design constraints and ecological validity limitations. Online platforms have the capacity to enable rigorous study on diverse topics, but they rarely enable independent research---and when they do, they usually impose significant limitations on problem selection, methods, or results reporting.

Researchers, civil society groups, and policymakers have begun to respond. An interdisciplinary group of scholars recently proposed recognition of a new ``crisis discipline'' at the intersection of technology and society, where we currently ``lack the scientific framework we would need to answer even the most basic questions'' and novel empirical research methods are essential for making progress~\cite{bak-coleman_2021_stewardship-collective}. Civil society groups are exploring new models for research support~\cite{shapiro21}. Legislators in the U.S. are considering proposals to mandate forms of research data access, and the EU's new Digital Services Act includes analogous provisions~\cite{persily_2021_facebook-hides,engler_2021_platform-data}.

Our goal in this work is to advance a promising and complementary response to the technology and society research methods crisis: deploying studies in participant web browsers and devices.
The upside of browser- and device-based research is that it can reflect a person's vantage point on real online experiences, enabling observation and intervention in a field setting. Prior work has relied on browser and device instrumentation to study technology and society problems, but the methods remain underutilized owing to recurring implementation barriers and shortcomings.

We begin in Section~\ref{sec:motivation} by describing the promise of browser- and device-based studies in example research topic areas. We offer a high-level comparison of research methods for technology and society problems, highlighting the opportunities afforded by browser- and device-based methods.

Next, in Section~\ref{sec:survey}, we contribute a survey of academic scholarship in applied sciences and social sciences that uses browser- and device-based methods. Because the relevant literature is vast and scattered across disciplines, we use stratified sampling to characterize prior work. We screen an initial sample of 3,341 papers that contain relevant terms down to a dataset of 284 papers that include at least one browser- or device-based study. We annotate each paper and then analyze the dataset, finding that browser- and device-based methods have been applied to a range of technology and society problem areas but face significant headwinds in implementation burden, instrumentation sophistication, and participant recruiting.

We then turn to system design and evaluation. \Cref{sec:rally-and-webscience-overview,sec:rally,sec:websci,sec:privacy} contribute the Rally platform and the WebScience toolkit for conducting browser-based research. We set the design goals for these systems by distilling discrete challenges for browser-based research from the results of our literature review. Rally is a new platform for deploying browser-based studies, facilitating participant recruiting, data reporting, and data analysis. WebScience is a new toolkit for implementing browser-based studies, offering standardized browser instrumentation and a state-of-the-art model for understanding how people navigate and pay attention to web pages. Rally and WebScience were developed in a close collaboration between academic researchers and a web browser vendor, offering production-level scale and implementation quality. At the time of writing, Rally and WebScience are already in use at multiple academic institutions and an investigative newsroom.

\Cref{sec:eval} quantitatively and qualitatively evaluates Rally and WebScience against our design goals. We first present preliminary data from a pilot study on political and COVID-19 news engagement, which enrolled 2,709 participants over four months. This pilot study demonstrates the effectiveness of Rally at building a participant panel and of WebScience at precisely instrumenting online activity, but it also reveals limitations of panel composition. We next focus our analysis on a recent one-month subset of data, which includes 4,466,200 webpage visits by 1,817 participants. This analysis highlights the necessity of WebScience's fine-grained measurement abilities. Finally, we qualitatively assess the systems by interviewing academic researchers and investigative journalists who are currently using them.
\footnote{All components of the evaluation were approved by the Princeton University Institutional Review Board and all participants consented.}

\Cref{sec:future-work,sec:conclusion} discuss directions for future work and conclude.

\section{Motivation}
\label{sec:motivation}
For decades, academic researchers across disciplines have applied standard quantitative and qualitative research methods to understand problems at the intersection of information technology and society. These methods will remain valuable, but they face increasing limitations and are no longer sufficient for studying pressing sociotechnical challenges.

We begin by discussing a set of societally urgent research topics that exemplify the need for new methods. We select topics where prior work has exhaustively attempted conventional methods but has been stymied by the limitations of those methods: harmful content networks, black-box recommender systems, and targeted advertising/personalization. A throughline in the topics we discuss is the involvement of closed platforms, which have declined to provide data that would facilitate academic scholarship.

Next, we provide a high-level overview of the methodological landscape for descriptive research on online experiences and for causal inference research about beliefs and behaviors related to online activity. We describe the strengths and limitations of conventional data collection methods.
We then discuss methods for participant recruitment, which is a crosscutting challenge.

Our discussion highlights several desiderata for technology and society research methods: they should produce representative and real observations about online experiences, they should scale to new problems and populations, and they should allow research independence.




\subsection{State of critical sociotechnical research areas and methodologies}
Much of online life today is powered by proprietary services and products created by online platforms.
As this trend increases, policymakers and regulators are faced with a dearth of empirical information upon which to base policy recommendations.  We assert that there are several gaps in many areas of literature that result from similar methodological challenges.  We review three examples of sociotechnical inquiry relevant to policymakers and regulators and highlight the common measurement problems that inhibit progress in these areas.  

\subsubsection{Harmful Content Networks}
Harmful and abusive content abounds online, including misinformation
and disinformation~\cite{kumar_false_2018}, hate
speech and harassment~\cite{49786}, and violent content~\cite{conway_determining_2017}.
Research has examined these categories of content at the scope of
individual platforms~\cite{mondal_measurement_2017},
specific communities~\cite{ribeiro_evolution_2020}, or topics or events
of interest~\cite{guess_exposure_2020, wilson_cross-platform_2020}.
However, much of the previous literature on the effects of harmful content on society faces critical validity challenges.

In offline environments, media effects research typically takes the form of a field intervention design, where researchers use either experimental or observational methods in naturalistic settings to understand the effects of content on subsequent beliefs or behaviors.  For example, in order to understand the impact of a political messaging campaign on an electoral outcome, researchers may attempt an experimental design (e.g., expose people to content and then ask them about their likelihood to vote for a candidate) or an observational design (e.g., examine media consumption patterns in a region and then voting behaviors in the same region).

Translating these approaches to online environments requires access to data beyond the reach of most researchers today. The absence of accessible cross-platform data on ``who is exposed to what content, on which platform, and at what time'' means that critical questions presented by policy-makers and regulators cannot be answered with the state of most commonly available research methods. Consequently, lacking the capacity for field intervention studies, researchers struggle to measure the effects of harmful content on beliefs and behaviors. Additionally, because cross-platform data on content exposure remains inaccessible, researchers have not been able to measure incidence and engagement with harmful content at scale. Thus, it remains extremely challenging to evaluate the effectiveness of different policy proposals, such as content warnings or engagement limits, designed to mitigate the impact of harmful content.

\subsubsection{Black-box Recommender Systems}
Recommender systems (as well as other forms of decision-making systems) are typically based upon algorithmic models predicting engagement or consumption.  The goal of a recommender system is usually to match an outcome (such as video viewing or a product purchase) with latent user traits, such as political preferences or other personal interests.  This encourages the likelihood of user engagement, which underlies most business models of recommendation systems.  Much of the content that users encounter online derives from recommender systems including
social feeds~\cite{solsman_ever_2018,medvedev_powered_2019} and search results~\cite{hannak_measuring_2013}.

Researchers have posited that recommender systems may have deleterious consequences for users, and thus present compounding problems for society.  For example, recommender systems have been observed to homogenize user
behavior and content~\cite{chaney_how_2018, ciampaglia2018algorithmic}, 
exacerbate opinion polarization~\cite{perra_modelling_2019},
radicalize users~\cite{ribeiro_auditing_2020}, and spread
misinformation~\cite{fernandez_analysing_2021}.  However, research examining the systemic effect of recommender systems on aggregate social behaviors remains elusive.  Recommendation systems are adaptive to user behavior, meaning that the more information informing the algorithm about user preferences, the more users receive different experiences. Though ensuring ecological validity of research performed outside of the naturalistic environment is often a challenge, the adaptive nature of recommendation systems emphasizes the necessity for large, representative samples of online user experiences. To date, few studies have been able to analyze large, representative samples of real user
activity data~\cite{10.1145/2988450.2988454,hosseinmardi_evaluating_2020}.  

\subsubsection{Targeted Advertising/Personalization}

Research has demonstrated many instances of manipulative, misinformative,
and discriminatory online advertising
practices~\cite{ribeiro_microtargeting_2019,waddell_facebook_2020,pmlr-v81-speicher18a}.
Even after platforms added protections against discrimination, researchers
discovered that ads for credit, housing,
and employment could be targeted to users based on protected characteristics
like race~\cite{pmlr-v81-speicher18a}.
Measuring the incidence of discriminatory advertising is challenging,
however, because researchers must collect user
demographic data in addition to ad exposure data.
Studies have had to estimate user demographics using proxy variables from
auxiliary data~\cite{10.1145/3359301} or in the
case of Citizen Browser, leverage a panel of participants and custom
measurement tools~\cite{faife_credit_2021}.
Another challenge is that platform transparency reports are demonstrably
incomplete~\cite{edelson_analysis_2019}, but
independently tracking ads requires large-scale platform-independent data collection.

\subsection{Current methods and limitations for measuring user behavior}
\label{sec:prior-behavior}
Researchers have applied a diverse range of empirical methods in pursuit of observing user experiences within and across online platforms and services.  The portfolio of methodologies employed in studying user behavior online include simulations, web scraping, laboratory experiments, online surveys, browsing data (of which browsing history is only one form), and custom browser instrumentation.  These methods are used when researchers are unable to gain access to sufficient data from online platforms themselves, or when the data provided by platforms does not meet the scientific needs of the research design.  

Unfortunately, many of these methods only allow for partial examination of online user experiences, meaning that they address some but not all of the methodological challenges outlined in the previous section. For example, while some methods may allow for highly customized measurement of behavior, they may simultaneously lack the ability to collect data \textit{in-situ} -- in other words, high internal validity but low external validity.  Other methods may allow for \textit{in-situ} measurement, but fail to capture the surrounding context of user experience, which challenges the generalizability of resulting conclusions.  Finally, many of the most promising methods also have extensive infrastructure development costs.  While the resulting method may be possess high ecological, internal, and even external validity, the implementation and upkeep costs are far beyond the budgets of researchers. Thus, the scalability of these methods beyond a focused investigation remain undetermined.


\subsubsection{Simulations}
Researchers who seek to explicate the behavior of online services, but who lack access to the platforms or tools commonly applied in the field, will sometime create simulated versions of a
platform or system in order to establish the dynamics of system parameters on resulting behavior. Simulation can
be a valuable tool for experimenting with systems that are hard for researchers to
directly access, such as recommender
systems~\cite{lucherini_2021_t-recs, geschke_2019_triple-filter-bubble-modeling}.
Thus, while simulation provides complete independence for researchers from the constraints of platform partnerships, these studies fail to meet the representativity or scalability needs of socio-technical research.

\subsubsection{Web Scraping}

Web scraping, typically taking the form of instrumenting a browser to visit websites and collect desired content either through manual or automated means, is a popular technique employed in online research.

Web crawls are highly configurable and capture web content in a common form, usually structured information parsed from web documents.  This captures one aspect of online experience, the total sum of content possible to experience on the web. 
However, web crawls are not useful for studying personalized content, content behind
login walls, or how users interact with content~\cite{ribeiro_2019_comments-on-algorithmic-extremism}.
Finally, while there are tools that help researchers create web scrapers, web scraping requires substantial investment in development and operational resources, especially if the data is intended to represent content over time (required for any form of time-series analysis of behavior). In other words, there is the first-time cost and the on-going cost of web scraping, both carried by the researcher.

\subsubsection{Laboratory Experiments}
\label{sec:motivation:lab}
Laboratory studies of online behavior and experiences can be conducted either in-person or
through a remote platform.
In addition to observation, researchers may also conduct interviews or pose survey questions to participants, allowing for measurement of subjective reports that serve as proxies for otherwise unobservable traits.

The high degree of researcher control allows for experimental designs with extremely high internal validity, as these studies allow the researcher to carefully isolate variables of interest or change the standard behavior of platforms to observe how users respond to specific stimuli~\cite{kaiser_2021_adapting-security-warnings}.
However, the artificiality of the environment can limit the ecological validity of
findings~\cite{nichols_2010_good-subject-effect}, which makes it difficult to extend the conclusions of laboratory studies to general populations.

Further, laboratory experiments do not lend themselves well to studying user actions in the context of typical online experiences, due to many limitations including (but not exclusively): the duration of observable tasks, the total number of participants, the complexity of the task, and the technical skills of the research team.
Depending on the goals of a particular study, an experiment can range in start-up difficulty, from simply observing participants using existing systems in-person, to building new systems that allow for interventions, study participants remotely, or measure fine-grained aspects of behavior, such as by using eye-tracking hardware.
For this and many other examples of costly technical limitations (e.g., researchers must be physically present with each participant), laboratory studies fail to meet the scalability criteria for the current needs of sociotechnical research.

\subsubsection{Online Surveys}
Online surveys address the scalability challenge of laboratory studies for a highly restricted subset of subjective reporting.
Researchers can easily scale surveys to as large and diverse a population as they can recruit, which ensures high external validity as the resulting observations are assumed to generalize to the parent population (see~\Cref{sec:prior-recruiting}).  
Similar to laboratory experiments, researchers retain substantial control over the survey design, which allows for the investigation of a broad range of topics only limited to the methodological constraint of a survey instrument.  
Many intervention designs are also possible as survey experiments, such as by presenting participants with hypothesized scenarios and soliciting a predicted behavior, or by exposing participants to differing information via an engagement task and then asking for endorsements of beliefs or opinions.  
As businesses, survey platforms service many clients beyond researchers, including other commercial customers, which ensures that their tooling and infrastructure are robust and resilient.  Experiment development costs are thus typically contained within the level of questionnaires.
However, survey methods are severely limited in other critical areas of ecological and internal validity.  For example, while surveys may be embedded within the context of content engagement, the act of responding to a survey necessarily takes the subject out of context in order to respond.  Surveys are also prone to other measurement errors, in that responses to questions may not be always accurately report past experiences and that respondents asked about behavioral intentions may instead give idealized answers~\cite{araujo_2017_self-reported-internet-use, scharkow_2016_accuracy-self-reported-internet}.

\subsubsection{Platform Data}

In some ways, the gold standard for addressing the class of questions that can be characterized as ``who is exposed to what content and at what time'' is the platform's own instrumentation itself.  Platform-provided data typically consists of a representative sample of user behavior drawn directly from their own user base.  Typically, these datasets are produced through the same infrastructure and tooling used in day-to-day business.  In that sense, platform-provided data reflects the highest fidelity of instrumentation regarding the activity of real users on platforms.  
Platforms sometimes make this data available to researchers, either through private agreements with individual
research teams~\cite{bakshy_2015_exposure,vosoughi_spread_2018}, broadly shared
datasets~\cite{DVN/TDOAPG_2020}, or data access APIs~\cite{morstatter2013sample}.

However, platform data often does not meet the independence or generalizability criteria needed for sociotechnical research.  One challenge is that businesses may not be inclined to collect or distribute data that scrutinizes their own  practices~\cite{roose_2021_inside-fbs-data-wars}.
The data may also be proprietary, making replication studies difficult or impossible.  
In response to constraints on platform-provided data, some researchers have resorted to platform-specific datasets compiled by external organizations, such as the Pushshift archive of Reddit posts~\cite{baumgartner_2020_pushshift}.
However, whether from platforms themselves or efforts by other research teams, data consisting of a single platform only provides insight into usage of the services offered by that platform.  
Additionally, often these datasets are created to serve a general set of interests rather than allow for specific, purposeful investigation.  
Most critically, while some organizations may be transparent about data collection and processing, platform data is often generated through opaque means, and organizations often do not provide researchers visibility into the collection process to identify issues or nuances with resulting datasets~\cite{gaffney_2018_caveat-emptor}.
Facebook, for example, launched the Social Science One initiative in 2018 to facilitate academic study of democracy and elections issues. The project was plagued by delays and skewed data with limited research value, and implementation errors eventually required reissuing the data~\cite{timberg_2021_facebook-made-big-mistake,sso-blog-post_delays}.

\subsubsection{Browsing Data}
Commercial browsing panels, such as YouGov and Comscore, provide researchers with 
data about browsing habits of participants in
an ongoing research panel.  Crowdsourcing data represents a grassroots counterpart to commercial panel providers, where researchers can construct similar data to these panels by asking recruited participants to upload their own browsing data (generally by having users upload a file containing their browser's stored history), or access social media data by asking users to capture and upload posts from their
newsfeeds~\cite{255286,owen_how_2020,bentley_2019_understanding-online-news-behaviors}.

In both cases, this type of data shows actions taken by real users, provides data about surrounding actions (such
as the website visited before and after a visit of interest), and requires minimal infrastructure
development on the part of the researcher, since either the panel provider or the browser is
completing the data collection. However, researchers are typically not able to design interventions, and are limited
to the data collected by the provider, which may not include all the measures of interest. For example,
commercial browsing data typically consists of timestamped URLs visited by the participant, without information
about further user interaction with those sites.

\subsubsection{Custom Browser Instrumentation}
Some studies have instrumented participant's browsers directly,
typically using browser extensions~\cite{edelson_analysis_2019,wang_coming_2015}.
This approach has several benefits: it
collects data across platforms, allows for researcher-inserted interventions,
and can measure not only what users see online but also how they interact with content and services.
The naturalistic environment of browser-based studies is also
a benefit, as it aids ecological validity.
However, the cost of custom browser instrumentation is often extremely high, both in terms of development and operation.  Researchers who maintain custom browser instrumentation for their research designs face considerable challenges to scalability.  The NYU Online Political Ads Transparency Project created a browser
extension and recruited participants to install it with the goal to collect targeting data and corresponding ads that participants were exposed to on Facebook~\cite{adobserver}. 
The Markup's Citizen Browser is a another related project.  Although not a browser extension, participants install a standalone instrumented browser and log in to their Facebook accounts, enabling Citizen Browser to observe user's experiences on Facebook through an automated collection design: periodically it will
launch, navigate to Facebook, and collect data from participants' feeds.

The work described in this paper allows for simplifying the creation of studies like these in two ways: first,
the partnership with \mozilla{} enables researchers to more easily recruit a large pool of participants; second, \sdkName{} provides reusable functionality that reduces extension development effort.

\begin{table*}[t]
\footnotesize
	\newcolumntype{C}{>{\centering\arraybackslash}m{0.1156\linewidth}}
	\begin{tabular}{l C C C C C }
		&
		Shows \emph{in-situ} behavior &
		Allows for custom measurement &
		Allows for behavioral interventions &
		Shows surrounding context of user actions &
		Requires limited infrastructure development
		\\ \hline
		Lab experiments   &  
		  \harveyBallNone & 
		  \harveyBallFull & 
		  \harveyBallFull &
		  \harveyBallNone &
		  \harveyBallHalf
		  \\ \hline
		Online surveys    & 
		  \harveyBallNone & 
		  \harveyBallFull & 
		  \harveyBallFull &
		  \harveyBallNone &
		  \harveyBallFull
		  \\ \hline
		Platform data     & 
		  \harveyBallFull & 
		  \harveyBallNone & 
		  \harveyBallNone &
		  \harveyBallNone &
		  \harveyBallFull
		  \\ \hline
		Web scraping      &
		  \harveyBallFull & 
		  \harveyBallHalf & 
		  \harveyBallNone &
		  \harveyBallNone &
		  \harveyBallNone
		  \\ \hline
		Browsing data     &
		  \harveyBallFull & 
		  \harveyBallNone & 
		  \harveyBallNone &
		  \harveyBallFull &
		  \harveyBallFull
		  \\ \hline
		Custom browser instrumentation & 
		  \harveyBallFull & 
		  \harveyBallFull & 
		  \harveyBallFull &
		  \harveyBallFull &
		  \harveyBallNone
		  \\ \hline
	\end{tabular}
	\label{table:collection-methods}
	\caption{Methods used to study user behavior online and their ability to achieve certain goals.\\
		\protect\harveyBallFull{} : method can achieve goal\\
		\protect\harveyBallNone{} : method cannot achieve goal\\
		\protect\harveyBallHalf{} : goal achievement depends on specifics of measurement
		}
\end{table*}

\subsection{Current methods and limitations for recruiting participants}
\label{sec:prior-recruiting}

Once a method has been chosen for the measurement of user behavior, researchers face a second challenge -- the method of recruiting participants into a study.  
Commonly, researchers have enlisted an array of commercial services that allow for researchers to engage with users in research marketplaces, such as commercial survey panels or crowdworker platforms.  In this section, we go into more detail about these approaches, and review how each approach addresses the independence, representativity, and scalability criteria required for socio-technical research.  

\subsubsection{Commercial Panel Providers}
Commercial panel providers such as YouGov, Comscore, Nielsen, and Similarweb internalize many of the recruiting costs associated with recruiting participants for task-based online research.  These companies maintain a representative panel of people interested in participating in research studies.  A smaller subset of panelists have pre-consented to sharing their browsing data with third parties for research purposes.  To do so, panelists install browsing tracking software, which sends a pre-configured specification of data to the commercial panel providers.  
Commercial panel providers can be useful to researchers because of the
ease of scaling and representativeness of the panel-driven approaches, at least for researchers who want to study
populations covered by the commercial provider ~\cite{hosseinmardi_evaluating_2020,guess_2021_everything}.

However, commercial providers have substantial technical limitations that result in design constraints for socio-technical research. For example, researchers do not interact with the participants, which presents several challenges for verifying the data collection methods.  Additionally, the lack of ability to directly interact with the participants in a study renders many intervention designs impossible.

\subsubsection{Crowdworker Platforms}
Amazon Mechanical Turk, Figure Eight, Microworkers, Clickworkers,
and research-oriented platforms such as
Prolific enable researchers to recruit large samples of paid research participants.
Crowdworking platforms provide only rudimentary tools for collecting data
(e.g., survey templates), in order to
run more complex studies, researchers either ask users to collect and upload
their own data~\cite{fiesler_what_2017}
or ask users to install data collection tools~\cite{farke2021privacy}.

These platforms provide scale relatively easily, since recruiting more participants is a matter of cost.
Unlike when using commercial panel providers, researchers provide materials that the platform shows to each potential
participant, allowing researchers a high degree of control over onboarding, e.g. the presentation of informed consent documents, the description of the scope of the study, and any other number of question that can be administered via survey.  Many crowdworker platforms also allow researchers to filter participants by various demographics, allowing for user samples that meet specific representativity criteria. 

However, researchers who want to recruit a sample along a demographic not represented in the options provided by the platform will have to subsequently address this in follow-on analysis.  And if the desired population may not be present on the platform, this approach may also simply not be feasible.  

\subsubsection{Research Teams}
Researchers have many conventional options for recruiting participants, including
posting fliers, sending emails, posting on online
forums, purchasing advertisements, stopping people in public places, or asking participants
to refer friends.
These methods allow researchers to recruit participants
that match the population they aim to study and to control the
onboarding and consent processes, but recruiting a large sample of
such participants can be quite difficult. Often, researchers rely on an easy-to-reach
population, which may produce results that are not generalizable to other
populations~\cite{peterson_2001_use-of-college-students}.

\subsubsection{Platform Partnerships}
When researchers partner with a platform to study its users, they gain access to
an instantly-scaled and representative population: by definition, the entire population
of a platform's users are representative of themselves. However, these partnerships
are often exclusive, and usually do not allow external researchers to control the
onboarding process. Users may vary in their understanding of their role as research
subjects.

\begin{table*}[t]
\small
	\newcolumntype{C}{>{\centering\arraybackslash}m{0.13\linewidth}}
    \begin{tabular}{l C C C}
    & Scales easily &
      Representative of desired population &
      Researcher control over onboarding\\
      \hline
    Commercial panel providers       & \harveyBallFull &
                                       \harveyBallHalf &
                                       \harveyBallNone \\ \hline
    Crowdworker platforms           & \harveyBallFull &
                                      \harveyBallHalf &
                                      \harveyBallFull \\ \hline
    Research teams                  & \harveyBallNone &
                                      \harveyBallHalf &
                                      \harveyBallFull \\ \hline
    Platform partnership            & \harveyBallFull &
                                      \harveyBallFull &
                                      \harveyBallNone \\ \hline
    \end{tabular}
    \caption{
    Recruitment methods for studying user behavior online and their ability to achieve
    certain goals. Note that these recruitment methods are largely independent from
    the later data collection method (e.g., one could recruit users through a crowdworker
    platform, then collect data using a survey or by having participants upload
    browsing data (or both)).\\
		\protect\harveyBallFull{} : method can achieve goal\\
		\protect\harveyBallNone{} : method cannot achieve goal\\
		\protect\harveyBallHalf{} : goal achievement depends on specifics of measurement
		}
    \label{table:recruitment-methods}
\end{table*}


\section{Survey of Prior Work with Device-Based Research Methods}
\label{sec:survey}

In order to better understand both how browser-based, and, more broadly, device-based research methods have been applied so far and the barriers to greater use of these methods, we conducted a literature survey across academic disciplines.

We constructed a set of query terms for device-based research (see~\Cref{app:survey-search-terms} for the complete list)
and conducted Google Scholar searches for academic papers.
Because the queries yielded a large volume of papers, we grouped the papers based on the number of citations/year they had received
and took a sample of each stratum for our candidate set of papers.
\Cref{tab:survey-sampling-rate} lists the strata, number of papers, and sampling rates.

\begin{table}[]
	\newcolumntype{C}{>{\centering\arraybackslash}m{0.2\linewidth}}
    \centering
    \begin{tabular}{C C C C}
    Percentile range & Citations/year & Number of papers & Sampling rate
    \\ \hline
    95---100         & >=16.33         & 1147            & 100\%
    \\ \hline
    80---95          & >=3.77, < 16.33 & 3427            & 25\%
    \\ \hline
    50---80          & >= 0.52, <3.77  & 6862            & 10\%
    \\ \hline
    0---50           & < 0.52          & 13024           & 5\%
    \\ \hline
    \end{tabular}
    \caption{The citation-based strata and sampling rate for the literature survey.}
    \label{tab:survey-sampling-rate}
\end{table}

We then manually filtered each paper in the candidate set, applying five criteria. First, we required that the paper involve data about interactions between participants and computing devices.
Second, the paper had to include data from at least 10 individuals. Third, the data collection had to cover at least 30 days for at least one
participant.
Fourth, the participants had to use the device outside of a lab environment.
Fifth, the data described in the paper could not be usage statistics of a researcher-created tool.

We used a few criteria to exclude papers: papers we were unable to access, papers that were not written in English, and search results
that were actually books or unpublished dissertations, rather than academic papers.

At the conclusion of this process, our dataset included 284 papers. We then annotated each paper with the following attributes
(For the complete codebook, see~\Cref{app:survey-codebook}):
academic discipline, number of participants, duration of data collection, informed consent,
IRB approval, data source, collection mechanism, recruitment mechanism, compensation, interventions, qualitative data, web attention,
web navigation, and social media sharing.

When a paper included multiple distinct instances of qualifying data collection, we annotated them separately, yielding a final dataset of 290 studies.\footnote{The final annotated dataset is available at \url{https://github.com/CHI4450/submission4450/blob/main/literature_survey_dataset.csv}.}
Many of the papers in our dataset fell into one of two categories:
\begin{itemize}
    \item MOOC: Papers in this category used data from a MOOC (massive open online course) to predict or assist student success.
    \item E-Commerce: Papers in this category used data from an online store to predict or increase sales or browsing.
\end{itemize}

Within each of these two categories, the papers tended to be similar along the attributes we were coding, and neither category
was the primary focus of our literature review. Therefore, for these two categories of paper, we coded their publication venue,
academic discipline, topic, number of participants, and duration of data collection, but no other attributes.

Finally, as we did not read an equal sample from each stratum, we extrapolated our results from the sample we read to the full
population of papers as follows:

$$E_c = \sum_{i=0}^{n} \frac{D_{i_c}}{D_i} \frac{D_i}{S_i} U_i = \sum_{i=0}^n D_{i_c} \frac{U_i}{S_i}$$

where $E_c$ is the extrapolated count of papers meeting condition $c$, $i$ is the index of a stratum, $n$ is the number of strata,
$D_i$ is the number of papers in the dataset in stratum $i$, $D_{i_c}$ is the number of papers in the dataset in stratum $i$ meeting
condition $c$, $U_i$ is the number of papers in the original universe in stratum $i$, and $S_i$ is the number of papers sampled
from stratum $i$.

For numeric coding attributes, we calculated weighted percentiles and weighted means using the same weight as above:
$\frac{U_i}{S_i}$.

\begin{table*}[t]

\footnotesize
	\newcolumntype{C}{>{\centering\arraybackslash}m{0.2\linewidth}}
	\begin{tabular}{l C }
	    \textbf{Description} & \textbf{Number of papers}
		\\ \hline
	    Total papers resulting from search terms & 24584
		\\ \hline
		Sample of papers screened & 3341 (13.6\% of total)
		\\ \hline
		Papers containing qualifying studies &  284 (8.5\% of sample)
		\\ \hline
		Qualifying studies & 290
		\\ \hline
		Extrapolated count of qualifying studies in original total & 1276
		\\ \hline
		Studies in D not categorized as MOOC or e-commerce  & 177
		\\ \hline
		Extrapolated count of non-MOOC or e-commerce studies in U & 751
		\\ \hline
	\end{tabular}
\caption{Basic statistics about dataset. Note that papers may contain more than one study.}
\label{table:survey-overall-stats}
\end{table*}

Unless otherwise stated, results reported below are given on the extrapolated dataset resulting from this reweighting.
\Cref{table:survey-overall-stats} provides an overview of the number of studies at each step.

We offer the following observations from analyzing our literature review dataset.

\begin{table*}[t]

\footnotesize
	\newcolumntype{C}{>{\centering\arraybackslash}m{0.55\linewidth}}
	\begin{tabular}{l p{0.08\linewidth} p{0.65\linewidth} }
	    \textbf{Topic}                & \textbf{Number of studies}      & \textbf{Description} \\ \hline
	    Online sales and marketing    & 363 (28.4\%)                  & Studies about online sales \\ \hline
	    Education                     & 267 (20.9\%)                  & Studies about learning and teaching \\ \hline
	    Health                        & 155 (12.1\%)                  & Studies about individual and public health, both physical and mental \\ \hline
	    Behavior profiling            & 144 (11.3\%)                  & Studies on methods for sorting users into groups based on behavior, or predicting next actions \\ \hline
	    Security and privacy          & 68 (5.3\%)                    & Studies working to understand the state of security/privacy or evaluate new systems for improving security/privacy \\ \hline
	    Politics                      & 58 (4.5\%)                    & Studies about how people engage in political behavior or news-reading online. \\ \hline
	    HCI                           & 57 (4.5\%)                    & Studies on understanding the use of particular systems or tools \\ \hline
	    Network analysis              & 42 (3.3\%)                    & Studies on information flow in social networks \\ \hline
	    News consumption              & 29 (2.3\%)                    & Studies on patterns of non-political news consumption \\ \hline
	    Recommender systems           & 25 (2.0\%)                    & Studies evaluating systems that recommend products or media to users \\ \hline
	    Computer systems              & 19 (1.5\%)                    & Studies measuring or improving network speeds or operating system characteristics \\ \hline
	    Metadiscussion of collection  & 17 (1.3\%)                    & Studies building systems for collecting user data, or studies measuring consent and understanding of collection \\ \hline
	    Workplace                     & 15 (1.2\%)                    & Studies evaluating patterns of how employees work \\ \hline
	    Financial                     & 11 (0.9\%)                    & Studies using user data to predict loan repayment \\ \hline
	    Other                         & 6 (0.5\%)                     & \\ \hline
	\end{tabular}
\caption{Our inductively-coded study topics, the number of studies in each, and a brief description of the category.}
\label{table:survey-topics-descriptions}
\end{table*}

First, \textbf{device-based research methods are not exclusive to computer science}. The two topics best represented in our dataset are online sales and marketing (28.4\%), and education (20.9\%). Studies in health (11.3\%) and politics (4.5\%) also make up a significant proportion. \Cref{table:survey-topics-descriptions} describes the topics and their prevalence in the dataset.

\begin{figure}
\begin{tikzpicture}
\begin{axis}[
    ybar stacked,
	bar width=12pt,
	width = 300pt,
    enlarge x limits=0.10,
    enlarge y limits = 0.05,
    legend style={at={(0.95,0.8)},
      anchor=east,legend columns=1},
    ylabel={Number of studies},
    symbolic x coords={
         Online sales and marketing,Education,Behavior profiling,Health,Security and privacy,Politics, HCI,
         Network analysis,Computer systems,Recommender systems,News consumption,
		 Metadiscussion,Workplace,Financial,Other},
    xtick=data,
    x tick label style={rotate=45,anchor=east},
    ]
\addplot[ybar, fill=cb1, draw=black] plot coordinates { 
  (Behavior profiling,18) (Security and privacy, 0) (Other,1) 
  (Online sales and marketing,94) (Metadiscussion,4) (Health, 65) (Politics, 43) (News consumption, 3)
  (Network analysis, 1) (HCI, 20) (Recommender systems, 1) (Workplace, 0) (Financial, 0)
  (Education, 0) (Computer systems, 0)
  };
\addplot[ybar, fill=cb2, draw=black] plot coordinates { 
  (Behavior profiling, 83) (Security and privacy, 42) (Other, 4) (Online sales and marketing, 8) (Metadiscussion, 0) (Health, 19)
  (Politics, 4) (News consumption, 8) (Network analysis, 10) (HCI, 25) (Recommender systems, 15)
  (Workplace, 11) (Financial, 11) (Education, 1) (Computer systems, 10)
};
\addplot[ybar, fill=cb3, draw=black] plot coordinates { 
  (Behavior profiling, 34) (Security and privacy, 21) (Other, 0) (Online sales and marketing, 1) (Metadiscussion, 13) (Health, 58)
  (Politics, 7) (News consumption, 8) (Network analysis, 31) (HCI, 12) (Recommender systems, 0)
  (Workplace, 4) (Financial, 0) (Education, 15) (Computer systems, 19)
};
\addplot[ybar, fill=cb4, draw=black] plot coordinates { 
  (Behavior profiling, 15) (Security and privacy, 5) (Other, 1) (Online sales and marketing, 0) (Metadiscussion, 0) (Health, 2)
  (Politics, 4) (News consumption, 0) (Network analysis, 0) (HCI, 0) (Recommender systems, 0)
  (Workplace, 0) (Financial, 0) (Education, 0) (Computer systems, 0)
};
\addplot[ybar, fill=cb5, draw=black] plot coordinates { 
  (Behavior profiling, 0) (Security and privacy, 0) (Other, 0) (Online sales and marketing, 0) (Metadiscussion, 0) (Health, 0)
  (Politics, 0) (News consumption, 0) (Network analysis, 0) (HCI, 0) (Recommender systems, 0)
  (Workplace, 0) (Financial, 0) (Education, 251) (Computer systems, 0)
};
\addplot[ybar, fill=cb6, draw=black] plot coordinates { 
  (Behavior profiling, 5) (Security and privacy, 0) (Other, 0) (Online sales and marketing, 260) (Metadiscussion, 0) (Health, 0)
  (Politics, 0) (News consumption, 0) (Network analysis, 0) (HCI, 0) (Recommender systems, 9)
  (Workplace, 0) (Financial, 0) (Education, 0) (Computer systems, 0)
};
\legend{\strut Research data provider, Partnership, Custom instrumentation, Not stated, MOOC, E-commerce}
\end{axis}
\end{tikzpicture}
\caption{For each study topic, the data sources used by studies on that topic.}
\label{fig:survey-data-source-topic-barchart}
\end{figure}
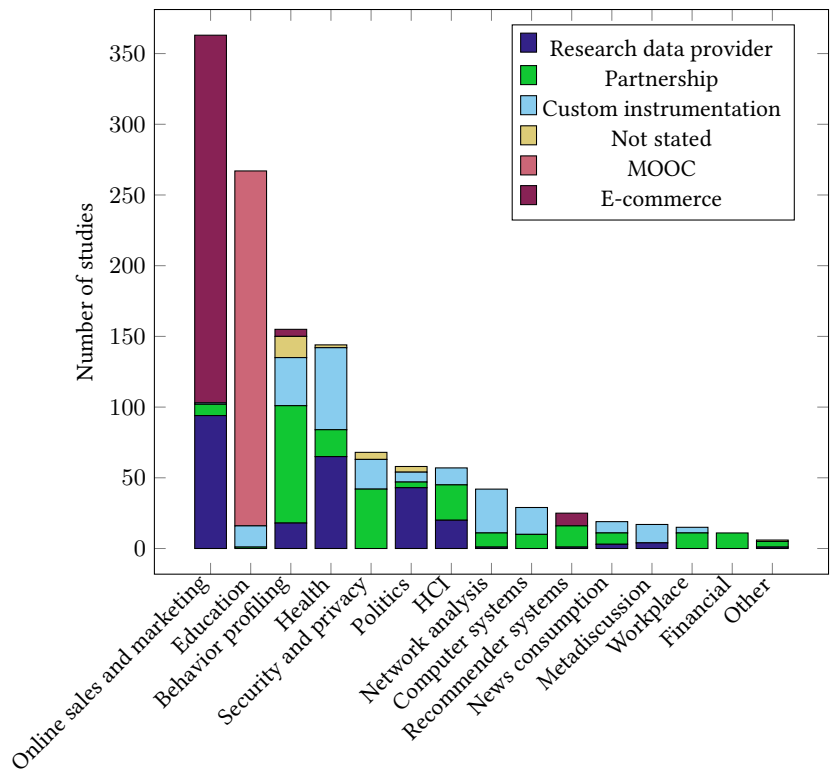

Second, \textbf{most device-based studies (80.5\%) rely on a research data provider or a partnership}. Just 17.5\% of studies build custom instrumentation to collect their data. \Cref{fig:survey-data-source-topic-barchart} shows the methods used by studies in each topic. Unsurprisingly, papers in online sales and marketing often use e-commerce methods to collect data, and education studies often use MOOCs, which are both forms of partnership.

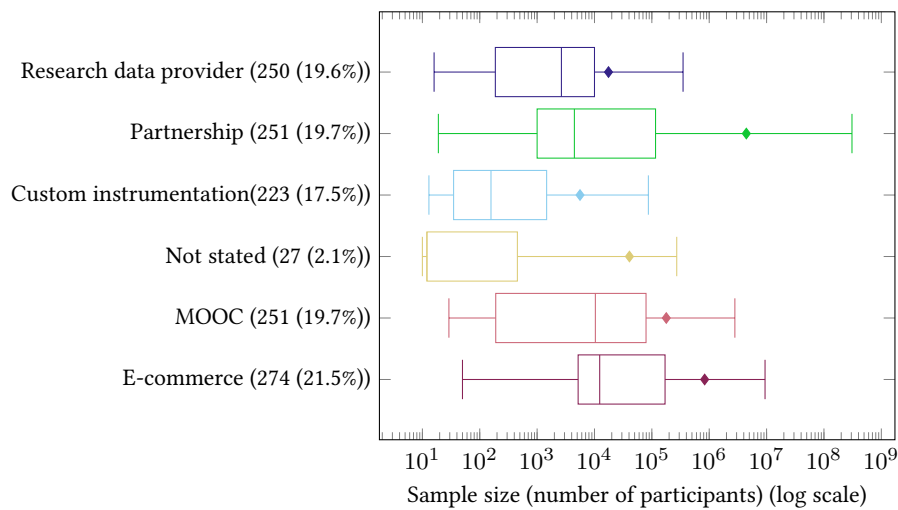
\begin{figure}
\begin{tikzpicture}
  \begin{axis}
    [
    ytick={1,2,3,4,5,6},
    xmode=log,
    yticklabels={
    E-commerce (274 (21.5\%)),
    MOOC (251 (19.7\%)),
    Not stated (27 (2.1\%)),
    Custom instrumentation(223 (17.5\%)),
    Partnership (251 (19.7\%)),
    Research data provider (250 (19.6\%)),
    },
    xlabel= Sample size (number of participants) (log scale),
    ]
    \addplot[cb6, 
    boxplot prepared={
      median=12330,
      upper quartile=169910,
      lower quartile=5187,
      upper whisker=9400000,
      lower whisker=50,
      average=832688
    },
    ] coordinates {};
    \addplot[cb5, 
    boxplot prepared={
      median=10359,
      upper quartile=79186,
      lower quartile=190,
      upper whisker=2800000,
      lower whisker=29,
      average=178538
    },
    ] coordinates {};
    \addplot[cb4, 
    boxplot prepared={
      median=12,
      upper quartile=453,
      lower quartile=12,
      upper whisker=271576,
      lower whisker=10,
      average=40678
    },
    ] coordinates {};
    \addplot[cb3, 
    boxplot prepared={
      median=157,
      upper quartile=1465,
      lower quartile=35,
      upper whisker=87000,
      lower whisker=13,
      average=5594
    },
    ] coordinates {};
    \addplot[cb2, 
    boxplot prepared={
      median=4462,
      upper quartile=115982,
      lower quartile=1000,
      upper whisker=308000000,
      lower whisker=19,
      average=4436570
    },
    ] coordinates {};
    \addplot[cb1, 
    boxplot prepared={
      median=2651,
      upper quartile=10000,
      lower quartile=187,
      upper whisker=350000,
      lower whisker=16,
      average=17553
    },
    ] coordinates {};
  \end{axis}
\end{tikzpicture}
\caption{For each source of data, the minimum, 25th, 50th, and 75 percentiles, and maximum number of participants across
the studies using that source. Mean is indicated by a diamond. Note the log scale. In the "not stated" category, the
25th and 50th percentiles are equal.}
\label{fig:survey-sources-participants}
\end{figure}

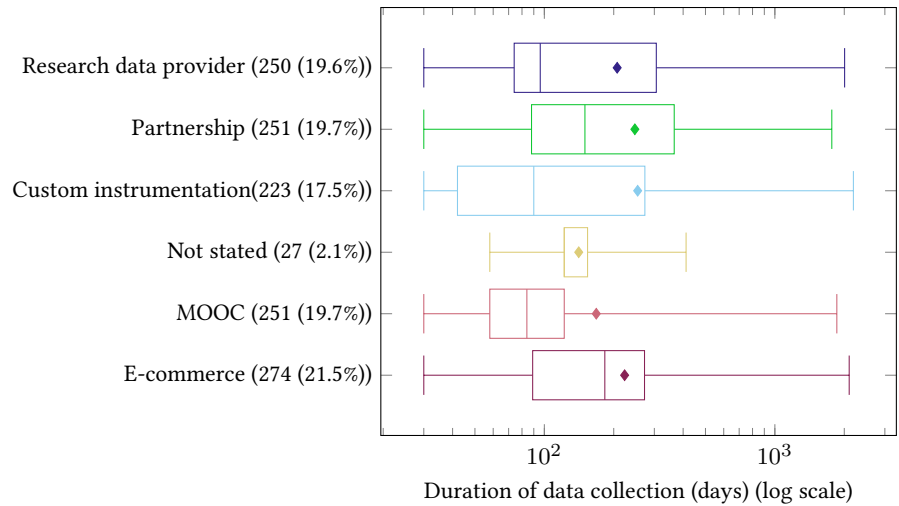
\begin{figure}
\begin{tikzpicture}
  \begin{axis}
    [
    ytick={1,2,3,4,5,6},
    xmode=log,
    yticklabels={
    E-commerce (274 (21.5\%)),
    MOOC (251 (19.7\%)),
    Not stated (27 (2.1\%)),
    Custom instrumentation(223 (17.5\%)),
    Partnership (251 (19.7\%)),
    Research data provider (250 (19.6\%)),
    },
    xlabel= Duration of data collection (days) (log scale),
    ]
    \addplot[cb6, 
    boxplot prepared={
      median=183,
      upper quartile=272,
      lower quartile=89,
      upper whisker=2101,
      lower whisker=30,
      average=223
    },
    ] coordinates {};
    \addplot[cb5, 
    boxplot prepared={
      median=84,
      upper quartile=122,
      lower quartile=58,
      upper whisker=1857,
      lower whisker=30,
      average=168
    },
    ] coordinates {};
    \addplot[cb4, 
    boxplot prepared={
      median=122,
      upper quartile=154,
      lower quartile=122,
      upper whisker=412,
      lower whisker=58,
      average=141
    },
    ] coordinates {};
    \addplot[cb3, 
    boxplot prepared={
      median=90,
      upper quartile=273,
      lower quartile=42,
      upper whisker=2191,
      lower whisker=30,
      average=254
    },
    ] coordinates {};
    \addplot[cb2, 
    boxplot prepared={
      median=150,
      upper quartile=366,
      lower quartile=88,
      upper whisker=1766,
      lower whisker=30,
      average=247
    },
    ] coordinates {};
    \addplot[cb1, 
    boxplot prepared={
      median=96,
      upper quartile=306,
      lower quartile=74,
      upper whisker=2007,
      lower whisker=30,
      average=207
    },
    ] coordinates {};
  \end{axis}
\end{tikzpicture}
\caption{For each source of data, the minimum, 25th, 50th, and 75 percentiles, and maximum number of days of data
collection across the studies using that source. Mean is indicated by a diamond. Note the log scale.
In the "not stated" category, the 25th and 50th percentiles are equal.}
\label{fig:survey-sources-duration}
\end{figure}

Third, \textbf{researchers face a trade-off between using sophisticated instrumentation and benefiting from a larger participant sample and longer data collection}. As seen in~\Cref{fig:survey-sources-participants,fig:survey-sources-duration}, the median number of participants for studies with custom instrumentation was 157, because these studies typically involved researchers developing their own recruiting strategy and methods. Researchers who recruited participants independently reported using multiple methods to advertise, often including emails, fliers, and word-of-mouth. The median number of participants for studies with research data providers, meanwhile, was 2651, and 4462 for studies using a partnership, because researchers could generally take advantage of an existing user base.
The difference in duration of data collection follows the same pattern, though less extreme: the median period of data collection for studies with custom instrumentation was 90 days, compared to 96 and 150, respectively for studies using research data providers or partnerships.
Commercial panels can incentivize participants to remain in the sample, and, since many studies use their data, often run for many years at a time, with researchers able to collect extensive retroactive data.
Ideally researchers would be able to benefit from both sophisticated, custom instrumentation and a larger, longer-lasting participant sample.

\begin{figure}
\begin{tikzpicture}
    \begin{axis}[
    xmin=0, xmax=3500,
    ymin=0, ymax=200,
    grid=major,
    legend style={at={(0.95,0.2)},
      anchor=east,legend columns=1},
    xlabel=Median number of participants,
    ylabel=Median days of data collection,
    ]
    \draw [cb2, fill, fill opacity=0.3] (624,91) ellipse [ 
    x radius=.0325\linewidth,
    y radius=.0325\linewidth,
    ];
    \draw [cb1, fill, fill opacity=0.3] (2654,150) ellipse [ 
    x radius=.0614\linewidth,
    y radius=.0614\linewidth,
    ];
    \draw [cb5, fill, fill opacity=0.3] (736,90) ellipse [ 
    x radius=.0055\linewidth,
    y radius=.0055\linewidth,
    ];
    \addplot [cb2, mark = star, ] coordinates {( 624, 91)};
    \addplot [cb1, mark = star, ] coordinates {( 2654, 150)};
    \addplot [cb5, mark = star, ] coordinates {( 736, 90)};
    \legend{Consent obtained, Not stated, Consent explained}
    \end{axis}
\end{tikzpicture}
\caption{The median number of participants and days of data collection, by the paper's description
of the consent process. The circle's radius indicates the number of studies in each category.
This figure excludes papers we categorized as MOOC or e-commerce.
}
\label{fig:survey-consent-circles}
\end{figure}
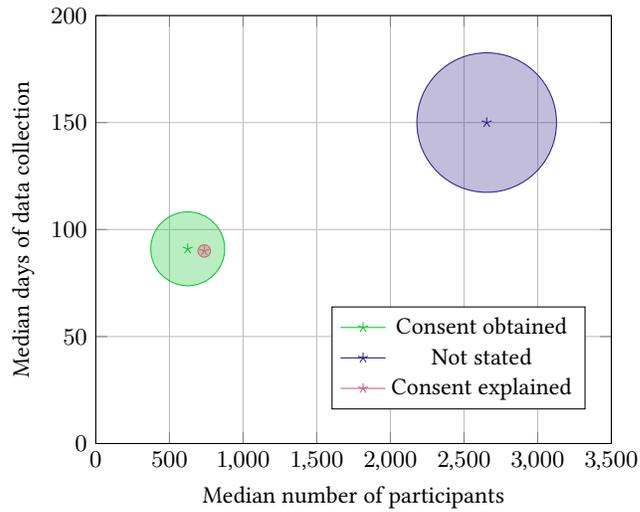

Fourth, \textbf{papers with larger numbers of participants and longer data collection were less specific about the consent of their participants}. This is illustrated in~\Cref{fig:survey-consent-circles}. The studies that we coded as "not stated" for consent were a diverse group: some very likely had received informed consent from their participants, but didn't mention it, and others appeared not to have notified users of the use of their data. Many studies using a partnership to collect data (80.1\% of which did not mention consent) clearly would have struggled to acquire consent, since their participants were often unaware that their data was being used.

\begin{figure}
\begin{tikzpicture}
\begin{axis}[
    ybar stacked,
	bar width=35pt,
	width = 180pt,
	height = 250pt,
    enlarge x limits=0.60,
    enlarge y limits = 0.05,
    legend style={at={(1.83,0.875)},
      anchor=east,legend columns=1,
      },
    ylabel={Number of studies},
    symbolic x coords={
        No, Yes
        },
    xtick=data,
    x tick label style={rotate=45,anchor=east},
    ]
\addplot[ybar, fill=cb1, draw=black] plot coordinates { 
  (Yes, 3)
  (No, 247)
  };
\addplot[ybar, fill=cb2, draw=black] plot coordinates { 
  (Yes, 5)
  (No, 246)
};
\addplot[ybar, fill=cb3, draw=black] plot coordinates { 
  (Yes, 50)
  (No, 173)
};
\addplot[ybar, fill=cb4, draw=black] plot coordinates { 
  (Yes, 4)
  (No, 23)
};
\legend{\strut Research data provider, Partnership, Custom instrumentation, Not stated}
\end{axis}
\end{tikzpicture}
\caption{Studies that did and did not use researcher-applied interventions, by data source.
This figure excludes studies that we categorized as MOOC or e-commerce.}
\label{fig:survey-data-source-interventions-barchart}
\end{figure}
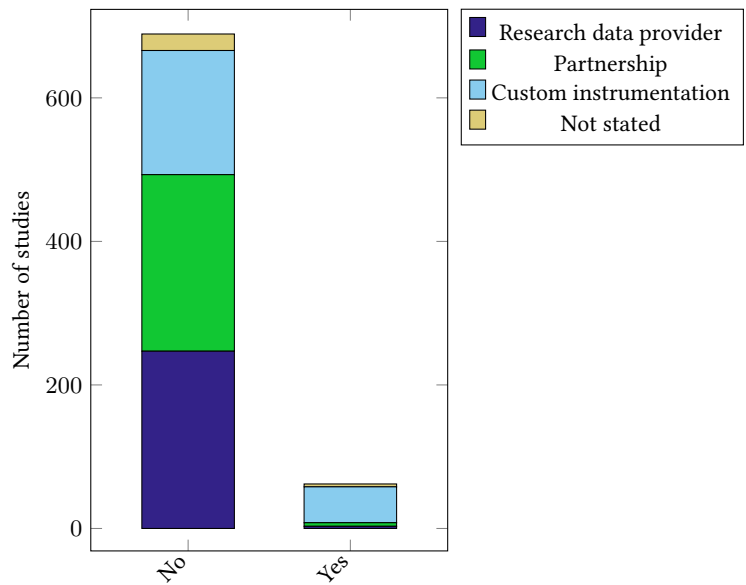

Fifth, \textbf{very few studies involved an experiment with user experience interventions} (8.3\% of the non-MOOC or e-commerce studies).
Nearly all of these studies used custom instrumentation, as seen in~\Cref{fig:survey-data-source-interventions-barchart}. Intervention study designs can be powerful for making progress on technology and society research problems, because they enable scholars to examine how individuals might respond to differently designed online services and experiences. Unfortunately, these methods appear to be out of reach except for teams with extensive technical backgrounds and resources.

Sixth, because of the limited data available, \textbf{many browser-based research studies use highly imprecise proxies to measure important aspects of user behavior}.
Of the 317 studies that used any method to measure attention, 246 (77.6\%) used simply the count of visits to different domains. 25 (7.9\%) used dwell time, 19 (6.0\%) used the interval between page loads, and 22 (7.0\%) did not state their methodology. The remaining 5 (1.6\%) used a mix of custom instrumentation and a measure supplied by the research data provider. Similarly, of the 102 papers that considered the navigation path between sites, 47 (46\%) used the chronologically-previous page load, 22 (22.6\%) relied on the HTTP referrer, and the other 33 (32.4\%) used a mix of custom instrumentation, most of which were for a specific platform or system, rather than being general purpose.

Seventh, \textbf{most papers do not describe their data in detail}. We planned to evaluate the overcollection of a study: whether it minimized the data collected, or swept up unnecessary detail in its collection. However, almost no papers gave a sufficient description of the data for such an analysis. From the descriptions provided, studies very rarely intentionally limited the data they collected, usually preferring to use everything available.

\section{An Overview of Rally and WebScience}
\label{sec:rally-and-webscience-overview}
We have demonstrated above that sociotechnical researchers are in need of new methods
to study users that are sophisticated, customizable, scalable, and that protect participant privacy.
We now present Rally and Webscience, new systems for browser-based research designed to meet this need,
and discuss how the systems work together to enableaa study.


\subsection{Rally Overview}
\Rally{} is a program run by \redact{Mozilla} that allows
researchers to use \firefox{} as a platform for studies.
Researchers using Rally partner with \mozilla{} to recruit users of \firefox{} as research participants,
giving researchers access to a large pool of potential participants.
A participant in a Rally study installs a browser extensions into the browser they already use, allowing
researchers to observe their everyday online activity and interactions with web services.
See~\Cref{sec:rally} for a full discussion of Rally.

\subsection{WebScience Overview}
Rally handles many aspects of a research study, including recruitment, participant onboarding, and
data storage. One of the central challenges of browser-based research still remains, however: researchers
need sophisticated, standardized, customizable instrumentation for carrying out a study.
\WebScience{} is an open-source SDK for creating a study, along with instrumentation
for many important but difficult-to-measure aspects of user behavior.
\WebScience{} is easy to customize, even for research teams with less technical
expertise, and easily extensible by research teams with more specialized needs.
See~\Cref{sec:websci} for a full discussion of WebScience.

\subsection{Study Overview}
\begin{figure}[t]
    \centering
    \resizebox{\textwidth}{!}{\includegraphics{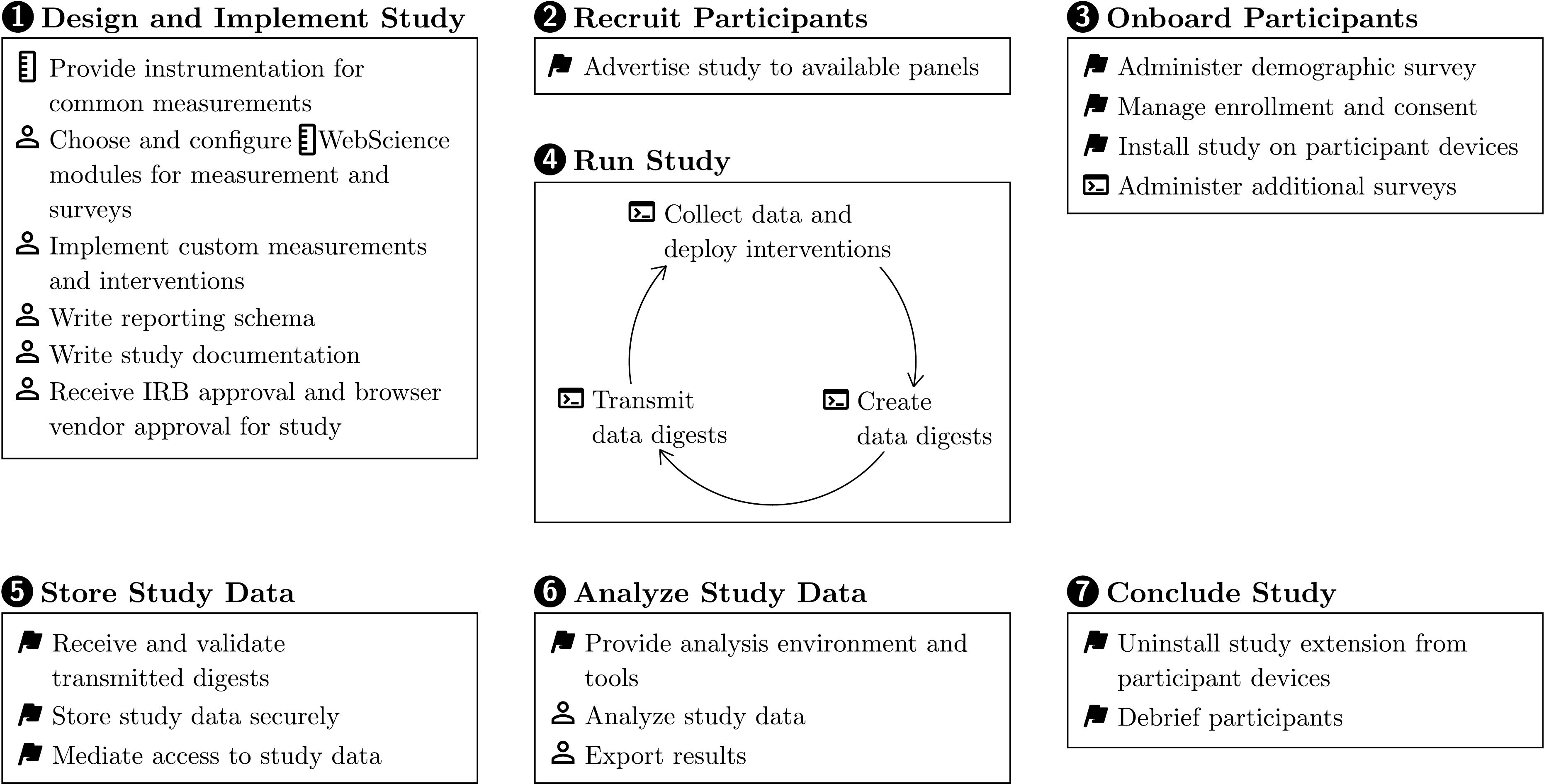}}
    \centering

\caption{A Rally study consists of seven stages. Using WebScience (\protect\webscienceicon{}), study authors (\protect\authoricon{}) design and implement the study (section~\protect\ref{subsubsec:designandimplement}). Rally (\rallyicon{}) recruits and onboards participants (sections~\protect\ref{subsubsec:recruit} and~\protect\ref{subsubsec:onboard}), then study code (\protect\codeicon{}) installed on participant devices executes the study (section~\protect\ref{subsubsec:run}). Rally securely stores data (section~\protect\ref{subsubsec:store}), permits study authors to analyze the data (section~\protect\ref{subsubsec:analyze}), and concludes the study (section~\protect\ref{subsubsec:conclude}).}

    \label{fig:study_lifecycle}
\end{figure}

\Cref{fig:study_lifecycle} shows how a typical research team uses Rally and WebScience to conduct a study.

\subsubsection{Design and Implement Study}
\label{subsubsec:designandimplement}
Research teams that wish to run a Rally study first receive approval from their
institution's IRB and from a team at \mozilla{} that evaluates studies. Once approved,
the researchers implement their study as a browser extension, likely utilizing \WebScience{}.
\Mozilla{} conducts a code review on the study extension, and researchers submit IRB-approved consent
forms and study explanations, as well as a schema describing the format of the data that the extension will report.

\subsubsection{Recruit Participants}
\label{subsubsec:recruit}
\Mozilla{} advertises the study to users of \firefox{}, using in-browser messaging and other methods.
Research teams may also conduct their own recruitment efforts.

\subsubsection{Onboard Participants}
\label{subsubsec:onboard}
Once a potential participant has indicated interest in enrolling, they first join Rally, and are asked to complete an optional
demographic survey.
Next, they can join a specific study. Rally conducts the informed consent process,
using the consent forms supplied by the research team. It then installs the study extension on the participant's
browser. Research teams can configure the study extension to ask participants to complete
a survey specific to their study, using Qualtrics or another survey platform.

\subsubsection{Run Study}
\label{subsubsec:run}
Studies run for a predetermined amount of time, ranging from a few weeks to more than a year. During this time, the study
extension collects measurements as participants browse, potentially deploys interventions, and periodically packages collected data into
digests, which it sends to the backend storage environment.

\subsubsection{Store Study Data}
\label{subsubsec:store}
\Mozilla{} maintains a pipeline for incoming Rally digests, using its browser telemetry infrastructure.
Digests are transmitted securely, then decrypted and validated against the study schema provided by the research team before being stored securely.

\subsubsection{Analyze Study Data}
\label{subsubsec:analyze}
Members of the research team can log in to the analysis environment to view reported data at any time.
The data is stored on machines administered by \mozilla{}, and researchers cannot transfer it elsewhere.
Instead, researchers run analyses in JupyterLab notebooks with access to the data.

\subsubsection{Conclude Study}
\label{subsubsec:conclude}
At the end of the predetermined study timeframe, Rally removes the study extension from participants' browsers.
If necessary, the research team can perform any debriefing (such as revealing deception) at this point.



\section{\Rally{} System Description}
\label{sec:rally}
We now describe how Rally addresses the barriers to sociotechnical research we established in~\Cref{sec:motivation,sec:survey}.

\subsection{Scale and Representativeness}
Our literature review showed that without a partnership or data provider,
most studies can only recruit a few hundred participants.
For Rally, \mozilla{} has accumulated a population of potential participants by
recruiting in-browser and on social media.
Approximately 6,500 users are currently signed up, and \mozilla{} plans to
significantly expand recruitment efforts in the near future. 
\Rally{} is public and any US-based 19-or-older user of \firefox{}
can join the program\redact{~\cite{mozilla-pioneer}} (\Cref{fig:recruiting}).

Of course, scale alone is not always enough for a useful sample.
Researchers often need samples whose demographics match some larger population.
Rally asks newly enrolled users to fill out a short demographic survey (\Cref{fig:demographics}),
and researchers running Rally studies can use that data to understand the demographics of their participants,
and if desired, rebalance their sample to achieve more representative results (see~\Cref{sec:representativity} for more).

\subsection{Participant privacy}
\begin{figure}[t]
    \centering
    \begin{subfigure}[b]{0.475\textwidth}
        \centering
        \captionsetup{justification=centering}
        \includegraphics[width=\textwidth, frame]{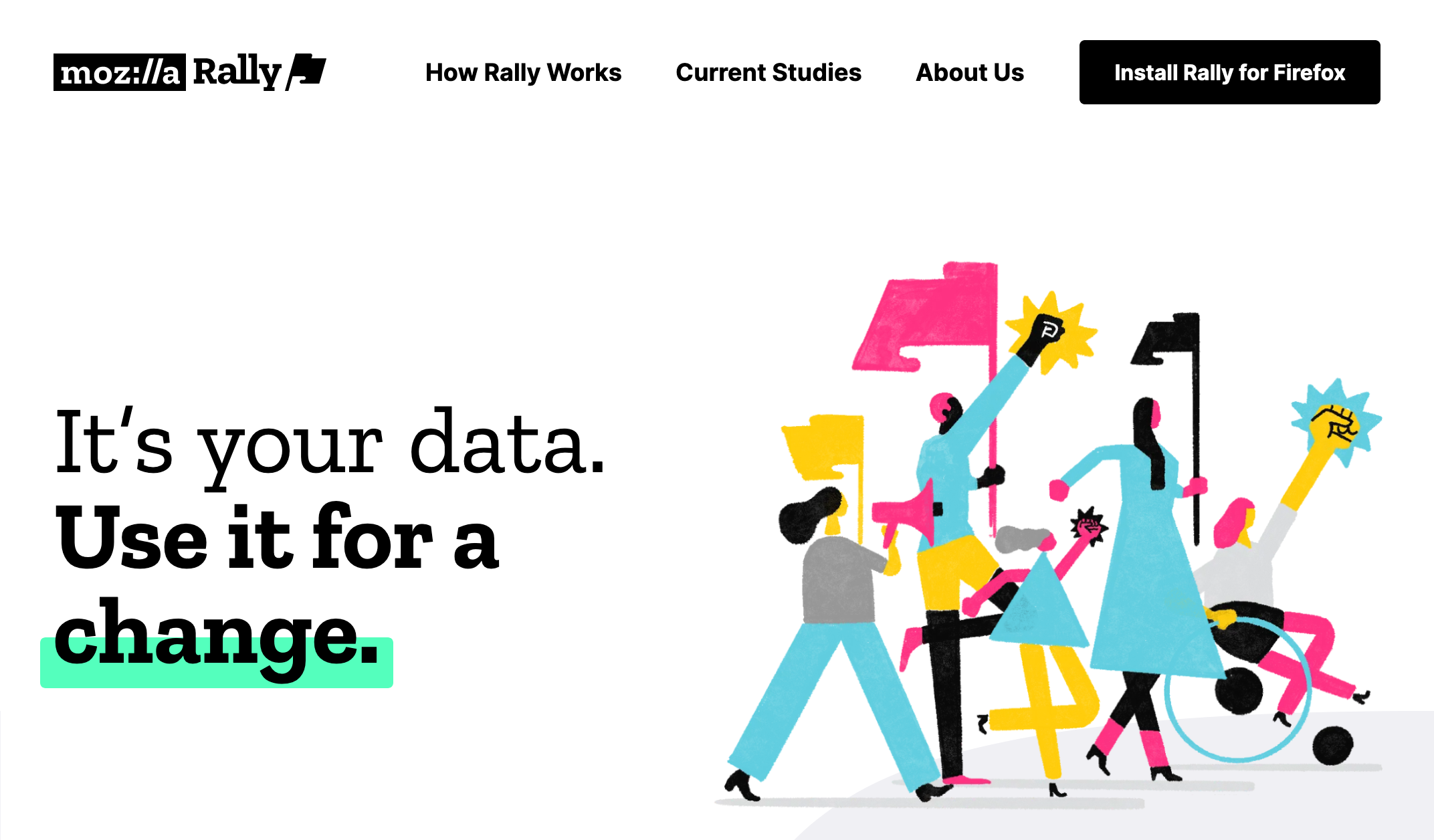}
        \caption[]
        {{\small A prospective user can learn about Rally and sign up through a website.}}    
        \label{fig:recruiting}
    \end{subfigure}
    \hfill
    \begin{subfigure}[b]{0.475\textwidth}  
        \centering 
        \captionsetup{justification=centering}
        \includegraphics[width=\textwidth, frame]{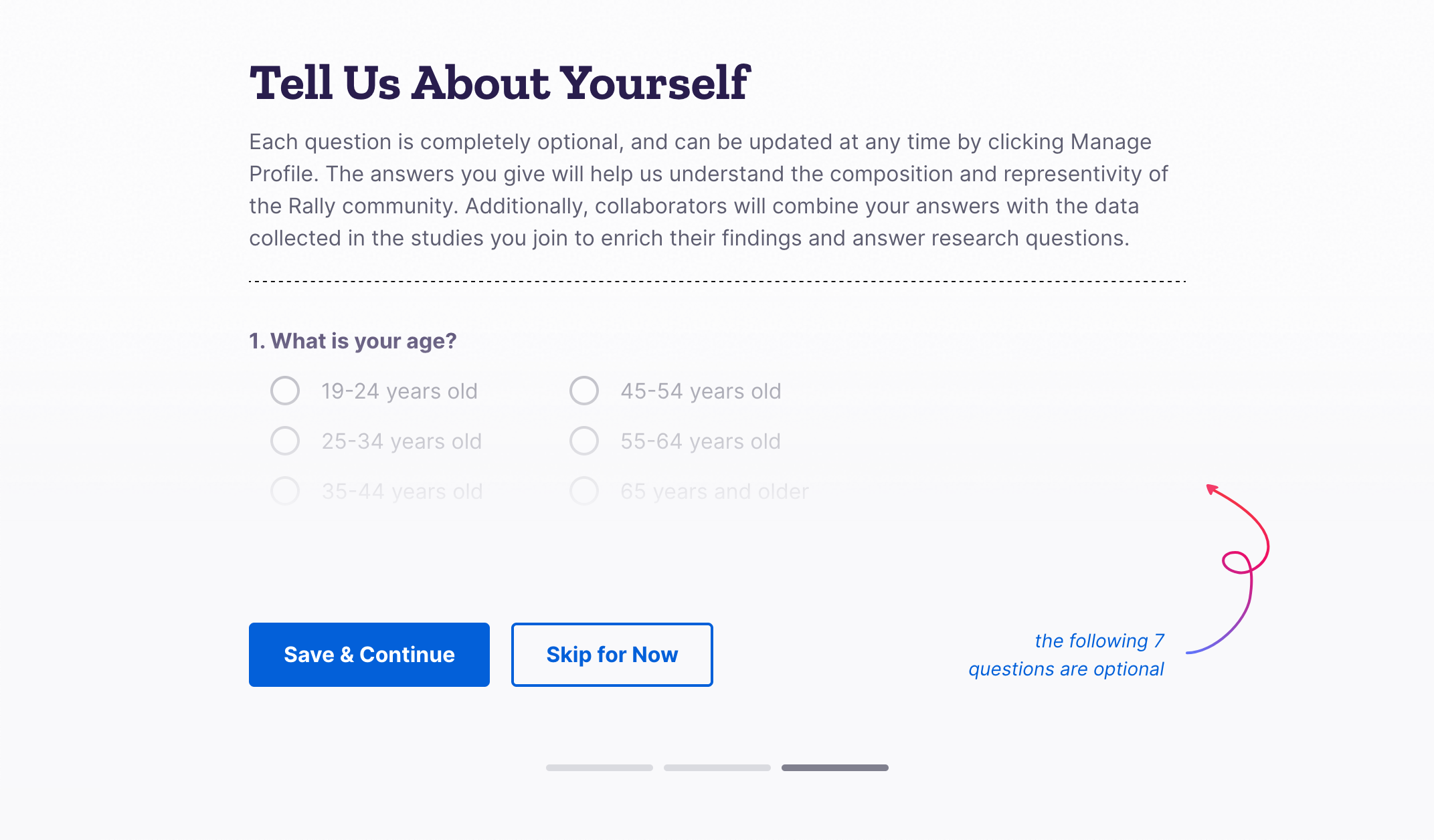}
        \caption[]%
        {{\small After joining, Rally asks the user to answer a demographic survey.}}    
        \label{fig:demographics}
    \end{subfigure}
    \vskip\baselineskip
    \begin{subfigure}[b]{0.475\textwidth}   
        \centering 
        \captionsetup{justification=centering}
        \includegraphics[width=\textwidth, frame]{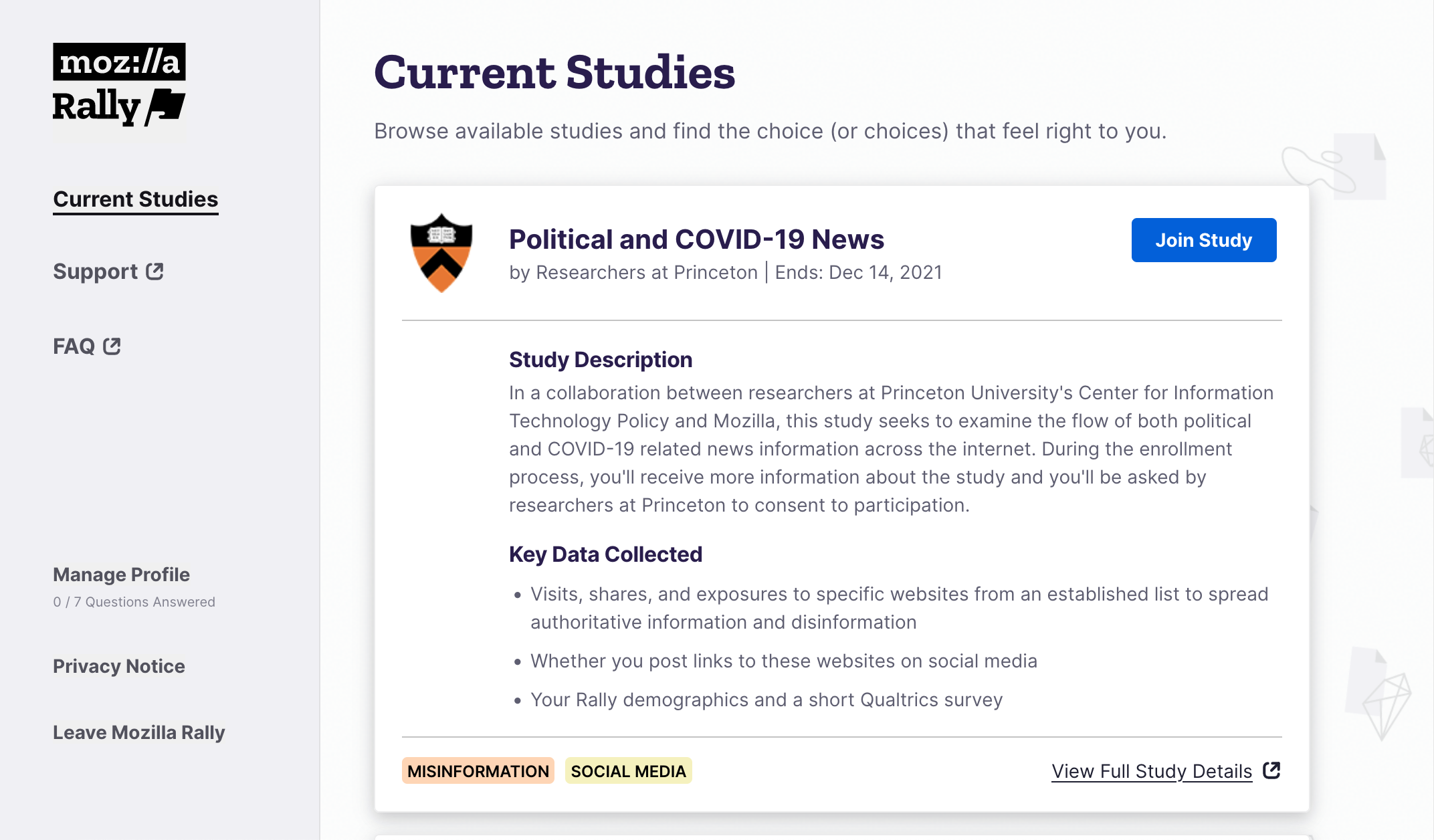}
        \caption[]%
        {{\small The user can browse available studies for which they are eligible.}}    
        \label{fig:current-studies}
    \end{subfigure}
    \hfill
    \begin{subfigure}[b]{0.475\textwidth}   
        \centering 
        \includegraphics[width=\textwidth, frame]{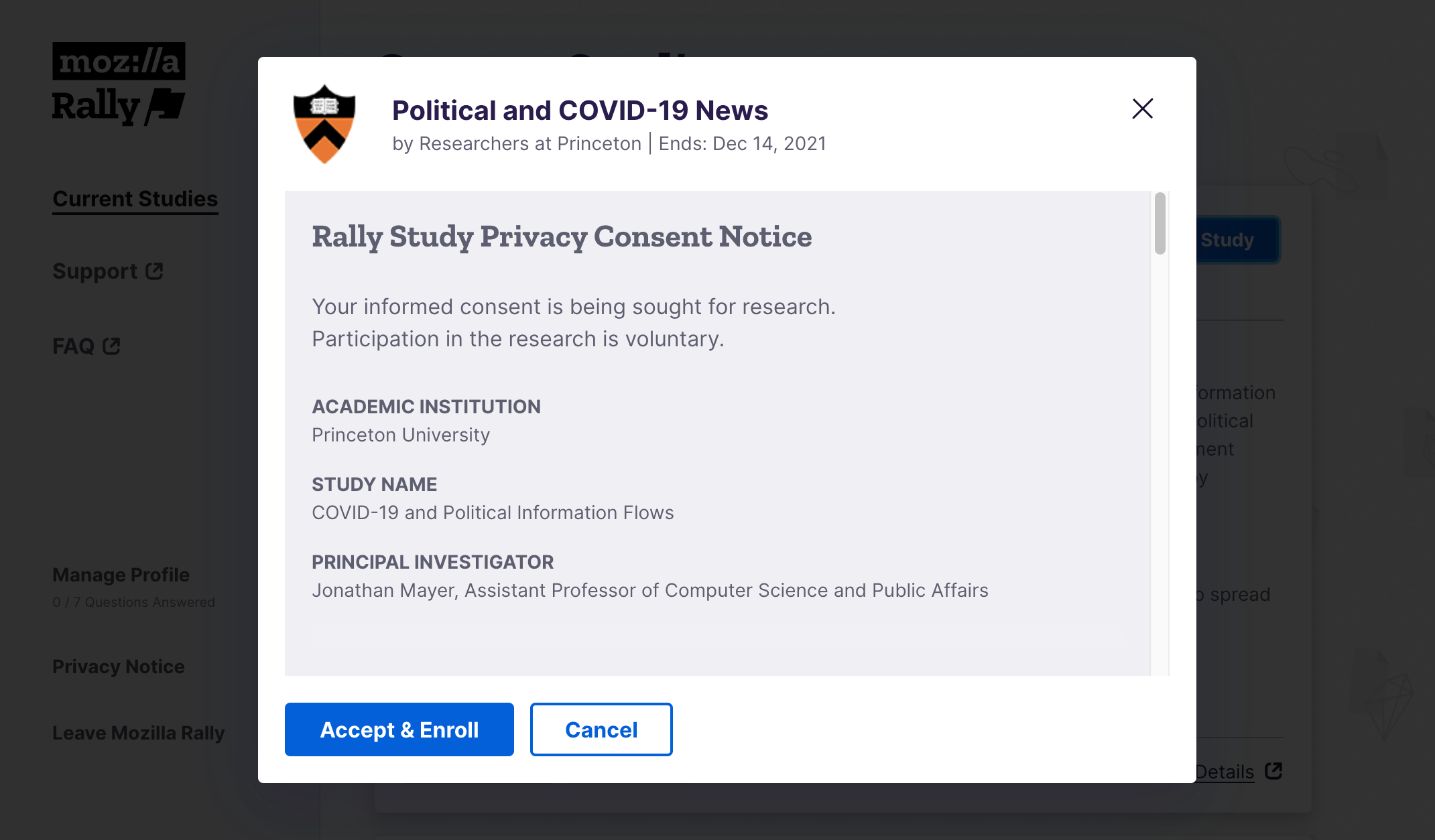}
        \captionsetup{justification=centering}
        \caption[]
        {{\small Before joining a study, the user must read and accept the study's consent notice.}}    
        \label{fig:consent}
    \end{subfigure}
    \caption[]
    {\small The signup flow for a new \Rally{} user.} 
    \label{fig:rally-signup}
\end{figure}

Rally both mitigates some privacy concerns of other methods and raises new privacy issues
that it must address. First, as Rally is a fully opt-in system, with users affirmatively
choosing to enroll first in Rally and then in each
individual study they join (\Cref{fig:current-studies,fig:consent}), it avoids
the privacy concerns of platform partnerships or commercial panel providers: researchers
control the language used and supply IRB-approved consent forms, so they can be
confident that participants know they are research subjects and have given
informed consent.

Given the sensitivity of the data that Rally studies can collect, researchers and
\mozilla{} must take special care to communicate clearly with potential participants,
to practice data minimization, to protect collected data in transit and in storage,
and to balance the goals and risks of a study.
Here, the involvement of a privacy-focused browser vendor helps research teams
accomplish these tasks. Studies are subject to reviews of their goals, code,
and of the specific fields of data collected. Each data point that research
teams plan to collect must be labeled and its risk to users classified.
The Rally data pipeline handles encrypting
the data in transit and shuttling it into an access-controlled analysis
environment, set up by \mozilla{}.


Privacy is addressed further in~\Cref{sec:privacy,sec:reporting-privacy}.

\subsection{Backend Infrastructure}
The backend infrastructure for a browser extension-based research study must
distribute the research extension to participants' browsers,
receive collected data from the extension, store it
securely, and mediate researcher access.
Building this infrastructure can be a barrier for independent teams.

Rally handles all of these tasks for researchers.
\Mozilla{} maintains the website that informs and enrolls participants,
distributes the study extension to enrolled participants, and
runs the servers that receive and store the data. The study extension reports data
through \firefox{}'s telemetry pipeline, with Rally data encrypted such that
only researchers have access.

\section{\sdkName{}}
\label{sec:websci}
Although custom instrumentation would be useful for studying many sociotechnical problems,
our survey identified that very few studies have created their own instrumentation, likely
due to technical complexity.
\WebScience{} significantly reduces the technical
burden of implementing measurement probes and includes instrumentation for aspects
of user attention and navigation that are otherwise difficult to measure correctly across
the breadth of the web.
\sdkName{} is available on
GitHub.
\footnote{\redact{\url{https://github.com/mozilla-rally/web-science}}}
It is open-source and permissively licensed.

\subsection{\sdkName{} implementation}

\sdkName{} follows the WebExtensions framework (a set of APIs for browser extensions,
supported by all major desktop browsers) and can be customized for each
research study that uses it~\cite{webextensions}.
It can be imported into studies as a group of ES6 modules.
Below, we provide brief background on WebExtensions, then describe \sdkName modules.

\subsubsection{WebExtensions Background}
We leave a detailed explanation of WebExtensions to its online documentation, but the following
terms will be useful in subsequent sections.
Extensions have three main types of processing units:
\textit{content scripts}, which run in the environment of a
particular webpage; \textit{background scripts}, which run persistently in the extension's own environment; and
\textit{web workers}, which run in separate threads and their own environment, and can be spun up and down as needed.
\textit{Match patterns} are regex-like strings that specify the URLs on which a content script will run.





\subsection{Navigation and attention}
\label{sec:websci:nav-attn}
Some of the most fundamental measurements associated with web browsing are which
webpages participants visit, how they reach them, and the time they spend
paying attention to them. While fundamental, these measurements are not as simple
as they may appear. \sdkName{} includes three modules that provide these measurements:
page navigation, page transitions, and attention tracking.

\subsubsection{Page Navigation}
\label{subsubsec:page-navigation}
Most page navigation takes place when a user clicks a link or enters an address in the URL bar, causing a new page to load.
Recording this type of page navigation is relatively straightforward, but not all websites follow this standard.
Some websites, including popular ones like YouTube, use the \texttt{History} API to change the URL shown to users
and manually change the DOM instead of performing traditional page loads.
This means that a user experiences an apparent page navigation (the URL and page content change in response to a click),
but browser instrumentation monitoring for page loads would not recognize the change as a new page. 

\sdkName{} handles this by monitoring for both types of page navigations, and
simplifies them into a single API for studies to use.

In keeping with our goal of making \sdkName{} easy to use, the code for a study to configure and run
this module is short:
\begin{Verbatim}[frame=single]
pageNavigation.onPageData.addListener(pageNavListener,
    { matchPatterns: [ "https://*.acm.org/*" ] });
\end{Verbatim}

The above line of code
configures the module to call the study's listener function (\texttt{pageNavListener}) for each
page the user visits on \url{acm.org} or its subdomains.
The module tracks the beginning and end of page visits (when the user navigates away,
or closes the tab or browser) and notifies the study code at the completion of the
visit. The notification includes the URL, the referrer, the starting and ending timestamps,
a unique ID, and information about the user's engagement with the
page (see~\cref{subsubsec:attention-tracking}).

Another issue addressed by this module is coordinating measurements across different
collection points. Some information, such as specific
page contents, is only available to content scripts. Other data, such as the content
of network requests, is accessible in background scripts. Understanding the full context
of a user action may require combining information from these environments and
associating measurements with the correct page visit (race conditions
around page or tab transitions add complexity to this task).
Unfortunately, the browser does not provide a mechanism to reliably link
measurements from the content script and background script environments.
The Page Navigation module assigns unique
IDs to each page visit, and makes those IDs available to other modules in both the content
and background script environments. This allows
study code to reliably correlate measurements from separate modules and collection points.


\begin{figure}[t]
    \centering
    \begin{subfigure}[b]{\textwidth}
    	\resizebox{\textwidth}{!}{  \begin{tikzpicture}[node distance=2.0in]
  	\tikzstyle{every node}=[square]
  	\node (page1)[label={[label distance=-0.8in, xshift=0.43in, yshift=-0.03in]150:http://www.a.com/}]{Browse 30 sec};
  	\node (page1tab)[tab, above=of page1.west, anchor=west, yshift=-1.57in]{1};
  	
  	\node (page2)[right of=page1,label={[label distance=-0.8in, xshift=0.43in, yshift=-0.03in]150:http://www.b.com/}]{Browse 17 sec};
  	\node (page2tab)[tab, above=of page2.west, anchor=west, yshift=-1.57in, xshift=0.2in]{2};
  	
  	\node (page3)[right of=page2,label={[label distance=-0.8in, xshift=0.43in, yshift=-0.03in]150:http://www.a.com/}]{Browse 45 sec};
  	\node (page3tab)[tab, above=of page3.west, anchor=west, yshift=-1.57in]{1};
  	
  	\node (page4)[right of=page3,label={[label distance=-0.8in, xshift=0.43in, yshift=-0.03in]150:http://www.c.com/}]{Browse 4 min\\then close browser};
  	\node (page4tab)[tab, above=of page4.west, anchor=west, yshift=-1.57in]{1};
  	
  	\node[left=0.7in of page1,draw=none,text width=0,minimum height=0] (start) {};
  	\node[above=of start,draw=none,minimum height=0,xshift=0.3in,yshift=-1.42in](activity-label){\Large{\textbf{User Activity}}};
  	
  	\node[above=of start,draw=none,minimum height=0,xshift=0.33in,yshift=-2.025in](arrow-1-lbl-top){Open browser};
  	\node[below=of start,draw=none,minimum height=0,xshift=0.33in,yshift=2.025in](arrow-1-lbl-bot){Type\\\scriptsize{http://www.a.com/}};
  	
  	\node[above=of page1,draw=none,minimum height=0,xshift=1in,yshift=-2.3in](arrow-2-lbl-top){Click link\\\scriptsize{http://www.b.com/}};
  	\node[below=of page1,draw=none,minimum height=0,xshift=1in,yshift=2.3in](arrow-2-lbl-bot){Open in\\new tab};
  	
  	\node[above=of page2,draw=none,minimum height=0,xshift=1in,yshift=-2.3in](arrow-3-lbl-top){Click tab 1};
  	
  	\node[above=of page3,draw=none,minimum height=0,xshift=1in,yshift=-2.3in](arrow-4-lbl-top){Click link\\\scriptsize{http://www.c.com/}};

  	\draw[-stealth] (start) -- (page1);
  	\draw[-stealth] (page1.east) -- (page2.west);
  	\draw[-stealth] (page2.east) -- (page3.west);
  	\draw[-stealth] (page3.east) -- (page4.west);
  \end{tikzpicture}

}
    	\label{fig:browsing-flow} 
    \end{subfigure}
    \begin{subfigure}[b]{\textwidth}
    	\resizebox{\textwidth}{!}{\begin{tabular}{|l|l|l|l|}
\multicolumn{4}{l}{\textbf{\large{ WebScience}}}                                                                                                                                                                                                                                                                                                                                                                                                                                                                                                                                                                                                                                                                                                                                                                                                                                                                           \\ 
\hline
\vcell{\begin{tabular}[b]{@{}l@{}}pageId: 1\\url: http://www.a.com/\\referrer: none\\priorPageId: none\\transitionType: typed\\transitionQualifier: from\_address\_bar\\startTime: 1/15/21 16:21:31\end{tabular}} & \vcell{\begin{tabular}[b]{@{}l@{}}pageId: 2\\url: http://www.b.com/\\referrer: http://www.a.com/\\priorPageId: 1\\transitionType: link\_click\\startTime: 1/15/21 16:22:01\\attentionDuration: 1700\\maxScrollDepth: 23\end{tabular}} & \vcell{\begin{tabular}[b]{@{}l@{}}pageId: 1\\url: http://www.a.com/\\referrer: none\\priorPageId:none\\startTime: 1/15/21 16:21:31\\stopTime: 1/15/21 16:23:03\\attentionDuration: 7500\\maxScrollDepth: 76\end{tabular}} & \vcell{\begin{tabular}[b]{@{}l@{}}pageId: 3\\url: http://www.c.com/\\referrer: http://www.a.com/\\priorPageId: 1\\transitionType: link\_click\\startTime: 1/15/21 16:23:03\\stopTime: 1/15/21 16:27:03\\attentionDuration: 24000\\maxScrollDepth: 91\end{tabular}}  \\[-\rowheight]
\printcelltop                                                                                                                                                                                             & \printcelltop                                                                                                                                                                                                         & \printcelltop                                                                                                                                                                                                  & \printcelltop                                                                                                                                                                                                                                       \\ 
\hline
\multicolumn{1}{l}{}                                                                                                                                                      & \multicolumn{1}{l}{}                                                                                                                                                                                                  & \multicolumn{1}{l}{}                                                                                                                                                                                           & \multicolumn{1}{l}{}                                                                                                                                                                                                                                \\ 
\multicolumn{1}{l}{\textbf{\large{ Conventional Measures}}}                                                                                                                                                      & \multicolumn{1}{l}{}                                                                                                                                                                                                  & \multicolumn{1}{l}{}                                                                                                                                                                                           & \multicolumn{1}{l}{}                                                                                                                                                                                                                                \\ 
\hline
\vcell{\begin{tabular}[b]{@{}l@{}}pageId: 1\\url: http://www.a.com/\\time: 1/15/21 16:21:31\\referrer: none\\priorPageLoaded: none\\timeToNextPageLoad: 3000\end{tabular}}                                                                  & \vcell{\begin{tabular}[b]{@{}l@{}}pageId: 2\\url: http://www.b.com/\\time: 1/15/21 16:22:01\\referrer: http://www.a.com/\\priorPageLoaded: 1\\timeToNextPageLoad: 62000\end{tabular}}                                              & \vcell{--}                                                                                                                                                                                                     & \vcell{\begin{tabular}[b]{@{}l@{}}pageId: 3\\url: http://www.c.com/\\time: 1/15/21 16:23:03\\referrer: http://www.a.com/\\priorPageLoaded: 2\end{tabular}}                                                                           \\[-\rowheight]
\printcelltop                                                                                                                                                                                             & \printcelltop                                                                                                                                                                                                         & \printcellmiddle                                                                                                                                                                                               & \printcelltop                                                                                                                                                                                                                                       \\
\hline
\end{tabular}}
    	\label{fig:browsing-table}
    \end{subfigure}
	\caption{In this example browsing session, a user visits three pages across two tabs. Each time they visit a page, a different type of transition occurs. WebScience records data about each visit and transition as a DAG that accurately models user activity. Conventional methods may use browser history logs or track \texttt{referrer} headers, which would incorrectly model the user activity in several ways (see ~\protect\Cref{subsubsec:page-transitions} and~\protect\Cref{subsubsec:attention-tracking}).}
    \label{fig:browsing}
\end{figure}


\subsubsection{Page Transitions}
\label{subsubsec:page-transitions}
Understanding user browsing often requires knowing how participants arrived at a
particular webpage. In the past, the \texttt{referrer} field set by browsers provided
this, listing the URL the user visited before the current page. However,
browsers have moved to allow sites to limit or omit this value,
and more recently have limited it by default, citing legitimate
privacy concerns~\cite{referrer-default, referrer-policy}. 
Referrers are therefore no longer necessarily an accurate source for understanding
navigation.

Researchers who collect browsing histories or page loads from users (as
commercial panel providers often supply) have often
taken the order of pages visited as a proxy for how a user travelled between pages.
However, since users may have many tabs open and move between them, knowing that page
A loaded before page B does not mean that the user clicked a link on A to get to B ---
instead, they may have been in separate and unrelated tabs.

\sdkName{} directly monitors the source of page navigations and provides the \sdkName{}-assigned
ID of the page visit where the participant followed a link.
This allows study code to consider
page navigations as a directed acyclic graph (DAG), where nodes are page visits and edges
connect a node to any page visits it generated.

As an example, consider the page visit to \texttt{http://www.c.com/} in~\Cref{fig:browsing}.
The participant navigated to \texttt{http://www.c.com/} from a link on \texttt{http://www.a.com/}.
A measurement that only considered page load times would incorrectly identify \texttt{http://www.b.com/} as the page that linked the user to \texttt{http://www.a.com/}, because the page load of \texttt{http://www.b.com/} directly preceded \texttt{http://www.a.com/}.
The Page Transitions module would correctly report that \texttt{http://www.a.com/} was the prior page for the \texttt{http://www.c.com/} visit.

\subsubsection{Attention Tracking}
\label{subsubsec:attention-tracking}

Another commonly needed measurement is user engagement with pages.

Previous methods of tracking attention often monitor dwell time: the length of
time between navigating to and leaving a site. Since users can pause browsing
and pick back up later, or can leave page open for a long time while working in another tab,
dwell time can both under- and over-estimate user attention.

Studies that collect browser histories for research have used the timespan between
page loads to estimate attention, positing that if a user loads site A at time $t_0$ and
later loads
site B at time $t_1$, they were on site A for $(t_1 - t_0)$ time. However,
users may leave sites
open in background tabs and return to them after navigating
elsewhere, so this method may underestimate attention to pages.

\sdkName{} measures attention by directly accumulating the time a user is actively
visiting a page.
We define the active page by checking whether a browser window is the selected window
on the system, and if so, finding the selected tab within that window.
For each page the participant visits, we accumulate the total time the page is
the active page, pausing the clock when the mouse and keyboard input indicate that
the user appears to have stepped away from
the computer or otherwise disengaged.

The importance of actively measuring attention is visible in ~\Cref{fig:browsing}. When the participant clicks tab 1 to switch back to that tab, \sdkName{} pauses attention tracking for \texttt{http://www.b.com/}, which is open in tab 2. Conventional methods would treat the loading of \texttt{http://www.b.com/} as the end of the user's visit to \texttt{http://www.a.com/} and record 30 seconds spent on \texttt{http://www.a.com/}. \sdkName{} correctly computes the attention duration for \texttt{http://www.a.com/} as 75 seconds. Conventional methods would include the additional 45 seconds spent on \texttt{http://www.a.com/} the next time a page load occurs, so they would model the user as spending 62 seconds on \texttt{http://www.b.com/} while \sdkName{} correctly records 17 seconds.

\subsection{Social Media Activity}
Many Internet phenomena occur on social media platforms, but these companies
can be particularly unwilling to share data with researchers,
and their public APIs are often
severely lacking~\cite{edelson_opinion_2021, perriam_2020_post-api}.
Creating instrumentation to measure activity on these sites is quite difficult,
as their code is usually obfuscated, and sometimes actively hostile to
comprehension. Many protections appear to be intended to defeat ad-blockers,
but also make instrumentation difficult.
\sdkName{} includes two modules that interface with social media sites
and allow researchers to collect data easily.
The Social Media Activity module provides a simple and powerful API for researchers
to collect data about
actions the user
takes on social media sites. The module can watch for events from
Facebook, where it tracks posts, shares, emoji reactions, and comments; Twitter, where it
tracks tweets (including quote tweets), retweets, and favorites; and Reddit, where it tracks posts, comments,
and votes.
When a researcher configures Social Media Activity,
they specify which platforms and events to track.

This module also provides functions for fetching more information about posts on social
media platforms.

\subsubsection{Social Media Link Sharing}
To simplify the process of tracking link shares on social media,
we provide
the Social Media Link Sharing module.
This module uses Social Media Activity to specifically
track ``link sharing'' events: any time a user shares a link on social media, including
making new posts that contain links as well as sharing existing posts that contain links.
Studies configure the module with a set of match patterns for URLs of interest.
The module tracks the total number of links shared as well as provides
full information (including URL, timestamp, platform,
sharing action, audience, and whether the post was original or a reshare)
about shared links in the match pattern set.

A selection of additional modules are briefly described in~\cref{app:webscience-modules}

\subsection{Reporting Data}
\label{sec:reporting-privacy}

Even once participants have understood the parameters of a study
and consented to participate, researchers have a responsibility to
collect the smallest amount of data possible to answer the research
questions. Many collection mechanisms simply give researchers
all of the potentially relevant (or even obviously irrelevant) data,
because determining relevancy in an automated fashion is difficult.

We have designed \WebScience{} to enable and encourage good data minimization.

Studies configure modules to collect their desired data, and then store the results of
that measurement in the browser to be later transmitted to the analysis environment.
Modules produce detailed information, but studies reduce the specificity
of that data before transmission
by completing analysis on the participant's computer and by
aggregating events.

While the exact analysis and aggregation steps are determined by individual studies,
we include two examples here to illustrate how
\Rally{} allows researchers to collect useful data while minimizing risks.

\subsubsection{Local Analysis}
Since studies run directly on participants' computers, they can complete some
data analysis before transmitting the data, and send only the result of the analysis.
As an example, one currently running
study uses news articles read by the participant as input to an NLP classifier
\emph{entirely in-browser}, and then reports only the binary
output of the classifier. 
Researchers receive the necessary information about the articles without ever collecting
the URL, title, or content.


\subsubsection{Aggregation}
The other way that studies can reduce the specificity of the data they transmit is
to aggregate individual data points into categories of related data points, and then
transmit information about the categories.





The study collects raw events for a period of time, perhaps several days,
then aggregates the events into categories and reports the categories and how many
events fell into each category.
By reducing the specificity of the data, we reduce the risk
that actions can be traced back to individual users.

In keeping with our goal of making \sdkName{} accessible to researchers across
disciplines, we include a module that automates aggregating data.
By providing a module with default
settings that protect participants' privacy, we make respecting privacy easier than
over-collecting data.

\section{Privacy}
\label{sec:privacy}
\Rally{} has the potential to collect a great deal of information, possibly
including sensitive
personal information, from its participants. 
\Rally{} does not collect names, email addresses, or other directly identifiable
information from participants. However, previous research has shown that simply
omitting identifiers from data is not sufficient to protect the identities
of the subjects~\cite{narayanan_2008_robust-deanonymization}.
Therefore, \Rally{} must take additional active steps to
reduce the risks to study participants, as described below.

\subsection{Opt-in}
\Rally{} is a fully opt-in program.
The entry point for \Rally{} is a webpage describing the
program where users can choose to enroll. Each study that
runs through \Rally{} requires an additional opt-in step
and participants can also leave any study, or
the \Rally{} program entirely, at any point.

\subsection{Data Deletion}
When a user leaves before the official end of the study, \mozilla{}
deletes raw data from that user from the storage servers.
\Rally{} retains raw data for researchers for a maximum of two years after
collection.
Researchers may retain higher level aggregated data for analysis.

\subsection{Informed Consent}
\Rally{} studies by academic researchers 
must receive IRB approval at the researcher's institution and develop an informed consent document.
Studies run by journalists or other non-academic researchers must also generate
an informed consent document for participants.
In both cases, the document describes the purpose of the study, the data collected, and how that data
will be stored and accessed. This document is presented to participants before
they can choose to join the study, and they must indicate agreement before being able
to join.
Participants can return to the informed consent document at any point during the study.

\subsection{Data Minimization}
\sdkName{} is designed to encourage minimizing the amount and granularity of the data collected.
All modules that collect data about websites must be configured by researchers,
who supply a set of match patterns for which the module should collect data. This limits
the collection to the set of sites that researchers care about for a particular study,
rather than automatically recording data for all sites. Many other methods of data collection
automatically collect all of the user's traffic, picking up sensitive page visits that
aren't relevant to the study's goals. Limiting collection to only the relevant data significantly
reduces the risk to participants.

\label{sec:privacy:aggregation}
Further, rather than transmitting raw data, such as full URLs and timestamps, studies can aggregate data
into broader categories, and send a description of the category and the number of
raw events that fell into it. By ensuring that raw data never leaves participants'
computers, we reduce the risk that they can be identified by researchers or have
personal information leaked.
\sdkName{} provides a module that automates aggregating
data, which researchers can customize for each study's needs.

\subsection{Lack of Persistent Identifiers}
Studies transmit data periodically to \mozilla{}'s servers, and researchers will
often need to connect reports from the same user throughout time. To facilitate
this need, participants are assigned a per-study identifier. However, this identifier
is unique for each study,
separate from any other systems in \firefox{}, and not available
to other parts of the browser.
This protects
users from being identified.

\subsection{Access Controls}
\label{sec:privacy:access}
Data is encrypted with a study-specific key before it is sent from the participant's
computer to \mozilla{}'s servers. It is only decrypted once it reaches the access-controlled
analysis environment controlled by \mozilla{} and open only to researchers involved with the study.
This restricts the audience at \mozilla{} that can
view it, as well as prevents snooping while the data is in transit. Automated exfiltration detection
prevents researchers or others from exporting large amounts of data.

\subsection{Study Review}
Finally, all studies are reviewed, first in concept and then in implementation, by
\mozilla{}. This review ensures that studies collect reasonable data from users, that
the informed consent document is correct and clear, and that the study's implementation
does not present security or usability risks.
This review also considers research ethics: whether the proposed study is
likely to produce
improvements to our understanding of technology and society problems,
while maintaining minimal privacy risks to participants.
\section{Evaluation}
\label{sec:eval}
\Rally{} is actively deployed. Four research teams have launched studies, all using \sdkName{}, and several more are in development. Based on this usage, we present three evaluations: a case study previewing an active \Rally{} study (\Cref{subsec:case-study}), a quantitative evaluation comparing \sdkName{} measurements to prior methods (\Cref{subsec:quant-eval}), and qualitative findings from semi-structured interviews of researchers, engineers, and investigative journalists who have used \Rally{} and/or \sdkName{} to launch studies (\Cref{subsec:qual-eval}).

\subsection{Study Preview: Pathways to Political and COVID-19 News}
\label{subsec:case-study}
Researchers, policymakers, and civil society groups have voiced concerns for over a decade about the proliferation of misinformation online and the potentially dangerous consequences for politics, democracy, and society~\cite{lewandowsky_beyond_2017}.
Concerns in the U.S. have focused on misinformation about politics and health, especially during the COVID-19 pandemic, when dangerous false claims about vaccine safety, off-label treatments, and other issues circulated widely and hindered public health efforts~\cite{cinelli_covid-19_2020}.

Prior work has measured how much news and misinformation users engage with by using data collected from platform APIs~\cite{grinberg_2019_fake, allcott2019trends, yang2021infodemic, singh_2020_understanding, bovet2019influence}, data scraped from platforms~\cite{samory2018conspiracies}, data received through platform partnerships~\cite{guess_2021_cracking, bailey2021interactions}, data purchased from commercial panel providers~\cite{altay2022quantifying, guess_exposure_2020}, and self-reported data from surveys and interviews with users.
As discussed in~\Cref{sec:prior-behavior}, these methods come with significant research design constraints and ecological validity limitations.
In particular, most studies have focused on just one or two platforms, because researchers have lacked practical methods to study engagement across the wide array of services through which users might access news and misinformation, including social media, search, email, web portals, news aggregators, and publishers' websites.

We designed a \Rally{} study using \sdkName{} to measure user engagement with political and health-related news and misinformation across the web. Our study addresses the questions:

\begin{enumerate}[label={RQ\arabic*}]
\item \label{RQ1} On what sites are desktop browser users being exposed to links to news and disinformation sites, and how frequent are these exposures compared to exposures to other sites?
\item How often do desktop browser users visit news websites and misinformation websites, what types of pages do they visit, and how long do they pay attention to those pages? 
\item How often do desktop browser users share links to news and disinformation websites on social media when using desktop browsers, what types of pages do they share, and to what audiences? 
\item How are exposure, visits, and sharing of news and misinformation correlated?
\item How do user demographics and political views correlate with news and misinformation engagement?
\end{enumerate}

\subsubsection{Study Design}
Using WebScience, we created a browser extension that collects data whenever the user shares, visits, or sees a link to a news, health, or misinformation website.
During six months of data collection between March 20, 2022 and September 14, 2022, 2,709 Rally users joined the study
and reported data at least once.
Below we briefly describe our methods and present sample data.
Note that this study is an independent research project, so a full explanation of the motivation, goals, methods, and results is left to the project's own paper.

\paragraph{Methods}
Drawing from public datasets and prior work, we compiled lists of domains for each category of website we planned to track: news outlets, health information websites, websites known to spread misinformation, social media platforms, search engines, webmail services, web portals, fact checkers, and news aggregators.
When a link to a tracked news, health, or misinformation website is exposed to a user, we collect the domain where the exposure occurred and the exposed domain.
When a user visits a tracked news, health, or misinformation website, we collect the source and destination domains and measures of user attention to the page.
We also run an in-browser classifier that derives features from the URL, page title, and page content to predict the type of the visited page: political news article, health-related article, an article about something other than politics or health, or a non-article.
Each time a user shares a tracked news, health, or misinformation link on Facebook, Twitter, or Reddit, we collect the platform name, the domain shared, and information about how and to what audience the link was shared.
We also store a flag indicating whether we observed the user visiting the link prior to sharing it.
For \textit{untracked} domains (those that do not appear on any of our lists), we collect aggregate counts of exposures, visits, and shares so that we can compute base rates. 
We never collect the domain name of an untracked website.
For social media, search, webmail, web portals, fact checkers, and news aggregators, we collect the domain name only if it is the source of a visit or exposure to a news, health, or misinformation website.


\subsubsection{Sample Study Data}
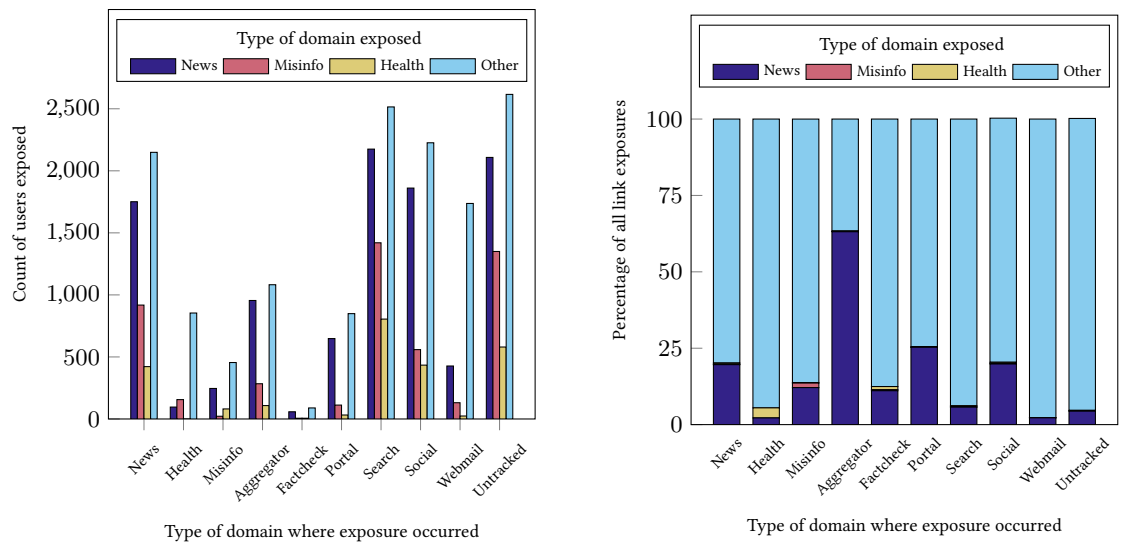
\begin{figure}
    \centering
    \begin{minipage}{0.48\textwidth}
        \centering
        \begin{tikzpicture}
\begin{axis}[
    ybar=0pt,
    ymin=0,
	bar width=2.5pt,
	width = \textwidth,
	height = 200pt,
	ymax = 3300,
	ytick={0,500,1000,1500,2000,2500},
    ylabel={\footnotesize Count of users exposed},
    xlabel={\footnotesize Type of domain where exposure occurred},
    symbolic x coords={
        News, Health, Misinfo, Aggregator, Factcheck, Portal, Search, Social, Webmail, Untracked
        },
    xtick=data,
    x tick label style={rotate=45, font=\scriptsize},
    xtick pos=left,
    ytick pos=left,
    legend columns=5,
    legend style={
            at={(0.5,0.9)},
            anchor= center,
            inner sep=1pt,
    },
    ]
\addlegendimage{empty legend}
\addlegendentry[yshift=14pt]{}
\addplot[ybar,fill=cb1,draw=black,area legend] coordinates
        {(News,1751) (Health,96) (Misinfo,246) (Aggregator,955) (Factcheck,58) (Portal,648) (Search,2175) (Social,1861) (Webmail,427) (Untracked,2108)};
\addlegendentry{\scriptsize News}
\addplot[ybar,fill=cb5,draw=black,area legend] coordinates
        {(News,918) (Health,156) (Misinfo,22) (Aggregator,284) (Factcheck,6) (Portal,112) (Search,1420) (Social,559) (Webmail,131) (Untracked,1350)};
\addlegendentry{\scriptsize Misinfo}
\addplot[ybar,fill=cb4,draw=black,area legend] coordinates
        {(News,422) (Health,0) (Misinfo,81) (Aggregator,108) (Factcheck,6) (Portal,32) (Search,805) (Social,434) (Webmail,24) (Untracked,580)};
\addlegendentry{\scriptsize Health}
\addplot[ybar,fill=cb3,draw=black,area legend] coordinates
        {(News,2149) (Health,854) (Misinfo,455) (Aggregator,1082) (Factcheck,89) (Portal,849) (Search,2515) (Social,2226) (Webmail,1737) (Untracked,2616)};
\addlegendentry{\scriptsize Other}

\end{axis}
\node[yshift=115pt,xshift=55pt] at (1,1) {\footnotesize Type of domain exposed};
\end{tikzpicture}
\subcaption{For each domain type, we report the number of users who were exposed at least once to a link to a news domain, a misinformation domain, a health domain, and any other domain.}
\label{fig:exposure_findings_1}
    \end{minipage}\hfill
    \begin{minipage}{0.48\textwidth}
        \centering
        \vspace{\parskip} 
\begin{tikzpicture}
\begin{axis}[
    ybar stacked,
    ymin=0,
	bar width=10pt,
	width = \textwidth,
	height = 200pt,
	ymax = 134,
	ytick={0,25,50,75,100},
    ylabel={\footnotesize Percentage of all link exposures},
    xlabel={\footnotesize Type of domain where exposure occurred},
    symbolic x coords={
        News, Health, Misinfo, Aggregator, Factcheck, Portal, Search, Social, Webmail, Untracked
        },
    xtick=data,
    x tick label style={rotate=45, font=\scriptsize},
    xtick pos=left,
    ytick pos=left,
    legend columns=5,
    legend style={
            at={(0.5,0.9)},
            anchor= center,
            inner sep=1pt,
    },
    ]
\addlegendimage{empty legend}
\addlegendentry[yshift=14pt]{}
\addplot[ybar,fill=cb1,draw=black,area legend] coordinates
        {(News,19.62) (Health,2.19) (Misinfo,12.11) (Aggregator,63.12) (Factcheck,11.09) (Portal,25.32) (Search,5.73) (Social,19.84) (Webmail,2.22) (Untracked,4.41)};
\addlegendentry{\scriptsize News}
\addplot[ybar,fill=cb5,draw=black,area legend] coordinates
        {(News,0.16) (Health,0) (Misinfo,1.49) (Aggregator,0.11) (Factcheck,0.34) (Portal,0.01) (Search,0.07) (Social,0.19) (Webmail,0.02) (Untracked,0.08)};
\addlegendentry{\scriptsize Misinfo}
\addplot[ybar,fill=cb4,draw=black,area legend] coordinates
        {(News,0.39) (Health,3.29) (Misinfo,0.11) (Aggregator,0.12) (Factcheck,1.03) (Portal,0.14) (Search,0.38) (Social,0.38) (Webmail,0.03) (Untracked,0.18)};
\addlegendentry{\scriptsize Health}
\addplot[ybar,fill=cb3,draw=black,area legend] coordinates
        {(News,79.83) (Health,94.52) (Misinfo,86.28) (Aggregator,36.65) (Factcheck,87.54) (Portal,74.53) (Search,93.82) (Social,79.88) (Webmail,97.73) (Untracked,95.53)};
\addlegendentry{\scriptsize Other}


\end{axis}
\node[yshift=115pt,xshift=55pt] at (1,1) {\footnotesize Type of domain exposed};
\end{tikzpicture}
\subcaption{For each domain type, we report the percentage of link exposures from that type of domain to news, health, misinformation, and other domains.}
\label{fig:exposure_findings_2}
    \end{minipage}
    \caption{Summaries of link exposure data from a preliminary deployment of the Pathways to Political and COVID-19 News study.}
    \label{fig:exposure_findings}
\end{figure}

\Cref{fig:exposure_findings} presents preliminary data on how often participants were exposed to links to news, health, misinformation, and other types of domains. 
Note that this preliminary data comes from a small convenience sample, and as such has significant unmeasured bias. 
For the full-scale study, we plan to construct a larger, more representative population sample.

\Cref{fig:exposure_findings_1} shows how many users were exposed to news, health, and misinformation domains, segmented by the type of domain where the exposure occurred.
The data indicates that search domains exposed more users to news and misinformation links than did any other type of domain, including social media.
It also shows that many users browsing news websites were exposed to misinformation links at least once.

\Cref{fig:exposure_findings_2} presents an alternate view of the same data: the proportion of exposures to each type of domain, segmented by the type of domain where the exposure occurred. 
Exposures to domains other than news, health, and misinformation dominate in all segments.
Misinformation domains are the only segment where misinformation exposures represent a noticeable proportion of link exposures (1.49\%); in all other categories they are less than 0.5\% and do not register visually on the bar chart.
This contextualizes~\Cref{fig:exposure_findings_1}: although in many segments users were likely to have at least one exposure to a misinformation domain, on average misinformation exposure was rare.

\subsection{Comparison to Conventional Browser Measurements}
\label{subsec:quant-eval}
Following our development of WebScience, we chose two user phenomena for which to compare our instrumentation in WebScience
to commonly used methods: user attention to webpages, and the navigation path between websites.

We evaluated several methods of measuring attention and navigation by implementing each method in the study described
above and reporting the measured value from each method for each page visit.

This section of the analysis is based on a sample of 1,817 users who submitted these measures from
August 16, 2022 until September 14, 2022, covering a total of 4,466,200 unique page visits.

\subsubsection{Attention Measures}
We compared four methods of measuring user attention (see~\cref{sec:websci:nav-attn} for more about these methods):
\begin{enumerate}
    \item Our attention model in \textbf{WebScience}, which tracks the focused tab and window, as well as whether the user is active or idle.
    \item A model using \textbf{dwell time}, the interval between a page loading and unloading.
    \item A model using the \textbf{interval} between subsequent page loads.
    \item A \textbf{simplified attention} model, which uses dwell time but accounts for intervals where the webpage is unfocused or not visible to the user.
\end{enumerate}



\textbf{Data format and pre-processing}.
Our study extension reported, for each page visit,
the amount of attention time measured by each method in milliseconds.

When comparing two methods, we first discarded any page visits for which we did not
receive a value from both methods.

For the measure based on load interval, we followed the convention in literature and capped measured attention at 30
minutes.

We then compared each method to WebScience by calculating the error ($e$) for each method
($m$) on each page visit ($p$). Below, $a_{p, m}$ is the attention time measured by method $m$ on page $p$.

$$ e_{p, m} = \frac{|a_{p, WebScience} - a_{p, m}|}{a_{p, WebScience}} * 100 $$

\textbf{Results}.
\Cref{tab:eval-mm-attn} shows
the proportion of visits with error at least 1\%, 10\%, and 25\%, for each method.
Both dwell time and load interval, the most common methods in conventional browser-based research,
differ from the WebScience method by at least 25\% for more than one third of page visits.

\begin{table}[]
    \centering
	\newcolumntype{C}{>{\centering\arraybackslash}m{0.15\linewidth}}
    \begin{tabular}{l C C C }
                 & \multicolumn{3}{c}{\emph{Proportion of visits where $e_{p, m} >= x$}} \\ 
         \textbf{Method ($m$)} & $x = 1\%$       & $x = 10\%$      & $x = 25\%$ \\ \hline
         Dwell time            & 66.6\%          & 48.3\%          & 38.2\% \\ \hline
         Load interval         & 63.7\%          & 46.2\%          & 36.7\% \\ \hline
         Simple attention      & 60.7\%          & 34.5\%          & 21.1\% \\ \hline
    \end{tabular}
    \caption{For each of the three methods, the proportion of page visits for which the method and WebScience differed, by the
    amount of the difference relative to WebScience's measure.}
    \label{tab:eval-mm-attn}
\end{table}

Dwell time and load interval are both more accurate when the user's browsing is linear: the user opens a page,
browses for a time, then closes the page
and opens another. In the days before browsers supported and encouraged multiple windows and many tabs,
users did tend to browse linearly. However,
the modern web browser is different.

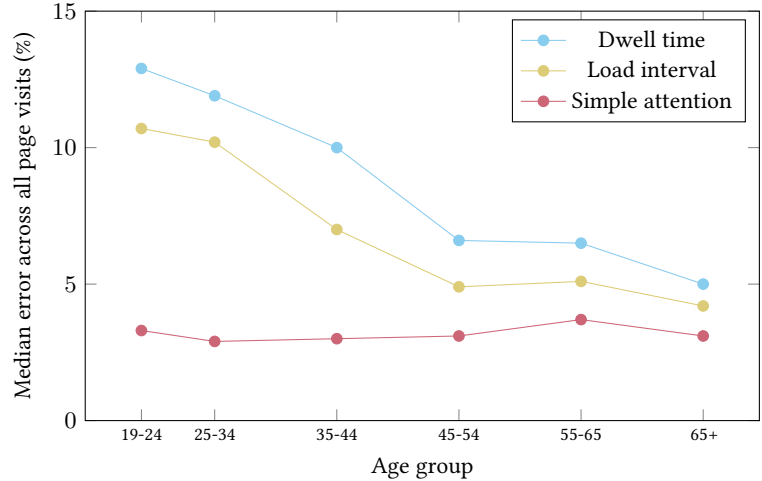
\begin{figure}
\begin{tikzpicture}
\begin{axis}
    [
    xtick={19, 25, 35, 45, 55, 65},
    xticklabels={19-24, 25-34, 35-44, 45-54, 55-65, 65+},
    width=300pt,
    height=200pt,
    ymin=0,
    ymax=15,
    xlabel={Age group},
    ylabel={Median error across all page visits (\%)},
    xticklabel style={align=center, font=\footnotesize, text width=17pt}
    ] 
    \addplot [mark=*,cb3,
    ]
    coordinates {
    (19, 12.9) (25, 11.9) (35, 10.0) (45, 6.6) (55, 6.5) (65, 5.0)
    };
    \addlegendentry{Dwell time}
    
    \addplot [mark=*,cb4] coordinates {
    (19, 10.7) (25, 10.2) (35, 7.0) (45, 4.9) (55, 5.1) (65, 4.2)
    };
    \addlegendentry{Load interval}
    
    \addplot [mark=*,cb5] coordinates {
    (19, 3.3) (25, 2.9) (35, 3.0) (45, 3.1) (55, 3.7) (65, 3.1)
    };
    \addlegendentry{Simple attention}
\end{axis}
\end{tikzpicture}
\caption{The median error for dwell time, load interval, and simple attention across age groups.}
\label{fig:eval-attn-ages}
\end{figure}

We hypothesized that younger users would make heavier use of this feature, resulting in more error.
We separated the page visits into categories
based on the age group that the participant provided in Rally's demographic survey. The amount of error was not consistent across groups,
and decreased as the participant's age increased, as seen in~\Cref{fig:eval-attn-ages}.
Further confirming that this effect is due to younger participants' use of tabs is the flat error for the simple attention model. By accounting for the time that a page is invisible to the participant, it avoids being led astray by background tabs.

Additional comparisons of attention methods are available in~\cref{app:attention-measurement}.

Attention measurements based on dwell time and load interval have been useful historically, but their
differential performance for different users points towards the growing need for more fine-grained measurements.

\subsubsection{Logical Referrers}
We compared four methods of measuring the logical referrer (i.e., the page on which the participant clicked a link
to load a new page) of a page visit:
\begin{enumerate}
    \item Our tracking method in \textbf{WebScience}, which considers multiple sources of information about how a page load occurred.
    \item A method based on page \textbf{load order}, which takes the chronologically-previous page as the logical referrer of the next.
    \item A method based on HTTP \textbf{referrers}, which uses the \textbf{referrer} listed in the header as the logical referrer of the page load.
    \item A method based on browser \textbf{history}, which searches the built-in history to find a page's logical referrer.
\end{enumerate}

For more detail, see~\cref{subsubsec:page-transitions}.

\textbf{Data format and pre-processing}. 
The study extension collected the logical referrer reported by each method on each page visit. To avoid collecting
the raw URLs that participants visited while preserving our ability to compare the methods, the extension
standardized the URLs and then reported pairwise comparisons of equality.

Again, we followed the convention in literature and dropped the logical referrer reported by the load order method
when the previous page had loaded more than 30 minutes prior.

Pages do not always have a logical referrer---for example, when the participant types a new URL, no page
can be reasonably identified as cause of the new page load.
Therefore, to compare each pair of methods, we divided the visits into categories based on whether neither, one, or both methods had measured a
logical referrer. For visits where both methods reported a value, we used our pairwise comparisons of equality to
measure how often the methods found the same value.

\begin{figure}
\begin{tikzpicture}
\begin{axis}[
    ybar stacked,
	bar width=30pt,
	width = 200,
	height = 200pt,
    enlarge x limits=0.40,
    enlarge y limits = 0.05,
    legend style={at={(1.8,0.795)},
      anchor=east,legend columns=1,
      },
    ylabel={Number of page visits},
    symbolic x coords={
        Referrer, Load interval, History
        },
    xtick=data,
    x tick label style={rotate=45,anchor=east},
    ]
\addplot[ybar, fill=cb1, draw=black] plot coordinates { 
  (Referrer, 185928)
  (History, 261700)
  (Load interval, 271770)
  };
\addplot[ybar, fill=cb2, draw=black] plot coordinates { 
  (Referrer, 223454) 
  (History, 147682)
  (Load interval, 137612)
};
\addplot[ybar, fill=cb3, draw=black] plot coordinates { 
  (Referrer, 0)
  (History, 1162663)
  (Load interval, 312779)
};
\addplot[ybar, fill=cb4, draw=black] plot coordinates { 
  (Referrer, 486463)
  (History, 2226163)
  (Load interval, 2825264)
};
\addplot[ybar, fill=cb6, draw=black] plot coordinates { 
  (Referrer, 3570355)
  (History, 667992)
  (Load interval, 918775)
};

\legend{\strut Neither present, Only other method present, Only WebScience present, Full URL match, Partial or no match}
\end{axis}
\end{tikzpicture}
\caption{Each method of measuring the logical referrer of a page compared to WebScience's method.}
\label{fig:eval-parent-barchart}
\end{figure}
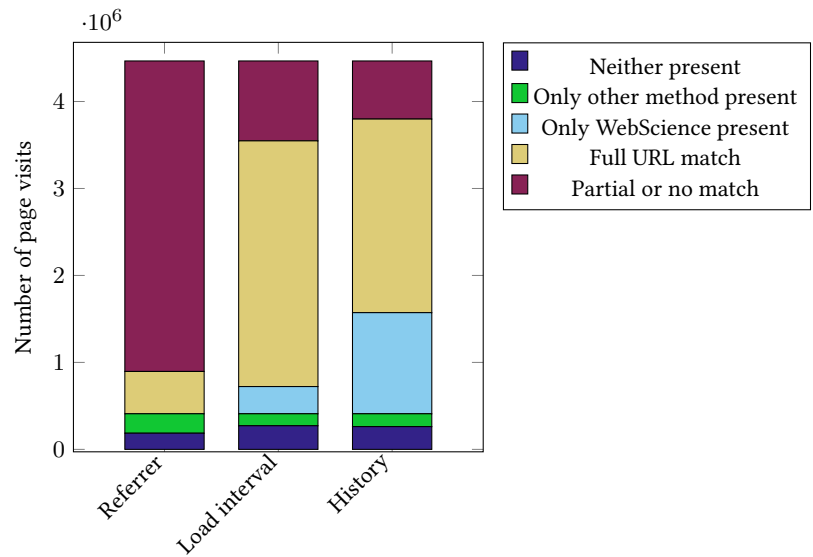

\textbf{Results}.
\Cref{fig:eval-parent-barchart} reports the results of the methods.

The privacy improvements made to the HTTP referrer result in a clear disadvantage when it is used to measure
logical referrers: the lack of path in the URL means it loses information when compared to methods without
this limitation. Both the load interval and history methods report the same URL as the WebScience instrumentation
most of the time: 63.3\% of the time for load interval, and 49.8\% for history.

\subsection{Qualitative User Study}
\label{subsec:qual-eval}
We conducted a user study of members of three other research teams that have used WebScience and Rally to launch browser-based studies.
We designed our study with the following goals:
\begin{enumerate}
    \item Understand why researchers chose Rally and WebScience for their studies over alternative tools and methods
    \item Evaluate the usability of Rally and how researchers view its strengths and limitations as a platform for browser-based studies
    \item Evaluate the usability and feature completeness of WebScience and its strengths and limitations as a library for implementing browser-based studies
\end{enumerate}

\subsubsection{Methods}
We recruited five participants: two researchers, a data scientist, an engineer, and a journalist, all of whom had used Rally and WebScience to launch studies.
We conducted 30-minute semi-structured interviews with each participant, asking questions about the participant's professional background, their study, and their experience with and opinions of various aspects of Rally and WebScience.\footnote{See~\Cref{app:interview-guide} for the interview guide.}
In each interview, one researcher led discussion and a second researcher took notes.
After all interviews were complete, two researchers independently open coded the notes from each interview using Taguette~\cite{rampin2021taguette}, then collaboratively reviewed and condensed codes, grouped codes into categories, and identified key themes within categories.

\subsubsection{Key Findings}
\paragraph{Studies} Participants' Rally studies examined self-preferencing by tech platforms, drivers of news consumption, and the pervasiveness of user tracking online.
All three studies relate to technology policy issues with significant ongoing debate and policy activity.
One participant stated that their intention for their Rally study was "grounding policy debate about this issue in evidence." 

\paragraph{Choosing Rally and WebScience}
When asked why they chose Rally and WebScience over alternative methods, four participants cited the vantage point of the browser as a key benefit.
One participant characterized this as allowing researchers to "see what the user is seeing," and another said Rally and WebScience allowed them to "get people's real-world data about how they use the internet."
They contrasted this vantage point with the limited perspectives of methods like network traffic interception, laboratory studies, and mathematical modeling. 

Two participants identified the granularity of the data they were able to collect with WebScience as a benefit over the aggregate data that would be typically provided by commercial browsing panels.
One participant had developed custom browser instrumentation for prior studies, and cited the rigor and high quality of WebScience's instrumentation and the availability of research infrastructure and support as benefits over purpose-built tooling.

\paragraph{Proposing a Rally study}
We asked participants to rate their satisfaction with the Rally study proposal process on a scale from 1 (extremely dissatisfied) to 5 (extremely satisfied).
Two participants rated the proposal process a 5.
Two other participants rated it a 3, citing the large number of stakeholders involved with studies and the long lead time before receiving approval as challenges.
One of these participant also stated that the proposal process strengthened the project by requiring a detailed research design up front and "making clear what data was necessary to collect."

\paragraph{Building a population sample}
Four participants noted limitations of their population samples.
They observed that the samples were demographically homogeneous, nonrepresentative, and limited by only including desktop browser users.
One participant said they had enrolled about 5,000 users despite hoping for 100,000, and that their recruitment campaigns were hindered by a "cumbersome" signup process for users. 

Two participants who noted limitations were nonetheless extremely satisfied with their sample.
Their study was designed to "surface leads" and find "things we should look into more," so a large, representative sample was not necessary.
One of these participants said "within a couple of weeks we had thousands of people" and that there was "not much to worry about."

\paragraph{Privacy and consent}
All participants said they were extremely satisfied with how Rally handled participant privacy.
Three participants cited [\mozilla's] expertise and reputation for privacy; one said they were "very happy with the level of thought [\mozilla] put in, because we would have to do it ourselves otherwise."
One participant noted that the data analysis environment was tightly controlled and "careful of data flows."
They also said they were encouraged by Rally's emphasis on client-side processing "to prevent stuff being transmitted to the server, as data we interact with is quite sensitive."

Two participants mentioned Rally's thorough process for informing users about study details and receiving consent; one said "it's very important, when looking at this with real users, to have actual informed consent where people have agreed to this unequivocally."

\paragraph{Rally infrastructure}
When asked to rate their satisfaction with Rally's reporting and analysis infrastructure on a scale from 1 (extremely dissatisfied) to 5 (extremely satisfied), two participants rated it a 5, one rated it a 4, and one rated it a 3.
The less satisfied participants noted that the analysis environment was difficult for non-technical team members to use and that using Jupyter Notebook was "frustrating" at times because it was "not meant for be-all, end-all access to data."
One participant suggested adding additional analysis tools to the environment, such as graphical dashboards for generating queries. 
 
\paragraph{Using WebScience}
Two participants had firsthand experience using WebScience to develop their studies, and both were extremely satisfied with the library, rating it a 5.
Both participants appreciated the reusable instrumentation; one said "timing and navigation measurements were straight out of the WebScience box and they worked really well."
The participants found the library easy to use, even though they had never developed a browser extension before.
When encountering development challenges, the participants said they used the library's documentation, examined the code directly to understand the APIs, and received prompt support when reaching out to WebScience developers with questions.
One participant said the study template was a valuable resource because they "got it up and running to watch data flow live and use that as a development environment."
Both participants also extended the library with custom measurements, which were "more complicated and time consuming."

\section{Future Work}
\label{sec:future-work}
We expect for the \Rally{} program to grow beyond the academic
research and journalism groups currently running studies, with more partnered researchers
and many more studies.

\subsection{Scale and Representativeness}
\label{sec:representativity}
To be of use to researchers across disciplines, Rally will need to provide large
samples that are representative of the populations researchers desire to study.

\subsubsection{Representativeness}
It is unlikely that the fully self-selected group of participants who join \Rally{}
will be representative of the target population across all
demographic
factors.
Some studies may not require a demographically-representative sample of participants.
For those that do, there are a few options.
First, studies can use data from the demographic survey collected by \mozilla{},
or an additional study-specific survey, to perform iterative proportional fitting,
resulting in data that is fitted to the target population~\cite{lomax_2016_ipf}.

Second, studies are not limited to recruiting users exclusively
from the \Rally{} userbase. Especially for studies that aim to research
particular groups, researchers can recruit users with other methods, such as
mailing lists or fliers.

Finally, as \Rally{} grows, \mozilla{} may recruit a nationally-representative panel
of users who are paid to participate in studies.
Studies could choose to pull from this pool of participants.

\subsubsection{Scale}
As noted above, \mozilla{} is exploring recruiting a paid panel of users to participate
in \pioneer{}. In addition, \mozilla{} is continuing to experiment with forms of
in-browser promotion for \pioneer{}.




\section{Conclusion}
\label{sec:conclusion}
Browser-based crowdsourced research allows researchers to collect data from real users, acting in their real-world environments, interacting with the platforms that shape society. We have developed \Rally{} and \sdkName{}, a platform and toolkit that simplify and expand access to creating and running crowdsourced browser-based research studies. Together, they drastically reduce the barriers to studying problems at the interaction of technology and society.

\bibliographystyle{ACM-Reference-Format}
\bibliography{paper}


\begin{thebibliography}{80}


\ifx \showCODEN    \undefined \def \showCODEN     #1{\unskip}     \fi
\ifx \showDOI      \undefined \def \showDOI       #1{#1}\fi
\ifx \showISBNx    \undefined \def \showISBNx     #1{\unskip}     \fi
\ifx \showISBNxiii \undefined \def \showISBNxiii  #1{\unskip}     \fi
\ifx \showISSN     \undefined \def \showISSN      #1{\unskip}     \fi
\ifx \showLCCN     \undefined \def \showLCCN      #1{\unskip}     \fi
\ifx \shownote     \undefined \def \shownote      #1{#1}          \fi
\ifx \showarticletitle \undefined \def \showarticletitle #1{#1}   \fi
\ifx \showURL      \undefined \def \showURL       {\relax}        \fi
\providecommand\bibfield[2]{#2}
\providecommand\bibinfo[2]{#2}
\providecommand\natexlab[1]{#1}
\providecommand\showeprint[2][]{arXiv:#2}

\bibitem[\protect\citeauthoryear{Ali, Sapiezynski, Bogen, Korolova, Mislove,
  and Rieke}{Ali et~al\mbox{.}}{2019}]%
        {10.1145/3359301}
\bibfield{author}{\bibinfo{person}{Muhammad Ali}, \bibinfo{person}{Piotr
  Sapiezynski}, \bibinfo{person}{Miranda Bogen}, \bibinfo{person}{Aleksandra
  Korolova}, \bibinfo{person}{Alan Mislove}, {and} \bibinfo{person}{Aaron
  Rieke}.} \bibinfo{year}{2019}\natexlab{}.
\newblock \showarticletitle{Discrimination through Optimization: How Facebook's
  Ad Delivery Can Lead to Biased Outcomes}.
\newblock \bibinfo{journal}{\emph{Proc. ACM Hum.-Comput. Interact.}}
  \bibinfo{volume}{3}, \bibinfo{number}{CSCW}, Article \bibinfo{articleno}{199}
  (\bibinfo{date}{Nov.} \bibinfo{year}{2019}), \bibinfo{numpages}{30}~pages.
\newblock
\urldef\tempurl%
\url{https://doi.org/10.1145/3359301}
\showDOI{\tempurl}


\bibitem[\protect\citeauthoryear{Allcott, Gentzkow, and Yu}{Allcott
  et~al\mbox{.}}{2019}]%
        {allcott2019trends}
\bibfield{author}{\bibinfo{person}{Hunt Allcott}, \bibinfo{person}{Matthew
  Gentzkow}, {and} \bibinfo{person}{Chuan Yu}.}
  \bibinfo{year}{2019}\natexlab{}.
\newblock \showarticletitle{Trends in the diffusion of misinformation on social
  media}.
\newblock \bibinfo{journal}{\emph{Research \& Politics}} \bibinfo{volume}{6},
  \bibinfo{number}{2} (\bibinfo{year}{2019}),
  \bibinfo{pages}{2053168019848554}.
\newblock
\urldef\tempurl%
\url{https://doi.org/10.1177/2053168019848554}
\showDOI{\tempurl}


\bibitem[\protect\citeauthoryear{Altay, Nielsen, and Fletcher}{Altay
  et~al\mbox{.}}{2022}]%
        {altay2022quantifying}
\bibfield{author}{\bibinfo{person}{Sacha Altay}, \bibinfo{person}{Rasmus~Kleis
  Nielsen}, {and} \bibinfo{person}{Richard Fletcher}.}
  \bibinfo{year}{2022}\natexlab{}.
\newblock \showarticletitle{Quantifying the “infodemic”: People turned to
  trustworthy news outlets during the 2020 coronavirus pandemic}.
\newblock \bibinfo{journal}{\emph{Journal of Quantitative Description: Digital
  Media}}  \bibinfo{volume}{2} (\bibinfo{year}{2022}).
\newblock


\bibitem[\protect\citeauthoryear{Araujo, Wonneberger, Neijens, and
  de~Vreese}{Araujo et~al\mbox{.}}{2017}]%
        {araujo_2017_self-reported-internet-use}
\bibfield{author}{\bibinfo{person}{Theo Araujo}, \bibinfo{person}{Anke
  Wonneberger}, \bibinfo{person}{Peter Neijens}, {and} \bibinfo{person}{Claes
  de Vreese}.} \bibinfo{year}{2017}\natexlab{}.
\newblock \showarticletitle{How Much Time Do You Spend Online? Understanding
  and Improving the Accuracy of Self-Reported Measures of Internet Use}.
\newblock \bibinfo{journal}{\emph{Communication Methods and Measures}}
  \bibinfo{volume}{11}, \bibinfo{number}{3} (\bibinfo{year}{2017}),
  \bibinfo{pages}{173--190}.
\newblock
\urldef\tempurl%
\url{https://doi.org/10.1080/19312458.2017.1317337}
\showDOI{\tempurl}


\bibitem[\protect\citeauthoryear{Bailey, Gregersen, and Roesner}{Bailey
  et~al\mbox{.}}{2021}]%
        {bailey2021interactions}
\bibfield{author}{\bibinfo{person}{Aydan Bailey}, \bibinfo{person}{Theo
  Gregersen}, {and} \bibinfo{person}{Franziska Roesner}.}
  \bibinfo{year}{2021}\natexlab{}.
\newblock \showarticletitle{Interactions with Potential Mis/Disinformation URLs
  Among US Users on Facebook, 2017-2019}. In
  \bibinfo{booktitle}{\emph{Proceedings of the ACM SIGCOMM 2021 Workshop on
  Free and Open Communications on the Internet}}. \bibinfo{pages}{16--26}.
\newblock


\bibitem[\protect\citeauthoryear{Bak-Coleman, Alfano, Barfuss, Bergstrom,
  Centeno, Couzin, Donges, Galesic, Gersick, Jacquet, Kao, Moran, Romanczuk,
  Rubenstein, Tombak, Van~Bavel, and Weber}{Bak-Coleman et~al\mbox{.}}{2021}]%
        {bak-coleman_2021_stewardship-collective}
\bibfield{author}{\bibinfo{person}{Joseph~B. Bak-Coleman},
  \bibinfo{person}{Mark Alfano}, \bibinfo{person}{Wolfram Barfuss},
  \bibinfo{person}{Carl~T. Bergstrom}, \bibinfo{person}{Miguel~A. Centeno},
  \bibinfo{person}{Iain~D. Couzin}, \bibinfo{person}{Jonathan~F. Donges},
  \bibinfo{person}{Mirta Galesic}, \bibinfo{person}{Andrew~S. Gersick},
  \bibinfo{person}{Jennifer Jacquet}, \bibinfo{person}{Albert~B. Kao},
  \bibinfo{person}{Rachel~E. Moran}, \bibinfo{person}{Pawel Romanczuk},
  \bibinfo{person}{Daniel~I. Rubenstein}, \bibinfo{person}{Kaia~J. Tombak},
  \bibinfo{person}{Jay~J. Van~Bavel}, {and} \bibinfo{person}{Elke~U. Weber}.}
  \bibinfo{year}{2021}\natexlab{}.
\newblock \showarticletitle{Stewardship of global collective behavior}.
\newblock \bibinfo{journal}{\emph{Proceedings of the National Academy of
  Sciences}} \bibinfo{volume}{118}, \bibinfo{number}{27}
  (\bibinfo{year}{2021}).
\newblock
\showISSN{0027-8424}
\urldef\tempurl%
\url{https://doi.org/10.1073/pnas.2025764118}
\showDOI{\tempurl}


\bibitem[\protect\citeauthoryear{Bakshy, Messing, and Adamic}{Bakshy
  et~al\mbox{.}}{2015}]%
        {bakshy_2015_exposure}
\bibfield{author}{\bibinfo{person}{Eytan Bakshy}, \bibinfo{person}{Solomon
  Messing}, {and} \bibinfo{person}{Lada~A. Adamic}.}
  \bibinfo{year}{2015}\natexlab{}.
\newblock \showarticletitle{Exposure to ideologically diverse news and opinion
  on Facebook}.
\newblock \bibinfo{journal}{\emph{Science}} \bibinfo{volume}{348},
  \bibinfo{number}{6239} (\bibinfo{year}{2015}), \bibinfo{pages}{1130--1132}.
\newblock
\showISSN{0036-8075}
\urldef\tempurl%
\url{https://doi.org/10.1126/science.aaa1160}
\showDOI{\tempurl}


\bibitem[\protect\citeauthoryear{Baumgartner, Zannettou, Keegan, Squire, and
  Blackburn}{Baumgartner et~al\mbox{.}}{2020}]%
        {baumgartner_2020_pushshift}
\bibfield{author}{\bibinfo{person}{Jason Baumgartner}, \bibinfo{person}{Savvas
  Zannettou}, \bibinfo{person}{Brian Keegan}, \bibinfo{person}{Megan Squire},
  {and} \bibinfo{person}{J. Blackburn}.} \bibinfo{year}{2020}\natexlab{}.
\newblock \showarticletitle{The Pushshift Reddit Dataset}. In
  \bibinfo{booktitle}{\emph{ICWSM}}.
\newblock


\bibitem[\protect\citeauthoryear{Bentley, Quehl, Wirfs-Brock, and Bica}{Bentley
  et~al\mbox{.}}{2019}]%
        {bentley_2019_understanding-online-news-behaviors}
\bibfield{author}{\bibinfo{person}{Frank Bentley}, \bibinfo{person}{Katie
  Quehl}, \bibinfo{person}{Jordan Wirfs-Brock}, {and} \bibinfo{person}{Melissa
  Bica}.} \bibinfo{year}{2019}\natexlab{}.
\newblock \showarticletitle{Understanding Online News Behaviors}. In
  \bibinfo{booktitle}{\emph{Proceedings of the 2019 CHI Conference on Human
  Factors in Computing Systems}} (Glasgow, Scotland Uk)
  \emph{(\bibinfo{series}{CHI '19})}. \bibinfo{publisher}{Association for
  Computing Machinery}, \bibinfo{address}{New York, NY, USA},
  \bibinfo{pages}{1–11}.
\newblock
\showISBNx{9781450359702}
\urldef\tempurl%
\url{https://doi.org/10.1145/3290605.3300820}
\showDOI{\tempurl}


\bibitem[\protect\citeauthoryear{Bovet and Makse}{Bovet and Makse}{2019}]%
        {bovet2019influence}
\bibfield{author}{\bibinfo{person}{Alexandre Bovet} {and}
  \bibinfo{person}{Hern{\'a}n~A Makse}.} \bibinfo{year}{2019}\natexlab{}.
\newblock \showarticletitle{Influence of fake news in Twitter during the 2016
  US presidential election}.
\newblock \bibinfo{journal}{\emph{Nature communications}} \bibinfo{volume}{10},
  \bibinfo{number}{1} (\bibinfo{year}{2019}), \bibinfo{pages}{1--14}.
\newblock


\bibitem[\protect\citeauthoryear{Chaney, Stewart, and Engelhardt}{Chaney
  et~al\mbox{.}}{2018}]%
        {chaney_how_2018}
\bibfield{author}{\bibinfo{person}{Allison J.~B. Chaney},
  \bibinfo{person}{Brandon~M. Stewart}, {and} \bibinfo{person}{Barbara~E.
  Engelhardt}.} \bibinfo{year}{2018}\natexlab{}.
\newblock \showarticletitle{How Algorithmic Confounding in Recommendation
  Systems Increases Homogeneity and Decreases Utility}. In
  \bibinfo{booktitle}{\emph{Proceedings of the 12th ACM Conference on
  Recommender Systems}} (Vancouver, British Columbia, Canada)
  \emph{(\bibinfo{series}{RecSys '18})}. \bibinfo{publisher}{Association for
  Computing Machinery}, \bibinfo{address}{New York, NY, USA},
  \bibinfo{pages}{224–232}.
\newblock
\showISBNx{9781450359016}
\urldef\tempurl%
\url{https://doi.org/10.1145/3240323.3240370}
\showDOI{\tempurl}


\bibitem[\protect\citeauthoryear{Cheng, Koc, Harmsen, Shaked, Chandra, Aradhye,
  Anderson, Corrado, Chai, Ispir, Anil, Haque, Hong, Jain, Liu, and Shah}{Cheng
  et~al\mbox{.}}{2016}]%
        {10.1145/2988450.2988454}
\bibfield{author}{\bibinfo{person}{Heng-Tze Cheng}, \bibinfo{person}{Levent
  Koc}, \bibinfo{person}{Jeremiah Harmsen}, \bibinfo{person}{Tal Shaked},
  \bibinfo{person}{Tushar Chandra}, \bibinfo{person}{Hrishi Aradhye},
  \bibinfo{person}{Glen Anderson}, \bibinfo{person}{Greg Corrado},
  \bibinfo{person}{Wei Chai}, \bibinfo{person}{Mustafa Ispir},
  \bibinfo{person}{Rohan Anil}, \bibinfo{person}{Zakaria Haque},
  \bibinfo{person}{Lichan Hong}, \bibinfo{person}{Vihan Jain},
  \bibinfo{person}{Xiaobing Liu}, {and} \bibinfo{person}{Hemal Shah}.}
  \bibinfo{year}{2016}\natexlab{}.
\newblock \showarticletitle{Wide and Deep Learning for Recommender Systems}. In
  \bibinfo{booktitle}{\emph{Proceedings of the 1st Workshop on Deep Learning
  for Recommender Systems}} (Boston, MA, USA) \emph{(\bibinfo{series}{DLRS
  2016})}. \bibinfo{publisher}{Association for Computing Machinery},
  \bibinfo{address}{New York, NY, USA}, \bibinfo{pages}{7–10}.
\newblock
\showISBNx{9781450347952}
\urldef\tempurl%
\url{https://doi.org/10.1145/2988450.2988454}
\showDOI{\tempurl}


\bibitem[\protect\citeauthoryear{Ciampaglia, Nematzadeh, Menczer, and
  Flammini}{Ciampaglia et~al\mbox{.}}{2018}]%
        {ciampaglia2018algorithmic}
\bibfield{author}{\bibinfo{person}{Giovanni~Luca Ciampaglia},
  \bibinfo{person}{Azadeh Nematzadeh}, \bibinfo{person}{Filippo Menczer}, {and}
  \bibinfo{person}{Alessandro Flammini}.} \bibinfo{year}{2018}\natexlab{}.
\newblock \showarticletitle{How algorithmic popularity bias hinders or promotes
  quality}.
\newblock \bibinfo{journal}{\emph{Scientific reports}} \bibinfo{volume}{8},
  \bibinfo{number}{1} (\bibinfo{year}{2018}), \bibinfo{pages}{1--7}.
\newblock


\bibitem[\protect\citeauthoryear{Cinelli, Quattrociocchi, Galeazzi, Valensise,
  Brugnoli, Schmidt, Zola, Zollo, and Scala}{Cinelli et~al\mbox{.}}{2020}]%
        {cinelli_covid-19_2020}
\bibfield{author}{\bibinfo{person}{Matteo Cinelli}, \bibinfo{person}{Walter
  Quattrociocchi}, \bibinfo{person}{Alessandro Galeazzi},
  \bibinfo{person}{Carlo~Michele Valensise}, \bibinfo{person}{Emanuele
  Brugnoli}, \bibinfo{person}{Ana~Lucia Schmidt}, \bibinfo{person}{Paola Zola},
  \bibinfo{person}{Fabiana Zollo}, {and} \bibinfo{person}{Antonio Scala}.}
  \bibinfo{year}{2020}\natexlab{}.
\newblock \showarticletitle{The {COVID}-19 social media infodemic}.
\newblock \bibinfo{journal}{\emph{Scientific Reports}} \bibinfo{volume}{10},
  \bibinfo{number}{1} (\bibinfo{date}{Oct.} \bibinfo{year}{2020}).
\newblock
\showISSN{2045-2322}
\urldef\tempurl%
\url{https://doi.org/10.1038/s41598-020-73510-5}
\showDOI{\tempurl}


\bibitem[\protect\citeauthoryear{Conway}{Conway}{2017}]%
        {conway_determining_2017}
\bibfield{author}{\bibinfo{person}{Maura Conway}.}
  \bibinfo{year}{2017}\natexlab{}.
\newblock \showarticletitle{Determining the {Role} of the {Internet} in
  {Violent} {Extremism} and {Terrorism}: {Six} {Suggestions} for {Progressing}
  {Research}}.
\newblock \bibinfo{journal}{\emph{Studies in Conflict \& Terrorism}}
  \bibinfo{volume}{40}, \bibinfo{number}{1} (\bibinfo{date}{Jan.}
  \bibinfo{year}{2017}), \bibinfo{pages}{77--98}.
\newblock
\showISSN{1057-610X}
\urldef\tempurl%
\url{https://doi.org/10.1080/1057610X.2016.1157408}
\showDOI{\tempurl}
\newblock
\shownote{Publisher: Routledge.}


\bibitem[\protect\citeauthoryear{de~Vreese, Bastos, Esser, Giglietto, Lecleher,
  Pfetsch, Puschmann, Tromble, King, and Persily}{de~Vreese
  et~al\mbox{.}}{2019}]%
        {sso-blog-post_delays}
\bibfield{author}{\bibinfo{person}{Claes de Vreese}, \bibinfo{person}{Marco
  Bastos}, \bibinfo{person}{Frank Esser}, \bibinfo{person}{Fabio Giglietto},
  \bibinfo{person}{Sophie Lecleher}, \bibinfo{person}{Barbara Pfetsch},
  \bibinfo{person}{Cornelius Puschmann}, \bibinfo{person}{Rebekah Tromble},
  \bibinfo{person}{Gary King}, {and} \bibinfo{person}{Nathaniel Persily}.}
  \bibinfo{year}{2019}\natexlab{}.
\newblock \showarticletitle{Public statement from the Co-Chairs and European
  Advisory Committee of Social Science One}.
\newblock  (\bibinfo{year}{2019}).
\newblock
\urldef\tempurl%
\url{https://socialscience.one/blog/public-statement-european-advisory-committee-social-science-one}
\showURL{%
\tempurl}


\bibitem[\protect\citeauthoryear{Edelson and McCoy}{Edelson and McCoy}{2021}]%
        {edelson_opinion_2021}
\bibfield{author}{\bibinfo{person}{Laura Edelson} {and} \bibinfo{person}{Damon
  McCoy}.} \bibinfo{year}{2021}\natexlab{}.
\newblock \showarticletitle{Opinion {\textbar} {We} {Research} {Misinformation}
  on {Facebook}. {It} {Just} {Disabled} {Our} {Accounts}.}
\newblock \bibinfo{journal}{\emph{The New York Times}} (\bibinfo{date}{Aug.}
  \bibinfo{year}{2021}).
\newblock
\showISSN{0362-4331}
\urldef\tempurl%
\url{https://www.nytimes.com/2021/08/10/opinion/facebook-misinformation.html}
\showURL{%
\tempurl}


\bibitem[\protect\citeauthoryear{Edelson, Nguyen, Goldstein, Goga, McCoy, and
  Lauinger}{Edelson et~al\mbox{.}}{2021}]%
        {edelson_2021_understanding-engagement}
\bibfield{author}{\bibinfo{person}{Laura Edelson}, \bibinfo{person}{Minh-Kha
  Nguyen}, \bibinfo{person}{Ian Goldstein}, \bibinfo{person}{Oana Goga},
  \bibinfo{person}{Damon McCoy}, {and} \bibinfo{person}{Tobias Lauinger}.}
  \bibinfo{year}{2021}\natexlab{}.
\newblock \showarticletitle{Understanding Engagement with U.S. (Mis)Information
  News Sources on Facebook}. In \bibinfo{booktitle}{\emph{Proceedings of the
  21st ACM Internet Measurement Conference}} (Virtual Event)
  \emph{(\bibinfo{series}{IMC '21})}. \bibinfo{publisher}{Association for
  Computing Machinery}, \bibinfo{address}{New York, NY, USA},
  \bibinfo{pages}{444–463}.
\newblock
\showISBNx{9781450391290}
\urldef\tempurl%
\url{https://doi.org/10.1145/3487552.3487859}
\showDOI{\tempurl}


\bibitem[\protect\citeauthoryear{Edelson, Sakhuja, Dey, and McCoy}{Edelson
  et~al\mbox{.}}{2019}]%
        {edelson_analysis_2019}
\bibfield{author}{\bibinfo{person}{Laura Edelson}, \bibinfo{person}{Shikhar
  Sakhuja}, \bibinfo{person}{Ratan Dey}, {and} \bibinfo{person}{Damon McCoy}.}
  \bibinfo{year}{2019}\natexlab{}.
\newblock \showarticletitle{An {Analysis} of {United} {States} {Online}
  {Political} {Advertising} {Transparency}}.
\newblock  (\bibinfo{date}{Feb.} \bibinfo{year}{2019}).
\newblock
\urldef\tempurl%
\url{https://arxiv.org/abs/1902.04385v1}
\showURL{%
\tempurl}


\bibitem[\protect\citeauthoryear{Engler}{Engler}{2021}]%
        {engler_2021_platform-data}
\bibfield{author}{\bibinfo{person}{Alex Engler}.}
  \bibinfo{year}{2021}\natexlab{}.
\newblock \showarticletitle{Platform data access is alynchpin of the EU's
  Digital Services Act}.
\newblock \bibinfo{journal}{\emph{Brookings}} (\bibinfo{year}{2021}).
\newblock
\urldef\tempurl%
\url{https://www.brookings.edu/blog/techtank/2021/01/15/platform-data-access-is-a-lynchpin-of-the-eus-digital-services-act/}
\showURL{%
\tempurl}


\bibitem[\protect\citeauthoryear{Faife and Ng}{Faife and Ng}{2021}]%
        {faife_credit_2021}
\bibfield{author}{\bibinfo{person}{Corin Faife} {and} \bibinfo{person}{Alfred
  Ng}.} \bibinfo{year}{2021}\natexlab{}.
\newblock \bibinfo{title}{Credit {Card} {Ads} {Were} {Targeted} by {Age},
  {Violating} {Facebook}’s {Anti}-{Discrimination} {Policy} – {The}
  {Markup}}.
\newblock
\newblock
\urldef\tempurl%
\url{https://themarkup.org/citizen-browser/2021/04/29/credit-card-ads-were-targeted-by-age-violating-facebooks-anti-discrimination-policy}
\showURL{%
\tempurl}


\bibitem[\protect\citeauthoryear{Farke, Balash, Golla, D{\"u}rmuth, and
  Aviv}{Farke et~al\mbox{.}}{2021}]%
        {farke2021privacy}
\bibfield{author}{\bibinfo{person}{Florian~M Farke}, \bibinfo{person}{David~G
  Balash}, \bibinfo{person}{Maximilian Golla}, \bibinfo{person}{Markus
  D{\"u}rmuth}, {and} \bibinfo{person}{Adam~J Aviv}.}
  \bibinfo{year}{2021}\natexlab{}.
\newblock \showarticletitle{Are Privacy Dashboards Good for End Users?
  Evaluating User Perceptions and Reactions to Google’s My Activity}. In
  \bibinfo{booktitle}{\emph{30th $\{$USENIX$\}$ Security Symposium
  ($\{$USENIX$\}$ Security 21)}}. \bibinfo{pages}{483--500}.
\newblock


\bibitem[\protect\citeauthoryear{Fernández, Bellogín, and
  Cantador}{Fernández et~al\mbox{.}}{2021}]%
        {fernandez_analysing_2021}
\bibfield{author}{\bibinfo{person}{Miriam Fernández},
  \bibinfo{person}{Alejandro Bellogín}, {and} \bibinfo{person}{Iván
  Cantador}.} \bibinfo{year}{2021}\natexlab{}.
\newblock \showarticletitle{Analysing the {Effect} of {Recommendation}
  {Algorithms} on the {Amplification} of {Misinformation}}.
\newblock \bibinfo{journal}{\emph{arXiv:2103.14748 [cs]}}
  (\bibinfo{date}{March} \bibinfo{year}{2021}).
\newblock
\urldef\tempurl%
\url{http://arxiv.org/abs/2103.14748}
\showURL{%
\tempurl}
\newblock
\shownote{arXiv: 2103.14748.}


\bibitem[\protect\citeauthoryear{Fiesler, Dye, Feuston, Hiruncharoenvate,
  Hutto, Morrison, Khanipour~Roshan, Pavalanathan, Bruckman, De~Choudhury, and
  Gilbert}{Fiesler et~al\mbox{.}}{2017}]%
        {fiesler_what_2017}
\bibfield{author}{\bibinfo{person}{Casey Fiesler}, \bibinfo{person}{Michaelanne
  Dye}, \bibinfo{person}{Jessica~L. Feuston}, \bibinfo{person}{Chaya
  Hiruncharoenvate}, \bibinfo{person}{C.J. Hutto}, \bibinfo{person}{Shannon
  Morrison}, \bibinfo{person}{Parisa Khanipour~Roshan},
  \bibinfo{person}{Umashanthi Pavalanathan}, \bibinfo{person}{Amy~S. Bruckman},
  \bibinfo{person}{Munmun De~Choudhury}, {and} \bibinfo{person}{Eric Gilbert}.}
  \bibinfo{year}{2017}\natexlab{}.
\newblock \showarticletitle{What (or {Who}) {Is} {Public}? {Privacy} {Settings}
  and {Social} {Media} {Content} {Sharing}}. In
  \bibinfo{booktitle}{\emph{Proceedings of the 2017 {ACM} {Conference} on
  {Computer} {Supported} {Cooperative} {Work} and {Social} {Computing}}}
  \emph{(\bibinfo{series}{{CSCW} '17})}. \bibinfo{publisher}{Association for
  Computing Machinery}, \bibinfo{address}{New York, NY, USA},
  \bibinfo{pages}{567--580}.
\newblock
\showISBNx{978-1-4503-4335-0}
\urldef\tempurl%
\url{https://doi.org/10.1145/2998181.2998223}
\showDOI{\tempurl}


\bibitem[\protect\citeauthoryear{Gaffney and Matias}{Gaffney and
  Matias}{2018}]%
        {gaffney_2018_caveat-emptor}
\bibfield{author}{\bibinfo{person}{Devin Gaffney} {and}
  \bibinfo{person}{J.~Nathan Matias}.} \bibinfo{year}{2018}\natexlab{}.
\newblock \showarticletitle{Caveat emptor, computational social science:
  Large-scale missing data in a widely-published Reddit corpus}. In
  \bibinfo{booktitle}{\emph{PLoS ONE}}.
\newblock
\urldef\tempurl%
\url{https://journals.plos.org/plosone/article?id=10.1371/journal.pone.0200162}
\showURL{%
\tempurl}


\bibitem[\protect\citeauthoryear{Geschke, Lorenz, and Holtz}{Geschke
  et~al\mbox{.}}{2019}]%
        {geschke_2019_triple-filter-bubble-modeling}
\bibfield{author}{\bibinfo{person}{Daniel Geschke}, \bibinfo{person}{Jan
  Lorenz}, {and} \bibinfo{person}{Peter Holtz}.}
  \bibinfo{year}{2019}\natexlab{}.
\newblock \showarticletitle{The triple-filter bubble: Using agent-based
  modelling to test a meta-theoretical framework for the emergence of filter
  bubbles and echo chambers}.
\newblock \bibinfo{journal}{\emph{British Journal of Social Psychology}}
  \bibinfo{volume}{58}, \bibinfo{number}{1} (\bibinfo{year}{2019}),
  \bibinfo{pages}{129--149}.
\newblock
\urldef\tempurl%
\url{https://doi.org/10.1111/bjso.12286}
\showDOI{\tempurl}
\showeprint{https://bpspsychub.onlinelibrary.wiley.com/doi/pdf/10.1111/bjso.12286}


\bibitem[\protect\citeauthoryear{Grinberg, Joseph, Friedland, Swire-Thompson,
  and Lazer}{Grinberg et~al\mbox{.}}{2019}]%
        {grinberg_2019_fake}
\bibfield{author}{\bibinfo{person}{Nir Grinberg}, \bibinfo{person}{Kenneth
  Joseph}, \bibinfo{person}{Lisa Friedland}, \bibinfo{person}{Briony
  Swire-Thompson}, {and} \bibinfo{person}{David Lazer}.}
  \bibinfo{year}{2019}\natexlab{}.
\newblock \showarticletitle{Fake news on Twitter during the 2016 U.S.
  presidential election}.
\newblock \bibinfo{journal}{\emph{Science}} \bibinfo{volume}{363},
  \bibinfo{number}{6425} (\bibinfo{year}{2019}), \bibinfo{pages}{374--378}.
\newblock
\urldef\tempurl%
\url{https://doi.org/10.1126/science.aau2706}
\showDOI{\tempurl}
\showeprint{https://www.science.org/doi/pdf/10.1126/science.aau2706}


\bibitem[\protect\citeauthoryear{Guess, Aslett, Tucker, Bonneau, and
  Nagler}{Guess et~al\mbox{.}}{2021}]%
        {guess_2021_cracking}
\bibfield{author}{\bibinfo{person}{Andy Guess}, \bibinfo{person}{Kevin Aslett},
  \bibinfo{person}{Joshua Tucker}, \bibinfo{person}{Richard Bonneau}, {and}
  \bibinfo{person}{Jonathan Nagler}.} \bibinfo{year}{2021}\natexlab{}.
\newblock \showarticletitle{Cracking open the news feed: Exploring what us
  Facebook users see and share with large-scale platform data}.
\newblock \bibinfo{journal}{\emph{Journal of Quantitative Description: Digital
  Media}}  \bibinfo{volume}{1} (\bibinfo{year}{2021}).
\newblock


\bibitem[\protect\citeauthoryear{Guess}{Guess}{2021}]%
        {guess_2021_everything}
\bibfield{author}{\bibinfo{person}{Andrew~M. Guess}.}
  \bibinfo{year}{2021}\natexlab{}.
\newblock \showarticletitle{(Almost) Everything in Moderation: New Evidence on
  Americans' Online Media Diets}.
\newblock  (\bibinfo{year}{2021}).
\newblock


\bibitem[\protect\citeauthoryear{Guess, Nyhan, and Reifler}{Guess
  et~al\mbox{.}}{2020}]%
        {guess_exposure_2020}
\bibfield{author}{\bibinfo{person}{Andrew~M. Guess}, \bibinfo{person}{Brendan
  Nyhan}, {and} \bibinfo{person}{Jason Reifler}.}
  \bibinfo{year}{2020}\natexlab{}.
\newblock \showarticletitle{Exposure to untrustworthy websites in the 2016 {US}
  election}.
\newblock \bibinfo{journal}{\emph{Nature Human Behaviour}} \bibinfo{volume}{4},
  \bibinfo{number}{5} (\bibinfo{date}{May} \bibinfo{year}{2020}),
  \bibinfo{pages}{472--480}.
\newblock
\showISSN{2397-3374}
\urldef\tempurl%
\url{https://doi.org/10.1038/s41562-020-0833-x}
\showDOI{\tempurl}


\bibitem[\protect\citeauthoryear{Hannak, Sapiezynski, Molavi~Kakhki,
  Krishnamurthy, Lazer, Mislove, and Wilson}{Hannak et~al\mbox{.}}{2013}]%
        {hannak_measuring_2013}
\bibfield{author}{\bibinfo{person}{Aniko Hannak}, \bibinfo{person}{Piotr
  Sapiezynski}, \bibinfo{person}{Arash Molavi~Kakhki},
  \bibinfo{person}{Balachander Krishnamurthy}, \bibinfo{person}{David Lazer},
  \bibinfo{person}{Alan Mislove}, {and} \bibinfo{person}{Christo Wilson}.}
  \bibinfo{year}{2013}\natexlab{}.
\newblock \showarticletitle{Measuring personalization of web search}. In
  \bibinfo{booktitle}{\emph{Proceedings of the 22nd international conference on
  {World} {Wide} {Web} - {WWW} '13}}. \bibinfo{publisher}{ACM Press},
  \bibinfo{address}{Rio de Janeiro, Brazil}, \bibinfo{pages}{527--538}.
\newblock
\showISBNx{978-1-4503-2035-1}
\urldef\tempurl%
\url{https://doi.org/10.1145/2488388.2488435}
\showDOI{\tempurl}


\bibitem[\protect\citeauthoryear{Hosseinmardi, Ghasemian, Clauset, Rothschild,
  Mobius, and Watts}{Hosseinmardi et~al\mbox{.}}{2020}]%
        {hosseinmardi_evaluating_2020}
\bibfield{author}{\bibinfo{person}{Homa Hosseinmardi}, \bibinfo{person}{Amir
  Ghasemian}, \bibinfo{person}{Aaron Clauset}, \bibinfo{person}{David~M.
  Rothschild}, \bibinfo{person}{Markus Mobius}, {and}
  \bibinfo{person}{Duncan~J. Watts}.} \bibinfo{year}{2020}\natexlab{}.
\newblock \showarticletitle{Evaluating the scale, growth, and origins of
  right-wing echo chambers on {YouTube}}.
\newblock \bibinfo{journal}{\emph{arXiv:2011.12843 [cs]}} (\bibinfo{date}{Nov.}
  \bibinfo{year}{2020}).
\newblock
\urldef\tempurl%
\url{http://arxiv.org/abs/2011.12843}
\showURL{%
\tempurl}


\bibitem[\protect\citeauthoryear{Jeffries}{Jeffries}{2020}]%
        {jeffries_how_2020}
\bibfield{author}{\bibinfo{person}{Adrianne Jeffries}.}
  \bibinfo{year}{2020}\natexlab{}.
\newblock \bibinfo{title}{How to {Stop} {Google} {Self}-{Preferencing}?
  {Europe} {May} {Not} {Be} the {Model} – {The} {Markup}}.
\newblock
\newblock
\urldef\tempurl%
\url{https://themarkup.org/google-the-giant/2020/10/15/big-tech-antitrust-google-nondiscrimination-enforcement}
\showURL{%
\tempurl}


\bibitem[\protect\citeauthoryear{Kaiser, Wei, Lucherini, Lee, Matias, and
  Mayer}{Kaiser et~al\mbox{.}}{2021}]%
        {kaiser_2021_adapting-security-warnings}
\bibfield{author}{\bibinfo{person}{Ben Kaiser}, \bibinfo{person}{Jerry Wei},
  \bibinfo{person}{Eli Lucherini}, \bibinfo{person}{Kevin Lee},
  \bibinfo{person}{J.~Nathan Matias}, {and} \bibinfo{person}{Jonathan Mayer}.}
  \bibinfo{year}{2021}\natexlab{}.
\newblock \showarticletitle{Adapting Security Warnings to Counter Online
  Disinformation}. In \bibinfo{booktitle}{\emph{30th {USENIX} Security
  Symposium ({USENIX} Security 21)}}. \bibinfo{publisher}{{USENIX}
  Association}, \bibinfo{pages}{1163--1180}.
\newblock
\showISBNx{978-1-939133-24-3}
\urldef\tempurl%
\url{https://www.usenix.org/conference/usenixsecurity21/presentation/kaiser}
\showURL{%
\tempurl}


\bibitem[\protect\citeauthoryear{Kumar and Shah}{Kumar and Shah}{2018}]%
        {kumar_false_2018}
\bibfield{author}{\bibinfo{person}{Srijan Kumar} {and} \bibinfo{person}{Neil
  Shah}.} \bibinfo{year}{2018}\natexlab{}.
\newblock \showarticletitle{False {Information} on {Web} and {Social} {Media}:
  {A} {Survey}}.
\newblock \bibinfo{journal}{\emph{arXiv:1804.08559 [cs]}}
  (\bibinfo{date}{April} \bibinfo{year}{2018}).
\newblock
\urldef\tempurl%
\url{http://arxiv.org/abs/1804.08559}
\showURL{%
\tempurl}


\bibitem[\protect\citeauthoryear{Lazer, Hargittai, Freelon, Gonzalez-Bailon,
  Munger, Ognyanova, and Radford}{Lazer et~al\mbox{.}}{2021}]%
        {lazer2021meaningful}
\bibfield{author}{\bibinfo{person}{David Lazer}, \bibinfo{person}{Eszter
  Hargittai}, \bibinfo{person}{Deen Freelon}, \bibinfo{person}{Sandra
  Gonzalez-Bailon}, \bibinfo{person}{Kevin Munger}, \bibinfo{person}{Katherine
  Ognyanova}, {and} \bibinfo{person}{Jason Radford}.}
  \bibinfo{year}{2021}\natexlab{}.
\newblock \showarticletitle{Meaningful measures of human society in the
  twenty-first century}.
\newblock \bibinfo{journal}{\emph{Nature}} \bibinfo{volume}{595},
  \bibinfo{number}{7866} (\bibinfo{year}{2021}), \bibinfo{pages}{189--196}.
\newblock


\bibitem[\protect\citeauthoryear{Lee and Kerschbaumer}{Lee and
  Kerschbaumer}{2021}]%
        {referrer-default}
\bibfield{author}{\bibinfo{person}{Dimi Lee} {and} \bibinfo{person}{Christoph
  Kerschbaumer}.} \bibinfo{year}{2021}\natexlab{}.
\newblock \showarticletitle{Firefox 87 trims HTTP Referrers by default to
  protect user privacy}.
\newblock  (\bibinfo{year}{2021}).
\newblock
\newblock
\shownote{\url{https://blog.mozilla.org/security/2021/03/22/firefox-87-trims-http-referrers-by-default-to-protect-user-privacy/}.}


\bibitem[\protect\citeauthoryear{Lewandowsky, Ecker, and Cook}{Lewandowsky
  et~al\mbox{.}}{2017}]%
        {lewandowsky_beyond_2017}
\bibfield{author}{\bibinfo{person}{Stephan Lewandowsky},
  \bibinfo{person}{Ullrich~K.H. Ecker}, {and} \bibinfo{person}{John Cook}.}
  \bibinfo{year}{2017}\natexlab{}.
\newblock \showarticletitle{Beyond Misinformation: Understanding and Coping
  with the “Post-Truth” Era}.
\newblock \bibinfo{journal}{\emph{Journal of Applied Research in Memory and
  Cognition}} \bibinfo{volume}{6}, \bibinfo{number}{4} (\bibinfo{year}{2017}),
  \bibinfo{pages}{353--369}.
\newblock
\showISSN{2211-3681}
\urldef\tempurl%
\url{https://doi.org/10.1016/j.jarmac.2017.07.008}
\showDOI{\tempurl}


\bibitem[\protect\citeauthoryear{Lomax and Norman}{Lomax and Norman}{2016}]%
        {lomax_2016_ipf}
\bibfield{author}{\bibinfo{person}{Nik Lomax} {and} \bibinfo{person}{Paul
  Norman}.} \bibinfo{year}{2016}\natexlab{}.
\newblock \showarticletitle{Estimating Population Attribute Values in a Table:
  “Get Me Started in” Iterative Proportional Fitting}.
\newblock \bibinfo{journal}{\emph{The Professional Geographer}}
  \bibinfo{volume}{68}, \bibinfo{number}{3} (\bibinfo{year}{2016}),
  \bibinfo{pages}{451--461}.
\newblock
\urldef\tempurl%
\url{https://doi.org/10.1080/00330124.2015.1099449}
\showDOI{\tempurl}


\bibitem[\protect\citeauthoryear{Lucherini, Sun, Winecoff, and
  Narayanan}{Lucherini et~al\mbox{.}}{2021}]%
        {lucherini_2021_t-recs}
\bibfield{author}{\bibinfo{person}{Eli Lucherini}, \bibinfo{person}{Matthew
  Sun}, \bibinfo{person}{Amy~A. Winecoff}, {and} \bibinfo{person}{Arvind
  Narayanan}.} \bibinfo{year}{2021}\natexlab{}.
\newblock \showarticletitle{{T-RECS:} {A} Simulation Tool to Study the Societal
  Impact of Recommender Systems}.
\newblock \bibinfo{journal}{\emph{CoRR}}  \bibinfo{volume}{abs/2107.08959}
  (\bibinfo{year}{2021}).
\newblock
\showeprint[arXiv]{2107.08959}
\urldef\tempurl%
\url{https://arxiv.org/abs/2107.08959}
\showURL{%
\tempurl}


\bibitem[\protect\citeauthoryear{Mathur, Mayer, and Kshirsagar}{Mathur
  et~al\mbox{.}}{2021}]%
        {mathur21}
\bibfield{author}{\bibinfo{person}{Arunesh Mathur}, \bibinfo{person}{Jonathan
  Mayer}, {and} \bibinfo{person}{Mihir Kshirsagar}.}
  \bibinfo{year}{2021}\natexlab{}.
\newblock \bibinfo{booktitle}{\emph{What Makes a Dark Pattern... Dark? Design
  Attributes, Normative Considerations, and Measurement Methods}}.
\newblock \bibinfo{publisher}{Association for Computing Machinery},
  \bibinfo{address}{New York, NY, USA}.
\newblock
\showISBNx{9781450380966}
\urldef\tempurl%
\url{https://doi.org/10.1145/3411764.3445610}
\showURL{%
\tempurl}


\bibitem[\protect\citeauthoryear{MDN Web Docs}{MDN Web Docs}{2021a}]%
        {webextensions}
MDN Web Docs \bibinfo{year}{2021}\natexlab{a}.
\newblock \bibinfo{title}{Browser Extensions}.
\newblock
\newblock
\newblock
\shownote{\url{https://developer.mozilla.org/en-US/docs/Mozilla/Add-ons/WebExtensions}.}


\bibitem[\protect\citeauthoryear{MDN Web Docs}{MDN Web Docs}{2021b}]%
        {referrer-policy}
MDN Web Docs \bibinfo{year}{2021}\natexlab{b}.
\newblock \bibinfo{title}{Referrer-Policy}.
\newblock
\newblock
\newblock
\shownote{\url{https://developer.mozilla.org/en-US/docs/Web/HTTP/Headers/Referrer-Policy}.}


\bibitem[\protect\citeauthoryear{Medvedev, Wu, and Gordon}{Medvedev
  et~al\mbox{.}}{2019}]%
        {medvedev_powered_2019}
\bibfield{author}{\bibinfo{person}{Ivan Medvedev}, \bibinfo{person}{Haotian
  Wu}, {and} \bibinfo{person}{Taylor Gordon}.} \bibinfo{year}{2019}\natexlab{}.
\newblock \bibinfo{title}{Powered by {AI}: {Instagram}’s {Explore}
  recommender system}.
\newblock
\newblock
\urldef\tempurl%
\url{https://ai.facebook.com/blog/powered-by-ai-instagrams-explore-recommender-system/}
\showURL{%
\tempurl}


\bibitem[\protect\citeauthoryear{Messing, DeGregorio, Hillenbrand, King,
  Mahanti, Mukerjee, Nayak, Persily, State, and Wilkins}{Messing
  et~al\mbox{.}}{2020}]%
        {DVN/TDOAPG_2020}
\bibfield{author}{\bibinfo{person}{Solomon Messing}, \bibinfo{person}{Christina
  DeGregorio}, \bibinfo{person}{Bennett Hillenbrand}, \bibinfo{person}{Gary
  King}, \bibinfo{person}{Saurav Mahanti}, \bibinfo{person}{Zagreb Mukerjee},
  \bibinfo{person}{Chaya Nayak}, \bibinfo{person}{Nate Persily},
  \bibinfo{person}{Bogdan State}, {and} \bibinfo{person}{Arjun Wilkins}.}
  \bibinfo{year}{2020}\natexlab{}.
\newblock \bibinfo{title}{{Facebook Privacy-Protected Full URLs Data Set}}.
\newblock
\newblock
\urldef\tempurl%
\url{https://doi.org/10.7910/DVN/TDOAPG}
\showDOI{\tempurl}


\bibitem[\protect\citeauthoryear{Mondal, Silva, and Benevenuto}{Mondal
  et~al\mbox{.}}{2017}]%
        {mondal_measurement_2017}
\bibfield{author}{\bibinfo{person}{Mainack Mondal},
  \bibinfo{person}{Leandro~Araújo Silva}, {and} \bibinfo{person}{Fabrício
  Benevenuto}.} \bibinfo{year}{2017}\natexlab{}.
\newblock \showarticletitle{A {Measurement} {Study} of {Hate} {Speech} in
  {Social} {Media}}. In \bibinfo{booktitle}{\emph{Proceedings of the 28th {ACM}
  {Conference} on {Hypertext} and {Social} {Media}}}
  \emph{(\bibinfo{series}{{HT} '17})}. \bibinfo{publisher}{Association for
  Computing Machinery}, \bibinfo{address}{New York, NY, USA},
  \bibinfo{pages}{85--94}.
\newblock
\showISBNx{978-1-4503-4708-2}
\urldef\tempurl%
\url{https://doi.org/10.1145/3078714.3078723}
\showDOI{\tempurl}


\bibitem[\protect\citeauthoryear{Morstatter, Pfeffer, Liu, and
  Carley}{Morstatter et~al\mbox{.}}{2013}]%
        {morstatter2013sample}
\bibfield{author}{\bibinfo{person}{Fred Morstatter},
  \bibinfo{person}{J{\"u}rgen Pfeffer}, \bibinfo{person}{Huan Liu}, {and}
  \bibinfo{person}{Kathleen~M Carley}.} \bibinfo{year}{2013}\natexlab{}.
\newblock \showarticletitle{Is the Sample Good Enough? Comparing Data from
  Twitter's Streaming API with Twitter's Firehose}. In
  \bibinfo{booktitle}{\emph{International AAAI Conference on Weblogs and Social
  Media (ICWSM)}}.
\newblock
\urldef\tempurl%
\url{https://ojs.aaai.org/index.php/ICWSM/article/download/14401/14250/17919}
\showURL{%
\tempurl}


\bibitem[\protect\citeauthoryear{Mozilla Rally}{Mozilla Rally}{2020}]%
        {mozilla-pioneer}
Mozilla Rally \bibinfo{year}{2020}\natexlab{}.
\newblock \bibinfo{title}{{A}bout {F}irefox {P}ioneer}.
\newblock
\newblock
\newblock
\shownote{\url{https://rally.mozilla.org/how-rally-works/faqs/}.}


\bibitem[\protect\citeauthoryear{{Narayanan} and {Shmatikov}}{{Narayanan} and
  {Shmatikov}}{2008}]%
        {narayanan_2008_robust-deanonymization}
\bibfield{author}{\bibinfo{person}{A. {Narayanan}} {and} \bibinfo{person}{V.
  {Shmatikov}}.} \bibinfo{year}{2008}\natexlab{}.
\newblock \showarticletitle{Robust De-anonymization of Large Sparse Datasets}.
  In \bibinfo{booktitle}{\emph{2008 IEEE Symposium on Security and Privacy (sp
  2008)}}. \bibinfo{pages}{111--125}.
\newblock
\urldef\tempurl%
\url{https://doi.org/10.1109/SP.2008.33}
\showURL{%
\tempurl}


\bibitem[\protect\citeauthoryear{Nichols and Maner}{Nichols and Maner}{2008}]%
        {nichols_2010_good-subject-effect}
\bibfield{author}{\bibinfo{person}{Austin~Lee Nichols} {and}
  \bibinfo{person}{Jon~K. Maner}.} \bibinfo{year}{2008}\natexlab{}.
\newblock \showarticletitle{The Good-Subject Effect: Investigating Participant
  Demand Characteristics}.
\newblock \bibinfo{journal}{\emph{The Journal of General Psychology}}
  \bibinfo{volume}{135}, \bibinfo{number}{2} (\bibinfo{year}{2008}),
  \bibinfo{pages}{151--166}.
\newblock
\urldef\tempurl%
\url{https://doi.org/10.3200/GENP.135.2.151-166}
\showDOI{\tempurl}
\newblock
\shownote{PMID: 18507315.}


\bibitem[\protect\citeauthoryear{{Online Political Transparency Project at New
  York University}}{{Online Political Transparency Project at New York
  University}}{2020}]%
        {adobserver}
\bibfield{author}{\bibinfo{person}{{Online Political Transparency Project at
  New York University}}.} \bibinfo{year}{2020}\natexlab{}.
\newblock \bibinfo{title}{{AdObserver}}.
\newblock
\newblock


\bibitem[\protect\citeauthoryear{Owen}{Owen}{2020}]%
        {owen_how_2020}
\bibfield{author}{\bibinfo{person}{Laura~Hazard Owen}.}
  \bibinfo{year}{2020}\natexlab{}.
\newblock \bibinfo{title}{How much political news do people see on {Facebook}?
  {I} went inside 173 people’s feeds to find out}.
\newblock
\newblock
\urldef\tempurl%
\url{https://www.niemanlab.org/2020/11/how-much-political-news-do-people-see-on-facebook-i-went-inside-173-peoples-feeds-to-find-out/}
\showURL{%
\tempurl}


\bibitem[\protect\citeauthoryear{Perra and Rocha}{Perra and Rocha}{2019}]%
        {perra_modelling_2019}
\bibfield{author}{\bibinfo{person}{Nicola Perra} {and} \bibinfo{person}{Luis
  E.~C. Rocha}.} \bibinfo{year}{2019}\natexlab{}.
\newblock \showarticletitle{Modelling opinion dynamics in the age of
  algorithmic personalisation}.
\newblock \bibinfo{journal}{\emph{Scientific Reports}} \bibinfo{volume}{9},
  \bibinfo{number}{1} (\bibinfo{date}{May} \bibinfo{year}{2019}),
  \bibinfo{pages}{7261}.
\newblock
\showISSN{2045-2322}
\urldef\tempurl%
\url{https://doi.org/10.1038/s41598-019-43830-2}
\showDOI{\tempurl}


\bibitem[\protect\citeauthoryear{Perriam, Birkbak, and Freeman}{Perriam
  et~al\mbox{.}}{2020}]%
        {perriam_2020_post-api}
\bibfield{author}{\bibinfo{person}{Jessamy Perriam}, \bibinfo{person}{Andreas
  Birkbak}, {and} \bibinfo{person}{Andy Freeman}.}
  \bibinfo{year}{2020}\natexlab{}.
\newblock \showarticletitle{Digital methods in a post-API environment}.
\newblock \bibinfo{journal}{\emph{International Journal of Social Research
  Methodology}} \bibinfo{volume}{23}, \bibinfo{number}{3}
  (\bibinfo{year}{2020}), \bibinfo{pages}{277--290}.
\newblock
\urldef\tempurl%
\url{https://doi.org/10.1080/13645579.2019.1682840}
\showDOI{\tempurl}


\bibitem[\protect\citeauthoryear{Persily}{Persily}{2021}]%
        {persily_2021_facebook-hides}
\bibfield{author}{\bibinfo{person}{Nathan Persily}.}
  \bibinfo{year}{2021}\natexlab{}.
\newblock \showarticletitle{Facebook hides data showing it harms users. Outside
  scholars need access.}
\newblock \bibinfo{journal}{\emph{The Washington Post}} (\bibinfo{year}{2021}).
\newblock
\urldef\tempurl%
\url{https://www.washingtonpost.com/outlook/2021/10/05/facebook-research-data-haugen-congress-regulation/}
\showURL{%
\tempurl}


\bibitem[\protect\citeauthoryear{Peterson}{Peterson}{2001}]%
        {peterson_2001_use-of-college-students}
\bibfield{author}{\bibinfo{person}{Robert~A. Peterson}.}
  \bibinfo{year}{2001}\natexlab{}.
\newblock \showarticletitle{{On the Use of College Students in Social Science
  Research: Insights from a Second-Order Meta-analysis}}.
\newblock \bibinfo{journal}{\emph{Journal of Consumer Research}}
  \bibinfo{volume}{28}, \bibinfo{number}{3} (\bibinfo{date}{12}
  \bibinfo{year}{2001}), \bibinfo{pages}{450--461}.
\newblock
\showISSN{0093-5301}
\urldef\tempurl%
\url{https://doi.org/10.1086/323732}
\showDOI{\tempurl}


\bibitem[\protect\citeauthoryear{Phadke and Mitra}{Phadke and Mitra}{2020}]%
        {phadke_many_2020}
\bibfield{author}{\bibinfo{person}{Shruti Phadke} {and}
  \bibinfo{person}{Tanushree Mitra}.} \bibinfo{year}{2020}\natexlab{}.
\newblock \showarticletitle{Many {Faced} {Hate}: {A} {Cross} {Platform} {Study}
  of {Content} {Framing} and {Information} {Sharing} by {Online} {Hate}
  {Groups}}.
\newblock In \bibinfo{booktitle}{\emph{Proceedings of the 2020 {CHI}
  {Conference} on {Human} {Factors} in {Computing} {Systems}}}.
  \bibinfo{publisher}{Association for Computing Machinery},
  \bibinfo{address}{New York, NY, USA}, \bibinfo{pages}{1--13}.
\newblock
\showISBNx{978-1-4503-6708-0}
\urldef\tempurl%
\url{https://doi.org/10.1145/3313831.3376456}
\showURL{%
\tempurl}


\bibitem[\protect\citeauthoryear{Rampin and Rampin}{Rampin and Rampin}{2021}]%
        {rampin2021taguette}
\bibfield{author}{\bibinfo{person}{R{\'e}mi Rampin} {and}
  \bibinfo{person}{Vicky Rampin}.} \bibinfo{year}{2021}\natexlab{}.
\newblock \showarticletitle{Taguette: open-source qualitative data analysis}.
\newblock \bibinfo{journal}{\emph{Journal of Open Source Software}}
  \bibinfo{volume}{6}, \bibinfo{number}{68} (\bibinfo{year}{2021}),
  \bibinfo{pages}{3522}.
\newblock


\bibitem[\protect\citeauthoryear{Ribeiro, Saha, Babaei, Henrique, Messias,
  Benevenuto, Goga, Gummadi, and Redmiles}{Ribeiro et~al\mbox{.}}{2019a}]%
        {ribeiro_microtargeting_2019}
\bibfield{author}{\bibinfo{person}{Filipe~N. Ribeiro}, \bibinfo{person}{Koustuv
  Saha}, \bibinfo{person}{Mahmoudreza Babaei}, \bibinfo{person}{Lucas
  Henrique}, \bibinfo{person}{Johnnatan Messias}, \bibinfo{person}{Fabricio
  Benevenuto}, \bibinfo{person}{Oana Goga}, \bibinfo{person}{Krishna~P.
  Gummadi}, {and} \bibinfo{person}{Elissa~M. Redmiles}.}
  \bibinfo{year}{2019}\natexlab{a}.
\newblock \showarticletitle{On {Microtargeting} {Socially} {Divisive} {Ads}:
  {A} {Case} {Study} of {Russia}-{Linked} {Ad} {Campaigns} on {Facebook}}. In
  \bibinfo{booktitle}{\emph{Proceedings of the {Conference} on {Fairness},
  {Accountability}, and {Transparency}}} \emph{(\bibinfo{series}{{FAT}* '19})}.
  \bibinfo{publisher}{Association for Computing Machinery},
  \bibinfo{address}{New York, NY, USA}, \bibinfo{pages}{140--149}.
\newblock
\showISBNx{978-1-4503-6125-5}
\urldef\tempurl%
\url{https://doi.org/10.1145/3287560.3287580}
\showDOI{\tempurl}


\bibitem[\protect\citeauthoryear{Ribeiro, Blackburn, Bradlyn, De~Cristofaro,
  Stringhini, Long, Greenberg, and Zannettou}{Ribeiro et~al\mbox{.}}{2020a}]%
        {ribeiro_evolution_2020}
\bibfield{author}{\bibinfo{person}{Manoel~Horta Ribeiro},
  \bibinfo{person}{Jeremy Blackburn}, \bibinfo{person}{Barry Bradlyn},
  \bibinfo{person}{Emiliano De~Cristofaro}, \bibinfo{person}{Gianluca
  Stringhini}, \bibinfo{person}{Summer Long}, \bibinfo{person}{Stephanie
  Greenberg}, {and} \bibinfo{person}{Savvas Zannettou}.}
  \bibinfo{year}{2020}\natexlab{a}.
\newblock \showarticletitle{The {Evolution} of the {Manosphere} {Across} the
  {Web}}.
\newblock  (\bibinfo{date}{Jan.} \bibinfo{year}{2020}).
\newblock
\urldef\tempurl%
\url{https://arxiv.org/abs/2001.07600v5}
\showURL{%
\tempurl}


\bibitem[\protect\citeauthoryear{Ribeiro, Ottoni, West, Almeida, and
  Meira}{Ribeiro et~al\mbox{.}}{2020b}]%
        {ribeiro_auditing_2020}
\bibfield{author}{\bibinfo{person}{Manoel~Horta Ribeiro},
  \bibinfo{person}{Raphael Ottoni}, \bibinfo{person}{Robert West},
  \bibinfo{person}{Virgílio A.~F. Almeida}, {and} \bibinfo{person}{Wagner
  Meira}.} \bibinfo{year}{2020}\natexlab{b}.
\newblock \showarticletitle{Auditing radicalization pathways on {YouTube}}. In
  \bibinfo{booktitle}{\emph{Proceedings of the 2020 {Conference} on {Fairness},
  {Accountability}, and {Transparency}}} \emph{(\bibinfo{series}{{FAT}* '20})}.
  \bibinfo{publisher}{Association for Computing Machinery},
  \bibinfo{address}{New York, NY, USA}, \bibinfo{pages}{131--141}.
\newblock
\showISBNx{978-1-4503-6936-7}
\urldef\tempurl%
\url{https://doi.org/10.1145/3351095.3372879}
\showDOI{\tempurl}


\bibitem[\protect\citeauthoryear{Ribeiro, Zannettou, Cristofaro, Stringhini,
  and Blackburn}{Ribeiro et~al\mbox{.}}{2019b}]%
        {ribeiro_2019_comments-on-algorithmic-extremism}
\bibfield{author}{\bibinfo{person}{Manoel~Horta Ribeiro},
  \bibinfo{person}{Savvas Zannettou}, \bibinfo{person}{Emiliano~De Cristofaro},
  \bibinfo{person}{Gianluca Stringhini}, {and} \bibinfo{person}{Jeremy
  Blackburn}.} \bibinfo{year}{2019}\natexlab{b}.
\newblock \showarticletitle{Comments on “Algorithmic Extremism: Examining
  YouTube’s Rabbit Hole of Radicalization”}.
\newblock  (\bibinfo{year}{2019}).
\newblock
\newblock
\shownote{\url{https://idrama.science/posts/2019/12/youtube-radicalization-study/}.}


\bibitem[\protect\citeauthoryear{Roose}{Roose}{2021}]%
        {roose_2021_inside-fbs-data-wars}
\bibfield{author}{\bibinfo{person}{Kevin Roose}.}
  \bibinfo{year}{2021}\natexlab{}.
\newblock \showarticletitle{Inside Facebook’s Data Wars}.
\newblock  (\bibinfo{year}{2021}).
\newblock
\newblock
\shownote{\url{https://www.nytimes.com/2021/07/14/technology/facebook-data.html}.}


\bibitem[\protect\citeauthoryear{Samory and Mitra}{Samory and Mitra}{2018}]%
        {samory2018conspiracies}
\bibfield{author}{\bibinfo{person}{Mattia Samory} {and}
  \bibinfo{person}{Tanushree Mitra}.} \bibinfo{year}{2018}\natexlab{}.
\newblock \showarticletitle{Conspiracies online: User discussions in a
  conspiracy community following dramatic events}. In
  \bibinfo{booktitle}{\emph{Proceedings of the International AAAI Conference on
  Web and Social Media}}, Vol.~\bibinfo{volume}{12}.
\newblock


\bibitem[\protect\citeauthoryear{Scharkow}{Scharkow}{2016}]%
        {scharkow_2016_accuracy-self-reported-internet}
\bibfield{author}{\bibinfo{person}{Michael Scharkow}.}
  \bibinfo{year}{2016}\natexlab{}.
\newblock \showarticletitle{The Accuracy of Self-Reported Internet Use—A
  Validation Study Using Client Log Data}.
\newblock \bibinfo{journal}{\emph{Communication Methods and Measures}}
  \bibinfo{volume}{10}, \bibinfo{number}{1} (\bibinfo{year}{2016}),
  \bibinfo{pages}{13--27}.
\newblock
\urldef\tempurl%
\url{https://doi.org/10.1080/19312458.2015.1118446}
\showDOI{\tempurl}


\bibitem[\protect\citeauthoryear{Shapiro, Sugarman, Bermejo, and
  Zuckerman}{Shapiro et~al\mbox{.}}{2021}]%
        {shapiro21}
\bibfield{author}{\bibinfo{person}{Elizabeth~Hansen Shapiro},
  \bibinfo{person}{Michael Sugarman}, \bibinfo{person}{Fernando Bermejo}, {and}
  \bibinfo{person}{Ethan Zuckerman}.} \bibinfo{year}{2021}\natexlab{}.
\newblock \bibinfo{booktitle}{}.
\newblock \bibinfo{type}{{T}echnical {R}eport}.
\newblock
\urldef\tempurl%
\url{https://www.netgainpartnership.org/resources/2021/2/25/new-approaches-to-platform-data-research}
\showURL{%
\tempurl}


\bibitem[\protect\citeauthoryear{Singh, Bode, Budak, Kawintiranon, Padden, and
  Vraga}{Singh et~al\mbox{.}}{2020}]%
        {singh_2020_understanding}
\bibfield{author}{\bibinfo{person}{Lisa Singh}, \bibinfo{person}{Leticia Bode},
  \bibinfo{person}{Ceren Budak}, \bibinfo{person}{Kornraphop Kawintiranon},
  \bibinfo{person}{Colton Padden}, {and} \bibinfo{person}{Emily Vraga}.}
  \bibinfo{year}{2020}\natexlab{}.
\newblock \showarticletitle{Understanding high-and low-quality URL Sharing on
  COVID-19 Twitter streams}.
\newblock \bibinfo{journal}{\emph{Journal of computational social science}}
  \bibinfo{volume}{3}, \bibinfo{number}{2} (\bibinfo{year}{2020}),
  \bibinfo{pages}{343--366}.
\newblock


\bibitem[\protect\citeauthoryear{Solsman}{Solsman}{2018}]%
        {solsman_ever_2018}
\bibfield{author}{\bibinfo{person}{Joan~E. Solsman}.}
  \bibinfo{year}{2018}\natexlab{}.
\newblock \bibinfo{title}{Ever get caught in an unexpected hourlong {YouTube}
  binge? {Thank} {YouTube} {AI} for that}.
\newblock
\newblock
\urldef\tempurl%
\url{https://www.cnet.com/news/youtube-ces-2018-neal-mohan/}
\showURL{%
\tempurl}


\bibitem[\protect\citeauthoryear{Speicher, Ali, Venkatadri, Ribeiro,
  Arvanitakis, Benevenuto, Gummadi, Loiseau, and Mislove}{Speicher
  et~al\mbox{.}}{2018}]%
        {pmlr-v81-speicher18a}
\bibfield{author}{\bibinfo{person}{Till Speicher}, \bibinfo{person}{Muhammad
  Ali}, \bibinfo{person}{Giridhari Venkatadri}, \bibinfo{person}{Filipe~Nunes
  Ribeiro}, \bibinfo{person}{George Arvanitakis}, \bibinfo{person}{Fabrício
  Benevenuto}, \bibinfo{person}{Krishna~P. Gummadi}, \bibinfo{person}{Patrick
  Loiseau}, {and} \bibinfo{person}{Alan Mislove}.}
  \bibinfo{year}{2018}\natexlab{}.
\newblock \showarticletitle{Potential for Discrimination in Online Targeted
  Advertising}. In \bibinfo{booktitle}{\emph{Proceedings of the 1st Conference
  on Fairness, Accountability and Transparency}}
  \emph{(\bibinfo{series}{Proceedings of Machine Learning Research},
  Vol.~\bibinfo{volume}{81})}, \bibfield{editor}{\bibinfo{person}{Sorelle~A.
  Friedler} {and} \bibinfo{person}{Christo Wilson}} (Eds.).
  \bibinfo{publisher}{PMLR}, \bibinfo{pages}{5--19}.
\newblock
\urldef\tempurl%
\url{https://proceedings.mlr.press/v81/speicher18a.html}
\showURL{%
\tempurl}


\bibitem[\protect\citeauthoryear{Thomas, Akhawe, Bailey, Boneh, Bursztein,
  Consolvo, Dell, Durumeric, Kelley, Kumar, McCoy, Meiklejohn, Ristenpart, and
  Stringhini}{Thomas et~al\mbox{.}}{2021}]%
        {49786}
\bibfield{editor}{\bibinfo{person}{Kurt Thomas}, \bibinfo{person}{Devdatta
  Akhawe}, \bibinfo{person}{Michael Bailey}, \bibinfo{person}{Dan Boneh},
  \bibinfo{person}{Elie Bursztein}, \bibinfo{person}{Sunny Consolvo},
  \bibinfo{person}{Nicola Dell}, \bibinfo{person}{Zakir Durumeric},
  \bibinfo{person}{Patrick~Gage Kelley}, \bibinfo{person}{Deepak Kumar},
  \bibinfo{person}{Damon McCoy}, \bibinfo{person}{Sarah Meiklejohn},
  \bibinfo{person}{Thomas Ristenpart}, {and} \bibinfo{person}{Gianluca
  Stringhini}} (Eds.). \bibinfo{year}{2021}\natexlab{}.
\newblock \bibinfo{booktitle}{\emph{SoK: Hate, Harassment, and the Changing
  Landscape of Online Abuse}}.
\newblock


\bibitem[\protect\citeauthoryear{Timberg}{Timberg}{2021}]%
        {timberg_2021_facebook-made-big-mistake}
\bibfield{author}{\bibinfo{person}{Craig Timberg}.}
  \bibinfo{year}{2021}\natexlab{}.
\newblock \showarticletitle{Facebook made big mistake in data it provided to
  researchers, undermining academic work}.
\newblock  (\bibinfo{year}{2021}).
\newblock
\urldef\tempurl%
\url{https://www.washingtonpost.com/technology/2021/09/10/facebook-error-data-social-scientists/}
\showURL{%
\tempurl}


\bibitem[\protect\citeauthoryear{Tobin}{Tobin}{2019}]%
        {tobin_hud_2019}
\bibfield{author}{\bibinfo{person}{Ariana Tobin}.}
  \bibinfo{year}{2019}\natexlab{}.
\newblock \bibinfo{title}{{HUD} {Sues} {Facebook} {Over} {Housing}
  {Discrimination} and {Says} the {Company}’s {Algorithms} {Have} {Made} the
  {Problem} {Worse}}.
\newblock
\newblock
\urldef\tempurl%
\url{https://www.propublica.org/article/hud-sues-facebook-housing-discrimination-advertising-algorithms?token=MZ4huG2khovdFzdzgBUWYctqeKrXQgA5}
\showURL{%
\tempurl}


\bibitem[\protect\citeauthoryear{Vogels}{Vogels}{2021}]%
        {vogels_digital_2021}
\bibfield{author}{\bibinfo{person}{Emily~A. Vogels}.}
  \bibinfo{year}{2021}\natexlab{}.
\newblock \bibinfo{title}{Some digital divides persist between rural, urban and
  suburban {America}}.
\newblock
\newblock
\urldef\tempurl%
\url{https://www.pewresearch.org/fact-tank/2021/08/19/some-digital-divides-persist-between-rural-urban-and-suburban-america/}
\showURL{%
\tempurl}


\bibitem[\protect\citeauthoryear{Vosoughi, Roy, and Aral}{Vosoughi
  et~al\mbox{.}}{2018}]%
        {vosoughi_spread_2018}
\bibfield{author}{\bibinfo{person}{Soroush Vosoughi}, \bibinfo{person}{Deb
  Roy}, {and} \bibinfo{person}{Sinan Aral}.} \bibinfo{year}{2018}\natexlab{}.
\newblock \showarticletitle{The spread of true and false news online}.
\newblock \bibinfo{journal}{\emph{Science}} \bibinfo{volume}{359},
  \bibinfo{number}{6380} (\bibinfo{date}{March} \bibinfo{year}{2018}),
  \bibinfo{pages}{1146--1151}.
\newblock
\showISSN{0036-8075, 1095-9203}
\urldef\tempurl%
\url{https://doi.org/10.1126/science.aap9559}
\showDOI{\tempurl}


\bibitem[\protect\citeauthoryear{Waddell}{Waddell}{[n.d.]}]%
        {waddell_facebook_2020}
\bibfield{author}{\bibinfo{person}{Kaveh Waddell}.}
  \bibinfo{year}{[n.d.]}\natexlab{}.
\newblock \bibinfo{title}{Facebook {Approved} {Ads} {With} {Coronavirus}
  {Misinformation}}.
\newblock
\newblock
\urldef\tempurl%
\url{https://www.consumerreports.org/social-media/facebook-approved-ads-with-coronavirus-misinformation/}
\showURL{%
\tempurl}


\bibitem[\protect\citeauthoryear{Wang, Niiya, Mark, Reich, and Warschauer}{Wang
  et~al\mbox{.}}{2015}]%
        {wang_coming_2015}
\bibfield{author}{\bibinfo{person}{Yiran Wang}, \bibinfo{person}{Melissa
  Niiya}, \bibinfo{person}{Gloria Mark}, \bibinfo{person}{Stephanie~M. Reich},
  {and} \bibinfo{person}{Mark Warschauer}.} \bibinfo{year}{2015}\natexlab{}.
\newblock \showarticletitle{Coming of {Age} ({Digitally}): {An} {Ecological}
  {View} of {Social} {Media} {Use} among {College} {Students}}. In
  \bibinfo{booktitle}{\emph{Proceedings of the 18th {ACM} {Conference} on
  {Computer} {Supported} {Cooperative} {Work} \& {Social} {Computing}}}
  \emph{(\bibinfo{series}{{CSCW} '15})}. \bibinfo{publisher}{Association for
  Computing Machinery}, \bibinfo{address}{New York, NY, USA},
  \bibinfo{pages}{571--582}.
\newblock
\showISBNx{978-1-4503-2922-4}
\urldef\tempurl%
\url{https://doi.org/10.1145/2675133.2675271}
\showDOI{\tempurl}


\bibitem[\protect\citeauthoryear{Wei, Stamos, Veys, Reitinger, Goodman, Herman,
  Filipczuk, Weinshel, Mazurek, and Ur}{Wei et~al\mbox{.}}{2020}]%
        {255286}
\bibfield{author}{\bibinfo{person}{Miranda Wei}, \bibinfo{person}{Madison
  Stamos}, \bibinfo{person}{Sophie Veys}, \bibinfo{person}{Nathan Reitinger},
  \bibinfo{person}{Justin Goodman}, \bibinfo{person}{Margot Herman},
  \bibinfo{person}{Dorota Filipczuk}, \bibinfo{person}{Ben Weinshel},
  \bibinfo{person}{Michelle~L. Mazurek}, {and} \bibinfo{person}{Blase Ur}.}
  \bibinfo{year}{2020}\natexlab{}.
\newblock \showarticletitle{What Twitter Knows: Characterizing Ad Targeting
  Practices, User Perceptions, and Ad Explanations Through
  Users{\textquoteright} Own Twitter Data}. In \bibinfo{booktitle}{\emph{29th
  {USENIX} Security Symposium ({USENIX} Security 20)}}.
  \bibinfo{publisher}{{USENIX} Association}, \bibinfo{pages}{145--162}.
\newblock
\showISBNx{978-1-939133-17-5}
\urldef\tempurl%
\url{https://www.usenix.org/conference/usenixsecurity20/presentation/wei}
\showURL{%
\tempurl}


\bibitem[\protect\citeauthoryear{Wilson and Starbird}{Wilson and
  Starbird}{2020}]%
        {wilson_cross-platform_2020}
\bibfield{author}{\bibinfo{person}{Tom Wilson} {and} \bibinfo{person}{Kate
  Starbird}.} \bibinfo{year}{2020}\natexlab{}.
\newblock \showarticletitle{Cross-{Platform} {Disinformation} {Campaigns}:
  {Lessons} {Learned} and {Next} {Steps}}.
\newblock \bibinfo{journal}{\emph{Harvard Kennedy School Misinformation
  Review}} \bibinfo{volume}{1}, \bibinfo{number}{1} (\bibinfo{date}{Jan.}
  \bibinfo{year}{2020}).
\newblock
\urldef\tempurl%
\url{https://doi.org/10.37016/mr-2020-002}
\showDOI{\tempurl}


\bibitem[\protect\citeauthoryear{Yang, Pierri, Hui, Axelrod, Torres-Lugo,
  Bryden, and Menczer}{Yang et~al\mbox{.}}{2021}]%
        {yang2021infodemic}
\bibfield{author}{\bibinfo{person}{Kai-Cheng Yang}, \bibinfo{person}{Francesco
  Pierri}, \bibinfo{person}{Pik-Mai Hui}, \bibinfo{person}{David Axelrod},
  \bibinfo{person}{Christopher Torres-Lugo}, \bibinfo{person}{John Bryden},
  {and} \bibinfo{person}{Filippo Menczer}.} \bibinfo{year}{2021}\natexlab{}.
\newblock \showarticletitle{The COVID-19 Infodemic: Twitter versus Facebook}.
\newblock \bibinfo{journal}{\emph{Big Data \& Society}} \bibinfo{volume}{8},
  \bibinfo{number}{1} (\bibinfo{year}{2021}),
  \bibinfo{pages}{20539517211013861}.
\newblock
\urldef\tempurl%
\url{https://doi.org/10.1177/20539517211013861}
\showDOI{\tempurl}
\showeprint{https://doi.org/10.1177/20539517211013861}


\bibitem[\protect\citeauthoryear{York}{York}{2019}]%
        {york_how_2019}
\bibfield{author}{\bibinfo{person}{Jillian~C. York}.}
  \bibinfo{year}{2019}\natexlab{}.
\newblock \bibinfo{title}{How {American} {Corporations} {Are} {Policing}
  {Online} {Speech} {Worldwide}}.
\newblock
\newblock
\urldef\tempurl%
\url{https://gizmodo.com/how-american-corporations-are-policing-online-speech-wo-1836010637}
\showURL{%
\tempurl}


\end{thebibliography}

\appendix
\section{Appendix}
\subsection{Literature Survey Search Terms}
\label{app:survey-search-terms}

Below, we provide the search terms used for the literature survey, and the number of papers resulting from each.
Each cell of the table is the number of papers resulting from concatenating the search term of the row with the
term heading the column. For example, in~\cref{tab:app-search-terms-1}, the top left cell indicates that the search query
\begin{quote}
    "computer use data" "behavior" 
\end{quote}
resulted in 252 papers. We combined the results from each query and then de-duplicated them, resulting in the count
reported in~\cref{table:survey-overall-stats}.

\begin{landscape}
\begin{table}[]
	\newcolumntype{C}{>{\centering\arraybackslash}m{0.0413\linewidth}}
    \centering
    \footnotesize
    \begin{tabular}{l C C C C C C C C C C C C C C C }
    & use data" "behavior"	& interaction data" "behavior"	& usage data" "behavior"	&  use" "participant" OR "participants" "in-situ data"	& usage" "participant" OR "participants" "in-situ data"	& interaction" "participant" OR "participants" "in-situ data"	&  * use data" "behavior"	&  * use * data" "behavior"	&  use * data" "behavior"	& * interaction data" "behavior"	&  * interaction * data" "behavior"	&  interaction * data" "behavior"	&  * usage data" "behavior"	&  * usage * data" "behavior"	&  usage * data" "behavior"
    \\ \hline
    "computer &	252	& 508 &	198 &	41 &	18 &	378 &	56 &	125 &	388 &	36 &	32 &	554 &	67 &	10 &	112
    \\ \hline
    "browser &	1 &	50 &	17 &	2 &	7 &	0 &	3 &	9 &	4 &	5 &	2 &	0 &	8 &	0 &	4
    \\ \hline
    "phone &	185	 &30 &	682 &	66 &	55 &	12 &	11 &	27 &	129 &	2 &	1 &	4 &	30 &	10 &	112
    \\ \hline
    "smartphone &	120 &	30 &	419 &	41 &	55 &	4 &	19 &	24 &	64 &	6 &	1 &	4 &	40 &	9 &	140
    \\ \hline
    "desktop &	0 &	1 &	1 &	4 &	0 &	2 &	4 &	4 &	6 &	2 &	0 &	0 &	2 &	1 &	3
    \\ \hline
    "laptop &	5 &	0 &	10 &	2 &	1 &	0 &	1 &	6 &	5 &	0 &	0 &	0 &	3 &	0 &	3
    \\ \hline
    \end{tabular}
    \caption{First set of literature survey search terms.}
    \label{tab:app-search-terms-1}
\end{table}

\begin{table}[]
	\newcolumntype{C}{>{\centering\arraybackslash}m{0.05\linewidth}}
    \centering
    \footnotesize
    \begin{tabular}{l C C C C C C C C C C}
    \\ \hline
    & extension" "behavior"	& add-on" "behavior"	&addon" "behavior"	& * extension" "behavior"	& * add-on" "behavior"	& * addon" "behavior"	&plugin" "behavior"	&plug-in" "behavior"	& * plugin" "behavior"	& * plug-in" "behavior"
    \\ \hline
    "chrome &	2580 &	61 &	4 &	665 &	16 &	5 &	248 &	110 &	111 &	43
    \\ \hline
    "firefox &	1480 &	701 &	898 &	381 &	140 &	34 &	479 &	362& 	127 &	78
    \\ \hline
    "safari &	24 &	5 &	0 &	7 &	2 &	0 &	6 &	7 &	4 &	7
    \\ \hline
    \end{tabular}
    \caption{Second set of literature survey search terms.}
    \label{tab:app-search-terms-2}
\end{table}

\end{landscape}

\begin{table}[]
	\newcolumntype{C}{>{\centering\arraybackslash}m{0.1\linewidth}}
    \centering
    \footnotesize
    \begin{tabular}{l C C C}
    & data" "behavior"	&		" "participant" OR "participants" "in-situ data"	&		* data" "behavior"
    \\ \hline
    "browsing &	3650	&		301	&		2750
    \\ \hline
    "clickstream &	7560	&		13	&		400
    \\ \hline
    \end{tabular}
    \caption{Third set of literature survey search terms.}
    \label{tab:app-search-terms-3}
\end{table}

\begin{table}[]
	\newcolumntype{C}{>{\centering\arraybackslash}m{0.1\linewidth}}
    \centering
    \footnotesize
    \begin{tabular}{l C C C C C C}
    & chrome" "behavior"	& firefox" "behavior"	& safari" "behavior"	& * chrome" "behavior"	& * firefox" "behavior"	& * safari" "behavior"
    \\ \hline
    "extension &	59 &	40 &	4 &	366 &	449 &	19
    \\ \hline
    "add-on &	1 &	10 &	1 &	43 &	211 &	3
    \\ \hline
    "addon &	1 &	3 &	0 &	9 &	29 &	1
    \\ \hline
    "plugin &	13 &	25 &	1 &	80 &	191 &	8
    \\ \hline
    "plug-in &	4 &	5 &	1 &	44 &	180 &	10
    \\ \hline
    \end{tabular}
    \caption{Fourth set of literature survey search terms.}
    \label{tab:app-search-terms-4}
\end{table}

\subsection{Literature Survey Codebook}
\label{app:survey-codebook}

\begin{table}[]
	\newcolumntype{C}{>{\centering\arraybackslash}m{0.7\linewidth}}
    \centering
    \footnotesize
    \begin{tabular}{l C}
         \textbf{Name of attribute} & \textbf{Description}
         \\ \hline
         Title (*) & The title of the paper.
         \\ \hline
         Authors (*) & The authors of the paper.
         \\ \hline
         Year (*) & The publication year of the paper.
         \\ \hline
         Citations (*) & The number of citations of the paper, at the time of our searches in February and March 2022.
         \\ \hline
         Duplicate & Indicates second studies from a single paper.
         \\ \hline
         Venue & The publication venue of the paper.
         \\ \hline
         Discipline & The academic discipline of the publication venue, or the apparent discipline of the authors, for papers at multidisciplinary venues.
         \\ \hline
         Participants & The total number of participants in the study. We used the largest number of participants from whom the researchers reported possessing data, even if some were excluded from the full analysis.
         \\ \hline
         Duration & The largest number of days that the researchers reported collecting data from participants.
         \\ \hline
         MOOC & Whether the paper used data from a MOOC (massive open online course), and used the data to predict or assist student success.
         \\ \hline
         E-Commerce & Whether the paper used data from an online store, and used the data to predict or promote sales or browsing behavior.
         \\ \hline
         Devices & The devices from which the researchers had data. Options were: Smartphone, cell phone, desktop/laptop, tablet, mixed server data (data retrieved from a computer serving different clients), and not stated.
         \\ \hline
         Researcher-provided & Whether the researchers provided participants with the devices from which they collected data. We set this field when the researchers provided any participants with a device, even if other participants brought their own devices.
         \\ \hline
         Explicit consent & Whether the paper explicitly says that researchers received informed consent from participants. Options were: not stated (the paper did not explicitly mention consent), consent obtained (the paper says consent was obtained), consent explained (consent was obtained, and the paper describes the process of obtaining consent), and consent not obtained.
         \\ \hline
         IRB approval & Whether the paper mentions receiving approval from an institutional review board or undergoing similar ethical review. Options were: not stated, IRB approved, IRB exempt (the researchers mention that the project was considered exempt from their institution's review process).
         \\ \hline
         Data source & The way the researchers obtained the data. Options were: custom (the authors of the paper wrote the instrumentation), research data provider (the researchers used a company or tool that collected data for research), and partnership (the researchers partnered with (or were employed by) the organization under study, and received data from that company, organization, or platform).
         \\ \hline
         Collection mechanism & Regardless of the above (who collected the data), the tool or method that measured the data. Options were: extension (a browser extension, add-on, plug-in, or toolbar), user upload (participants manually uploaded a file containing data), manual (researchers accessed devices and manually retrieved the data), VPN (researchers installed a VPN on the participant's device and collected data from the VPN server), mobile app (an app installed on a smartphone), desktop program (a program installed on a computer), and business data (data collected through the normal business operation of an organization under study).
         \\ \hline
         Recruitment mechanism & The method of enrolling participants. Options were: traditional (researchers posted fliers, went to community hubs, recruited students, sent emails to listservs, and so on), advertising (researchers purchased ads on websites), crowdworkers (researchers recruited participants from MTurk or similar platforms), business operation (participants were customers or users of the platform under study), and panel recruitment (participants were members of a research panel).
         \\ \hline
         Compensation & Whether participants received anything of value from participating, including money, devices, and access to services. Options were: yes, no, and not stated.
         \\ \hline
         Interventions & Whether the researchers intentionally intervened with the interaction between the user and device, not including the act of measuring itself.
         \\ \hline
         Qualitative & Whether the researchers had survey or interview data from the same participants whose interaction data they used.
         \\ \hline
         Web attention & Whether the researchers measured the amount of attention participants paid to different webpages, and the method used if so. Options were: NA (the researchers did not measure attention), count visits (the researchers used the number of visits to infer attention), dwell time, time to next page load, and precise instrumentation (the researchers used a browser-based method that tracked participant actions).
         \\ \hline
         Web navigation & Whether the researchers measured the path between webpages, and the method used if so. Options were: NA, referrer (researchers used the HTTP referrer, last loaded page (researchers used the chronologically-previous page as the logical referrer), and precise instrumentation (researchers used a browser-based method that tracked user actions).
         \\ \hline
         Web exposure & Whether the researchers measured content within webpages viewed by the participant.
         \\ \hline
         Social media sharing & Whether the researchers measured the participant's sharing of content on social media platforms.
         \\ \hline
         
    \end{tabular}
    \caption{The coding attributes for the literature survey. * indicates fields that were filled automatically from the Google Scholar
    searches.  }
\end{table}

\newpage
\subsection{Qualitative User Study Interview Guide}
\label{app:interview-guide}
\paragraph{Research background}
\begin{enumerate}
    \item What best describes your professional training? Is it as a researcher, an engineer, a data scientist, a journalist, or something else?
    \begin{enumerate}
        \item How many years of experience do you have in that role?
        \item What types of projects have you worked on in that time?
        \begin{enumerate}
            \item If research -- what fields and what types of methods?
        \end{enumerate}
    \end{enumerate}
    \item How much experience do you have:
    \begin{enumerate}
        \item Conducting user studies, including measurements, surveys, and interviews?
        \item Writing software?
        \begin{enumerate}
            \item Writing JavaScript code?
            \item Writing browser extensions?
            \begin{enumerate}
                \item Using the WebExtensions API?
            \end{enumerate}
        \end{enumerate}
        \item Conducting browser-based studies?
        \begin{enumerate}
            \item Conducting other device-based studies?
        \end{enumerate}
        \item Using commercial panel providers like YouGov, Nielsen, or Comscore?
        \item Collecting experimental data?
    \end{enumerate}
\end{enumerate}

\paragraph{Study information}
\begin{enumerate}
    \setcounter{enumi}{2}
    \item What is your Rally / WebScience study about?
    \item What are the key contributions of your study?
    \begin{enumerate}
        \item What benefits does the study have for society?
        \item What makes your study novel?
    \end{enumerate}
    \item Why are browser-based methods needed to study your research problem?
    \begin{enumerate}
        \item What alternatives have you considered? What are the tradeoffs?
    \end{enumerate}
    \item What was your role in the study? How many collaborators did you work with, and what were their roles?
\end{enumerate}

\paragraph{Experience with Rally}
\begin{enumerate}
    \setcounter{enumi}{6}
    \item On a scale of 1 to 5, where 1 is extremely dissatisfied and 5 is extremely satisfied: how do you feel about the process of proposing your Rally study and getting approval?
    \begin{enumerate}
        \item How was the process different from how you would typically design a study? 
        \item Did you see any benefits to the Rally study approval process? Were there any ways in which it helped you with your study design?
        \item Were there any pain points? Were any requirements or process steps confusing or difficult to follow?
        \item Do you have any suggestions for how the Rally study proposal process could be improved?
    \end{enumerate}
    \item On a scale of 1 to 5, where 1 is extremely dissatisfied and 5 is extremely satisfied: how do you feel about the processes for recruiting and onboarding participants through Rally?
    \begin{enumerate}
        \item What recruitment methods did you use?
        \begin{enumerate}
            \item Enrolling existing Rally users in the study?
            \item Paid external advertising, like Facebook ads?
            \item Advertising to your own social networks? 
            \item Other methods?
        \end{enumerate}
        \item On a scale of 1 to 5, how satisfied are you with the size and makeup of your population sample?
        \begin{enumerate}
            \item Why?
            \item How does it differ from samples you have previously recruited by other means?
            \item Were there any costs to you?
        \end{enumerate}
        \item On a scale of 1 to 5, how satisfied are you with the Rally processes for informing participants about studies and requesting consent?
        \begin{enumerate}
            \item Were there significant differences from other processes you have used to inform and request consent?
            \item Did you feel that participants were sufficiently informed about the study and able to give informed, uncoerced consent?
            \item Did you, your team, or the Rally team identify any other ethical issues around recruiting and onboarding participants for your study?
        \end{enumerate}
        \item Do you have any suggestions for how Rally processes for recruitment, consent, and other onboarding steps could be improved?
    \end{enumerate}
    \item On a scale of 1 to 5, how satisfied are you with how Rally handles participant privacy?
    \begin{enumerate}
        \item What steps related to privacy did you take while designing and implementing your study?
        \begin{enumerate}
            \item Data minimization?
            \item Configuring your analysis environment?
        \end{enumerate} 
        \item Were any steps related to privacy difficult or onerous to implement?
        \item Compared to other studies you have run, was it easier or harder to handle privacy on your Rally study? 
        \item Do you think your Rally participants have better or worse protections than participants in other studies you have run?
    \end{enumerate}
    \item On a scale of 1 to 5, how satisfied are you with the technical infrastructure Rally provides for deploying your study, reporting data, and analyzing data?
    \begin{enumerate}
        \item Did you encounter any limitations or challenges working with Rally infrastructure?
        \item Do you have any suggestions for how to improve Rally infrastructure?
    \end{enumerate}
    \item Do you see it as generally positive, negative, or neither that Rally is run by Mozilla?
    \begin{enumerate}
        \item Are there any specific benefits you see or concerns you have with Rally being run by Mozilla?
    \end{enumerate}
\end{enumerate}

\paragraph{Experience with WebScience}
\begin{enumerate}
    \setcounter{enumi}{11}
    \item How much did you use WebScience in your study?
    \item Why did you choose to use WebScience for your study?
    \begin{enumerate}
        \item What do you see as the biggest benefits WebScience provides compared to alternative options for developing browser-based studies?
        \item The biggest drawbacks?
    \end{enumerate}
    \item How long did it take to develop your study extension?
    \begin{enumerate}
        \item Did you use the study template, or create your study extension from scratch?
    \end{enumerate}
    \item Was WebScience generally easy to use?
    \begin{enumerate}
        \item Were the WebScience APIs easy to understand and use?
        \item Was WebScience sufficiently configurable for your purposes?
        \item Were there any pain points you encountered using WebScience?
        \item Was WebScience documentation helpful for understanding how to use the library?
        \item Did you use any other support resources, such as the research support Slack channel in the Rally workspace, or reaching out to other colleagues?
        \item Are there any support resources you would like to see made available?
    \end{enumerate} 
    \item Do you see it as generally positive, negative, or neither that WebScience is an open-source project?
    \begin{enumerate}
        \item Are there any specific benefits you see or concerns you have with WebScience being an open-source project?
    \end{enumerate}
    \item What do you think are the most important features or measurement capabilities that WebScience should add?
\end{enumerate}

\paragraph{Concluding questions}
\begin{enumerate}
    \setcounter{enumi}{11}
    \item What do you think are the biggest benefits to using Rally / WebScience over alternative methods of creating browser-based studies?
    \begin{enumerate}
        \item What are the biggest drawbacks?
    \end{enumerate}
    \item If you run another browser-based study in the future, would you use Rally / WebScience again? 
    \item Would you recommend Rally / WebScience to other researchers in your field? 
    \item What advice would you give a researcher in your field who is considering using Rally / WebScience to launch a browser-based study?
\end{enumerate}

\subsection{Additional WebScience Modules}
\label{app:webscience-modules}

Below, we briefly describe other modules that handle common tasks and measurements.

\subsubsection{User Survey}
Researchers often need to combine behavioral data with self-reported information.
The user survey modules allows researchers to prompt participants to
fill out a survey on an external platform.

\subsubsection{Link Resolution}
Shortening services, such as  Bitly and Tinyurl, mean that two URLs that are not
the same
may ultimately lead to the same page.
The link resolution module
accepts a URL from another module and unrolls
nested shorteners and redirects to obtain the final page
that a URL points to.
    
\subsubsection{Link Exposure}
Researchers studying exposure need to know which links a participant saw while browsing (but
didn't necessarily visit).
The link exposure modules provides lists of links
that were in the viewport at sufficient size and
for sufficient time that the participant likely noticed them.
Researchers configure the module with a list of desired
domains on which to measure, as well as about which exposed links to notify.
    
\subsubsection{Scheduling}
Studies need to run maintenance or reporting tasks at regular intervals. Some
heavyweight tasks (e.g. aggregating raw events into reportable data) may be
CPU-intensive, and therefore should run when the participant is not actively using
the system. 
The scheduling module provides a service to monitor user
activity and run code at approximately
regular intervals, preferring periods of low user activity.
    
\subsubsection{Clock}
    Studies that use timestamps to record the order of user events need monotonically
    increasing timestamps. The system clock can be changed by the user, or by the
    system adjusting for skew. Studies that use the system clock without accounting
    for the possibility of adjustments may collect corrupted data. In addition,
    browser implementations of the W3C recommended clock are not yet correct and stable.
The clock module provides a monotonically increasing clock that syncs with the system clock at browser
    startup, and then ignores shifts in that clock.

\subsection{Additional Attention Measurement Charts}
\label{app:attention-measurement}

\begin{figure}
\begin{tikzpicture}
\begin{axis}
    [
    ybar interval,
    xtick={-100, -90, -80, -70, -60, -50, -40, -30, -20, -10, 0, 10, 20, 30, 40, 50, 60, 70, 80, 90, 100, 110, 120, 130, 140, 150, 160},
    ymax=2800000,
    xticklabels={-100, -90, -80, -70, -60, -50, -40, -30, -20, -10, 0, 10, 20, 30, 40, 50, 60, 70, 80, 90, 100, 110, 120, 130, 140, > 150},
    x tick label style={rotate=90,},
    xlabel = {The lower bound of the error interval.},
    ylabel = {Number of visits in each error interval.},
    width=300pt,
    height=200pt,
    ] 
    \addplot [draw=black, fill=cb3] coordinates {
    (-100, 4) (-90, 17) (-80, 38) (-70, 79) (-60, 160) (-50, 280) (-40, 434) (-30, 532) (-20, 553) (-10, 622) (0, 2201626) (10, 327012) (20, 180215) (30, 118552) (40, 87169) (50, 67065) (60, 53853) (70, 43916) (80, 37877) (90, 32849) (100, 28927) (110, 25999) (120, 23355) (130, 21503) (140, 19741) (150, 984287) (160, 0)
    };
\end{axis}
\end{tikzpicture}
\caption{Histogram of difference between WebScience's measure of attention and a measure based on dwell time.}
\label{fig:eval-attn-ws-dt-histo}
\end{figure}
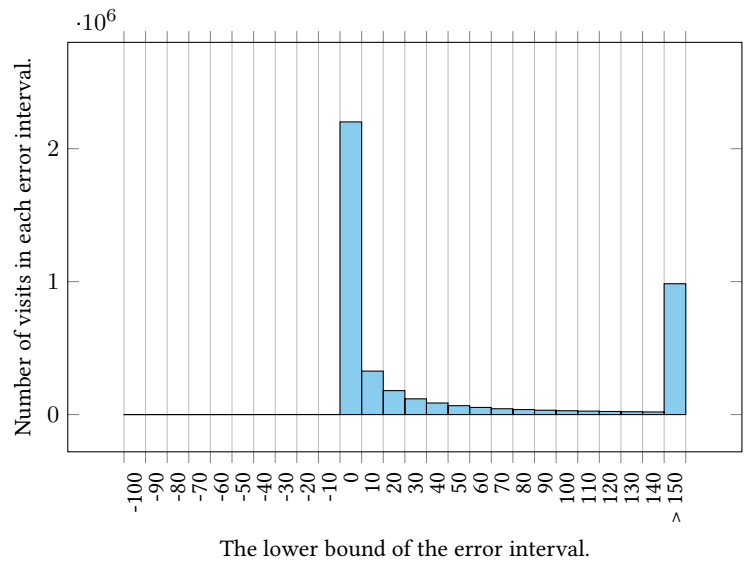

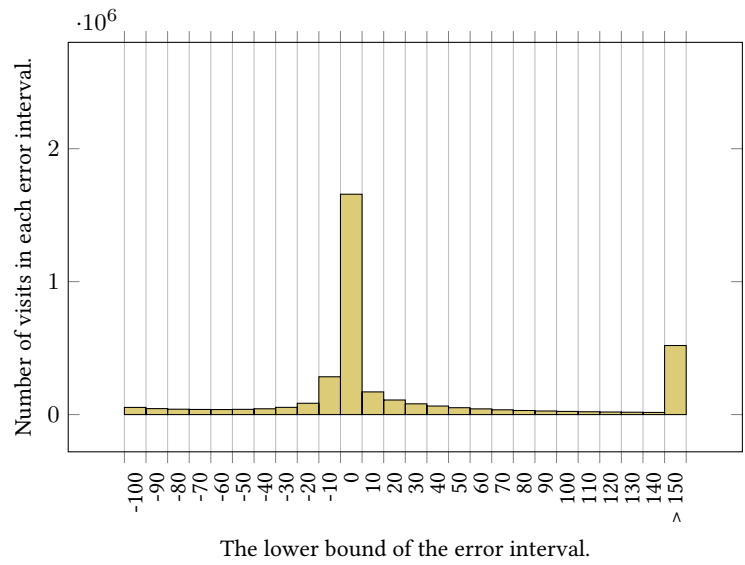
\begin{figure}
\begin{tikzpicture}
\begin{axis}
    [
    ybar interval,
    xtick={-100, -90, -80, -70, -60, -50, -40, -30, -20, -10, 0, 10, 20, 30, 40, 50, 60, 70, 80, 90, 100, 110, 120, 130, 140, 150, 160},
    ymax=2800000,
    xticklabels={-100, -90, -80, -70, -60, -50, -40, -30, -20, -10, 0, 10, 20, 30, 40, 50, 60, 70, 80, 90, 100, 110, 120, 130, 140, > 150},
    x tick label style={rotate=90,},
    xlabel = {The lower bound of the error interval.},
    ylabel = {Number of visits in each error interval.},
    width=300pt,
    height=200pt,
    ] 
    \addplot [draw=black, fill=cb4] coordinates {
    (-100, 53957) (-90, 44958) (-80, 40462) (-70, 38574) (-60, 37909) (-50, 39425) (-40, 43689) (-30, 54420) (-20, 85131) (-10, 284432) (0, 1657650) (10, 170641) (20, 109713) (30, 80962) (40, 64326) (50, 51713) (60, 42948) (70, 35850) (80, 30690) (90, 27071) (100, 23802) (110, 21227) (120, 19361) (130, 17736) (140, 16185)
    (150, 519623) (160, 0)
    };
\end{axis}
\end{tikzpicture}
\caption{Histogram of difference between WebScience's measure of attention and a measure based on load interval. }
\label{fig:eval-attn-ws-ttnl-histo}
\end{figure}

\begin{figure}
\begin{tikzpicture}
\begin{axis}
    [
    ybar interval,
    xtick={-100, -90, -80, -70, -60, -50, -40, -30, -20, -10, 0, 10, 20, 30, 40, 50, 60, 70, 80, 90, 100, 110, 120, 130, 140, 150, 160},
    ymax=2800000,
    xticklabels={-100, -90, -80, -70, -60, -50, -40, -30, -20, -10, 0, 10, 20, 30, 40, 50, 60, 70, 80, 90, 100, 110, 120, 130, 140, > 150},
    x tick label style={rotate=90,},
    xlabel = {The lower bound of the error interval.},
    ylabel = {Number of visits in each error interval.},
    width=300pt,
    height=200pt,
    ] 
    \addplot [draw=black, fill=cb5] coordinates {
    (-100, 55407) (-90, 45844) (-80, 45506) (-70, 50124) (-60, 59954) (-50, 79234) (-40, 115556) (-30, 188511) (-20, 364016) (-10, 1307181) (0, 1367883) (10, 58320) (20, 37925) (30, 27241) (40, 21352) (50, 17453) (60, 14414) (70, 12184) (80, 10456) (90, 9238) (100, 8282) (110, 7614) (120, 6521) (130, 6191) (140, 5559)
    (150, 162712) (160, 0)
    };
\end{axis}
\end{tikzpicture}
\caption{Histogram of difference between WebScience's measure of attention and a measure based on a simple attention model. }
\label{fig:eval-attn-ws-sa-histo}
\end{figure}
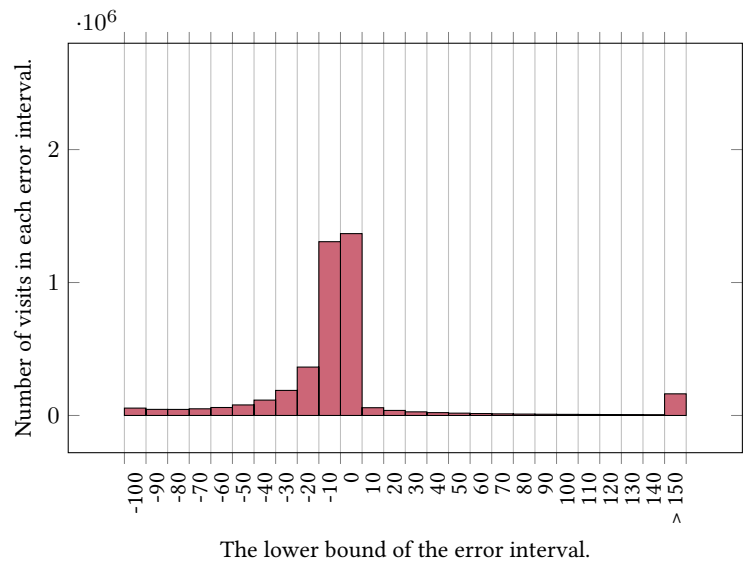

\Cref{fig:eval-attn-ws-dt-histo,fig:eval-attn-ws-ttnl-histo,fig:eval-attn-ws-sa-histo} show the percent
difference ($d$) each method had, compared to WebScience. 

$$ d_{p, m} = \frac{a_{p, WebScience} - a_{p, m}}{a_{p, WebScience}} * -100 $$

In all three cases, the distribution is centered around the zero mark: many page visits
result in the same attention measure under all four measures. However, all three distributions are long-tailed,
indicating that the methods sometimes report attention values that are far from our WebScience-based ground truth.

\end{document}